\documentclass[journal,keeplastbox]{IEEEtran}
\usepackage{subfigure}

\usepackage{afterpage}

\ifCLASSINFOpdf
\else
\fi
\usepackage{relsize}
\usepackage{enumitem}
 \setlist{leftmargin=10pt}

\usepackage{centernot}
\usepackage{cite}
\usepackage{amsfonts}
\usepackage{mathrsfs}
\usepackage{bbm}
\usepackage{pifont}
\newif\ifdraft
\draftfalse\usepackage{amsfonts}
\usepackage{amsmath, amsthm}
\usepackage{tabularx}
\usepackage{listings}
\usepackage{graphicx}
\usepackage{setspace}
\usepackage{float}
\usepackage{amssymb}
\usepackage{array}
\usepackage{fancyhdr}
\usepackage{cases}
 \usepackage{supertabular}
\usepackage{url}
\usepackage{fancyhdr}

\usepackage{color}
\usepackage{bbding}

\newcommand{\da}{\stackrel{\text{def}}{=}}
\newcommand{\de} {\textrm{d}\hspace{1pt}}

\newcommand{\bcup} {\hspace{2pt} \mathlarger{\cup}
\hspace{2pt}}

\newcommand{\expect}[1]{\bE{#1}}

\newcommand {\C} {{\rm I\kern-5.5pt C}}

\newcommand{\bE}[1]{{\mathbb{E}}\left[{#1}\right]}

\newcommand{\1}[1]{{\bf 1}\left[#1\right]}       

\newcommand{\fsquare}{\vrule height6pt width7pt depth1pt}   

\newcommand{\bd}[1]{{\mbox{\boldmath{$#1$}}}}

\def\centerhack#1{\hbox to 0pt{\hss\footnotesize #1\hss}}
\def\centerhackn#1{\hbox to 0pt{\hss #1\hss}}
\def\dchack#1{\vbox to 0pt{\vss{\hbox to 0pt{\hss#1\hss}}\vss}}

\newtheorem{lem}{Lemma}
\newtheorem{thm}{Theorem}

\newtheorem{rem}{Remark}
\newtheorem{cor}{Corollary}
\newtheorem{proposition}{Proposition}
\newtheorem*{proposition1.1}{Proposition 1.1}
\newtheorem*{proposition1.2}{Proposition 1.2}
\newtheorem*{proposition1.3}{Proposition 1.3}
\newtheorem*{proposition2.1}{Proposition 2.1}
\newtheorem*{proposition2.2}{Proposition 2.2}
\newcommand{\pf}{{\hfill \\ \bf Proof. \ }}           
\newcommand{\pfe}{\hfill\fsquare \\[0.1in]}             

\begin{document}


\title{\mbox{Preserving privacy enables ``co-existence equilibrium''} \mbox{of competitive diffusion in social networks}}








\author{~\\Jun~Zhao,~\IEEEmembership{Member,~IEEE,} and Junshan~Zhang,~\IEEEmembership{Fellow,~IEEE}
\thanks{The authors are with Arizona State University, Tempe, AZ 85281. Emails: junzhao@alumni.cmu.edu, junshan.zhang@asu.edu \protect\\ \indent This work was supported in part by the U.S. National Science Foundation (NSF) under Grants SaTC-1618768 and CNS-1422277, and in part by Army Research Office under Grant W911NF-16-1-0448.   }}

\maketitle 

 \begin{abstract}
 With the advent of social media,  different companies often promote \emph{competing} products simultaneously for word-of-mouth diffusion and adoption by users in  social networks. For such scenarios of \emph{competitive diffusion},  prior studies show that the weaker product will soon become
extinct (i.e., ``winner takes all''). It is intriguing to observe that  in practice, however,  competing products,  such as iPhone and Android phone, often co-exist in the market. This discrepancy may result from many factors such as the phenomenon that a user in the real world may not spread its use of a product due to dissatisfaction of the product or privacy protection. In this paper, we incorporate users' privacy for spreading behavior into competitive diffusion of two products and develop a problem formulation for \emph{\mbox{privacy-aware} competitive diffusion}. Then we prove that   privacy-preserving mechanisms can enable a ``co-existence equilibrium"  (i.e., two competing products co-exist in the equilibrium) in competitive diffusion over social networks. In addition
to the rigorous analysis, we also demonstrate our results with experiments
over real network topologies.


%
%

 \end{abstract}


%

%
%


\section{Introduction}

The advent of social media has generated tremendous interest in ways that companies use social networks to maximize product adoption by consumers~\cite{7451232,dinh2014cost,6517109}. One particular method of product promotion is viral marketing via
word-of-mouth effects \cite{trusov2009effects}. In word-of-mouth marketing applications, given users' limited attention span \cite{weng2013virality}, different companies often promote \emph{competing} products simultaneously in a social network using diffusion; i.e., there is a process of \emph{competitive diffusion} on the social network \cite{etesami2016complexity}.
For example, Apple may try to promote its new iPhone, while Samsung tries to advertise its new
Galaxy phone.

\begin{figure}[!t]
\begin{center}
  \includegraphics[width=.5\textwidth]{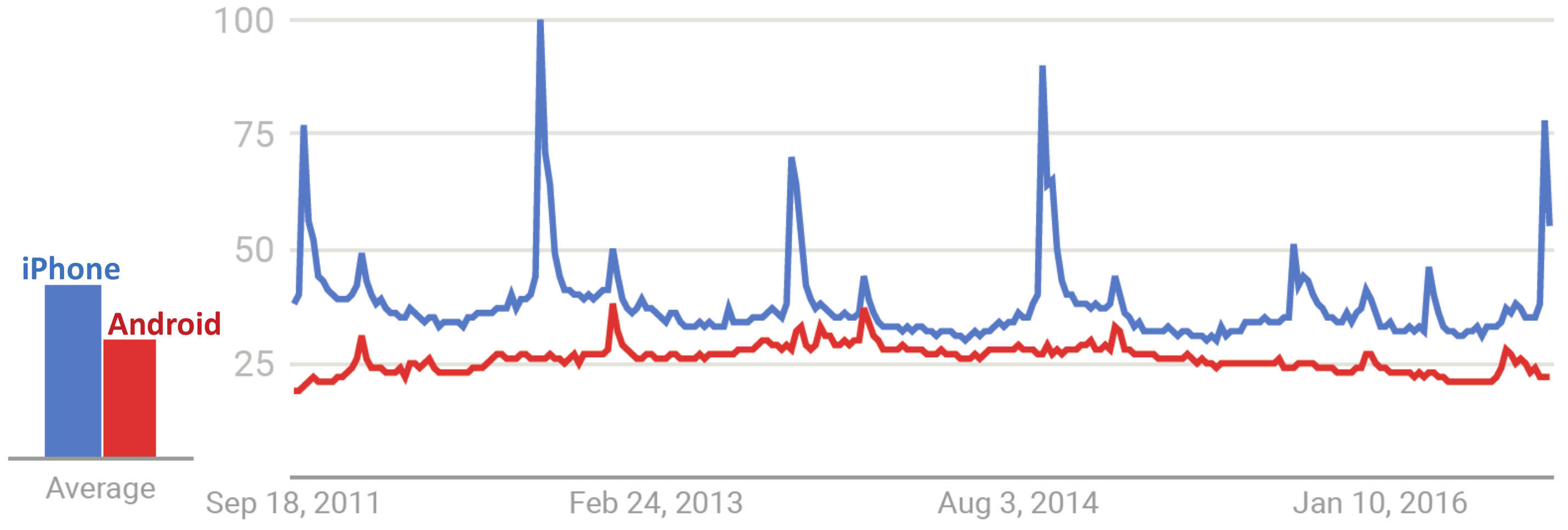}
  \caption{Global search-interests in
iPhone and Android over the past 5 years, generated from Google Trends$^{\text{\ref{footnote-google}}}$. Numbers represent search interest relative to the highest point on the chart. A value of 100 is the peak popularity for the term. A value of 50 means that the term is half as popular.}\label{fig-iphone}
  \end{center}
  \end{figure}

Recently, it has garnered much interest to study how competing products
spread in a social network \cite{bimpikis2016competitive,7451232,prakash2012winner,goyal2012competitive,apt2011diffusion,6517109}.
 In the diffusion of a pair of competing products, one significant result by Prakash \emph{et al.} \cite{prakash2012winner} shows that ``winner takes all'',
or, more accurately, the weaker product will soon become
extinct. A similar result is also given by Wei \emph{et al.} \cite{6517109}. The studies \cite{prakash2012winner,6517109} provide an insightful understanding of how competing products propagate among users via social networks.
It is intriguing to observe that  in practice, however,  competing products often co-exist in the market. For example, iPhone and Android phones both have a large number of users\footnote{\url{https://www.netmarketshare.com/operating-system-market-share.aspx}}. Figure~\ref{fig-iphone}, generated from Google Trends\footnote{\url{https://www.google.com/trends/explore?q=iphone,android} \label{footnote-google}}, shows the time evolution of worldwide search-interests in
iPhone and Android over the past 5 years. It is clear from the plot that  interests in
iPhone and Android co-exist in the market. We remark that this co-existence is the result of many real-world factors (e.g., companies can advertise the products directly to customers), so its analysis can be complex. Yet, the purpose of this example is to simply provide an intuitive understanding for the co-existence of competing products in the real world.

The ``winner-takes-all'' result by \cite{prakash2012winner,6517109} assumes that each user after adopting a product will positively spread the product to her social friends. In contrast, competing products co-exist
in the  real world, and one may attribute this co-existence to the fact that
a consumer using a product currently may be dissatisfied with the product and thus be unwilling to spread the product to her   friends. In fact, the user may even spread negative belief about the product that she is using \cite{luor2012tends}. Further, this can   be understood from a privacy perspective: a user may hesitate to disclose her product adoption due to privacy concerns so she may choose to not spread her product or even choose to spread the competing product with some probability. We now provide a simple example for illustration.
Consider  a couple's use of smart phones. The wife holds an iPhone, while the husband uses an Android phone. The husband feels disappointed about his  Android phone, and is told by the wife that she is happy with her iPhone. Hence, even when the husband still carries his  Android phone for some time and is not an iPhone user himself, he may recommend iPhone to his colleagues at work given the good experience of his wife. We can understand the husband's behavior from a privacy viewpoint: the husband would like to hide his use of an Android phone\footnote{The privacy concern here could happen owing to various reasons such as financial interests; for example, the husband may be doing some business with Apple so he would like to hide his use of an Android phone.} and thus recommend iPhone instead of Android to his colleagues.

Our model, building on the notion of privacy, is broad enough to represent the fact that a user may not spread her product adoption (even when the underlying reason is not because of privacy concern, but due to frustration of her product). Furthermore, users today are indeed concerned about privacy, especially for social networking where massive amounts of personal data are generated and   prone to adversarial attacks~\cite{prakash2012winner,heatherly2013preventing}.

Motivated to formally show that competing products can co-exist in the network, we  consider the  problem of \emph{\mbox{privacy-aware} competitive diffusion}, where two products  compete for  adoption by users
located in a social network. Individual users  are \mbox{privacy-aware} in the sense that an individual using a product may pretend adopting another product and spread the latter product to her social friends.


Our contributions can be
summarized as follows:
 \begin{itemize}
   \item We formulate a problem called \emph{\mbox{privacy-aware} competitive diffusion}, where products  compete for  adoption by \mbox{privacy-aware} users in a social network.
   \item We show that incorporating privacy into competitive diffusion can enable the co-existence of competing products (specifically, the probability of any user adopting any product is non-zero in the  equilibrium), while it is known that traditional (\mbox{privacy-oblivious}) competitive diffusion exhibits the ``winner-takes-all'' phenomenon.
 \end{itemize}

To provide more formal understanding, we present model details and main results below.

\textbf{\mbox{Models for privacy-aware competitive diffusion.}}  We consider the diffusion model of $\textrm{Susceptible-}  \textrm{Infected}_1  \textrm{-Infected}_2  \textrm{-Susceptible}$ (i.e., $S{I_1}{I_2}S$), which extends the widely studied ``flu-like'' Susceptible-Infected-Susceptible (i.e., $SIS$)
model from one product to the case of two competing products, so that each user in a social network can be in one of the following three
states: Susceptible (healthy), $I_1$ (using product 1), or $I_2$
(using product 2). Let $\sigma_1$ (resp., $\sigma_2$) denote the spreading strength (i.e., the infection rate divided by the healing rate) of product 1 (resp., product 2); more details can be found in Section \ref{sec-Traditional-Competitive-Diffusion} on Page \pageref{sec-Traditional-Competitive-Diffusion} later. A user's private data indicates which product she is using. To protect user's privacy, we consider:
\begin{itemize}
\item for $j\in\{1,2\}$, \textit{each user adopting product $j$} spreads product $1$ to its friends with probability $r_{j1}$,   spreads product $2$ to its friends with probability $r_{j2}$, and does not spread any product with probability $1-r_{j1}-r_{j2}$;
\item \textit{each user not yet adopting any product} does not advocate any product.
\end{itemize}

The above model is in the same spirit as the
randomized response technique \cite{warner1965randomized} that was introduced to protect users' privacy during survey interviews. We represent the privacy parameters by a matrix $
\begin{bmatrix}
    r_{11} & r_{12} & 1 - r_{11} -  r_{12} \\     r_{21} & r_{22}&  1 - r_{21} -  r_{22}
     \end{bmatrix}
$, and refer to this matrix as \vspace{1pt} the \emph{privacy scheme}. Note that we always require   each $r_{\bullet \bullet}$ in the matrix to be {strictly positive}, whereas $1-r_{j1}-r_{j2}$ can take $0$ or be greater than $0$. Formally, we always enforce
\begin{align}
\begin{Bmatrix}
 0 < r_{11} < 1,~~ &  ~~ 0 < r_{12} < 1, \\
 0 < r_{21} < 1,~~ &    ~~ 0 < r_{22} < 1, \\    r_{11} + r_{12} \leq 1,~~&~~    r_{21} + r_{22} \leq 1.
\end{Bmatrix} \label{rconditions}
\end{align}

A privacy scheme $
\begin{bmatrix}
    r_{11} & r_{12} & 1 - r_{11} -  r_{12} \\     r_{21} & r_{22}&  1 - r_{21} -  r_{22}
     \end{bmatrix}
$ is \textit{perfect} if $r_{11}=r_{21}$ and $r_{12}=r_{22}$ (i.e., the first row and the second row of the matrix are exactly the same). The intuition is that a user's private state, corresponding to adopting which product, is completely hidden from her spreading behavior. In a perfect privacy scheme, we write the same-valued $r_{11}$ and $r_{21}$ as $\gamma_{1}$, and write the same-valued $r_{12}$ and $r_{22}$ as $\gamma_{2}$ \vspace{1pt} so that the privacy scheme becomes $
\begin{bmatrix}
    \gamma_{1} & \gamma_{2} & 1 - \gamma_{1} -  \gamma_{2} \\     \gamma_{1} & \gamma_{2}&  1 - \gamma_{1} -  \gamma_{2}
     \end{bmatrix}
$ and the condition (\ref{rconditions}) becomes
\begin{align}
\begin{Bmatrix}
 0 < \gamma_{1} < 1,~~~~~~~   0 < \gamma_{2} < 1 ,~~~~~~~    \gamma_{1} + \gamma_{2} \leq 1.
\end{Bmatrix}. \label{rconditions-perfect}
\end{align}
We always enforce (\ref{rconditions-perfect}) for a perfect privacy scheme.

\textbf{Main results.} We summarize the main results as \ding{172} and \ding{173} below, and provide more details later. We emphasize that all results are for undirected social networks and the  $S{I_1}{I_2}S$ model described above. The conclusions can be different for other models.
\begin{itemize}
\item[\ding{172}] \textbf{For a connected social network with an \textit{\mbox{arbitrary}} topology, there exist {perfect privacy} schemes $
\begin{bmatrix}
    \gamma_{1}~& \gamma_{2}~ & 1 - \gamma_{1} -  \gamma_{2} \\     \gamma_{1}~ & \gamma_{2}~&  1 - \gamma_{1} -  \gamma_{2}
     \end{bmatrix}
$ satisfying  (\ref{rconditions-perfect}) to enable a \mbox{co-existence}  equilibrium of two competing products.}
\item[\ding{173}] \textbf{For a social network with a {complete graph}\footnote{A complete graph is an undirected graph in which any two nodes have an edge in between.} topology, there exist \textit{general} privacy schemes $
\begin{bmatrix}
    r_{11} & r_{12} & 1 - r_{11} -  r_{12} \\     r_{21} & r_{22}&  1 - r_{21} -  r_{22}
     \end{bmatrix}
$   satisfying (\ref{rconditions}) to enable a co-existence equilibrium of two competing products.}
\end{itemize}

Comparing \ding{172} and \ding{173} above, we see that \ding{172} considers a more general network topology while requiring a more restrictive privacy scheme, whereas \ding{173} considers a more general privacy scheme at the sacrifice of a more restrictive network topology. Below we explain  \ding{172} and \ding{173}, respectively.

\textbf{Further remarks on Result \ding{172}.}
We can formally state the above result \ding{172} as follows.  For a connected social network with an \textit{arbitrary} topology, let $\lambda$ be the largest eigenvalue of the adjacency matrix. Then \ding{172} means that under a privacy scheme $
\begin{bmatrix}
    \gamma_{1} & \gamma_{2} & 1 - \gamma_{1} -  \gamma_{2} \\     \gamma_{1} & \gamma_{2}&  1 - \gamma_{1} -  \gamma_{2}
     \end{bmatrix}
$ satisfying (\ref{rconditions-perfect}), if
\begin{align} \sigma_1 \gamma_{1}+\sigma_2 \gamma_{2}>1/\lambda,\label{completegraph-equi-condition-lambda-repeat-thmeq-introsb}
\end{align}   the system will reach a stable equilibrium where both products co-exist.  More specifically, in this equilibrium, the probability of any user adopting any product is positive. In contrast, if the sign ``$>$'' in the condition (\ref{completegraph-equi-condition-lambda-repeat-thmeq-introsb}) is replaced by ``$<$'', the system will reach a stable equilibrium where both products die out.


\textbf{Further remarks on Result \ding{173}.} We now explain the above result \ding{173}. Result \ding{173} applies to general privacy schemes and hence  overcomes the limitation of \ding{172} which requires privacy schemes to be perfect. However,
Result \ding{173} is only for complete graph topologies, while \ding{172} addresses arbitrary topologies. Formally, we have the following for result \ding{173}: for a social network of $n$ nodes with a complete graph topology, under a privacy scheme $
\begin{bmatrix}
    r_{11} & r_{12} & 1 - r_{11} -  r_{12} \\     r_{21} & r_{22}&  1 - r_{21} -  r_{22}
     \end{bmatrix}
$, if  \begin{align}
\frac{1}{2}\left[\begin{array}{l}
	\sigma_1 r_{11} + \sigma_2 r_{22} \\[2pt] + \sqrt{(\sigma_1 r_{11}-\sigma_2 r_{22})^2 + 4 \sigma_1 \sigma_2 r_{12} r_{21}}
\end{array}\right]
  >  \frac{1}{n-1}, \label{completegraph-equi-condition-lambda-repeat-thmeq-intro}
\end{align}
then the system will reach a stable {co-existence equilibrium}; if the sign ``$>$'' in the condition (\ref{completegraph-equi-condition-lambda-repeat-thmeq-intro}) is replaced by ``$<$'', the system will reach a stable equilibrium where both products die out. Compared with $1/\lambda$ on the right hand side of (\ref{completegraph-equi-condition-lambda-repeat-thmeq-introsb}), the right hand side of (\ref{completegraph-equi-condition-lambda-repeat-thmeq-intro}) is $\frac{1}{n-1}$ since the largest eigenvalue of the adjacency matrix for a complete graph of $n$ nodes is $n-1$, as explained in Footnote \ref{footnotelambdan1} later on Page \pageref{footnotelambdan1}.


Note that the above results all involve the largest eigenvalue of the adjacency matrix of the underlying social network. This comes from the eigenvalue-based approach of analyzing stable equilibria of dynamical systems discussed below.

\textbf{Technical approach.} \label{para-approach} We use the standard approach of analyzing stable equilibria of dynamical systems \cite{encyclopaedia} (Nevertheless, the analysis is still challenging as discussed in the next paragraph). With $\boldsymbol{x}$ denoting the state vector which contains the probability of each user adopting each product, we have a \textit{dynamical system} comprising differential equations in the form of
$\overset{\boldsymbol{\mathlarger{.}}}{\boldsymbol{x}} = G(\boldsymbol{x}, \boldsymbol{\alpha})$, where $\overset{\boldsymbol{\mathlarger{.}}}{\boldsymbol{x}}$ is the derivative of $\boldsymbol{x}$ (with respect to the time $t$), and $\boldsymbol{\alpha}$ is the parameter vector. Then the equilibria are defined to have a zero derivative and thus are obtained by solving  $G(\boldsymbol{x}, \boldsymbol{\alpha}) = \boldsymbol{0}$. Afterwards, an  equilibrium $\boldsymbol{x}_*$ is stable if each eigenvalue of the Jacobian matrix at $\boldsymbol{x}_*$ has a strictly negative real part, where the Jacobian matrix is an $m\times m$ matrix with the element in the $i$th row and $j$th column being $\frac{\partial\overset{\boldsymbol{\mathlarger{.}}}{{x_i}}}{\partial x_j}$, if the state vector $\boldsymbol{x}$ has $m$ dimensions: $x_1,x_2,\ldots,x_m$ (i.e., $\boldsymbol{x}=[x_1,x_2,\ldots,x_m]$). Finally, if the goal is to show that the system will reach a stable equilibrium where both products \textit{co-exist}, we prove that there is only one stable equilibrium and this equilibrium gives a state vector whose elements are all strictly positive. If the goal is to show that the system will reach a stable equilibrium where both products \textit{die out}, we prove that there is only one stable equilibrium and this equilibrium gives a state vector \mbox{whose elements are all zero.}


\textbf{Challenge.} Although the above approach is standard, the analysis is   nontrivial. One main difficulty lies in analyzing the Jacobian matrix and its eigenvalues to show the stability of an equilibrium. In our \mbox{privacy-aware} competitive diffusion problem, the Jacobian matrix has many non-zero elements, making it difficult to evaluate its eigenvalues. In our setting with arbitrary network topologies, dealing with the $2n \times 2n$ Jacobian matrix further complicates the analysis. In prior \mbox{privacy-oblivious} competitive diffusion studies, the Jacobian matrix has many zero elements and   is much easier to tackle.


%


The rest of the paper is organized as follows. Section \ref{section:Related} surveys related work.
Then we present the system model in Section \ref{section:SystemModel}, and the main results in Section \ref{section:Results}. Afterwards, we provide in Section \ref{section:Experiments} experiments to confirm the analytical results. In Sections \ref{section:Preliminaries}, we discuss the proof ideas for establishing the main results. We conclude the paper in Section~\ref{sec:Conclusion}. Many technical details are given in the appendices.

\section{Related Work}\label{section:Related}

\textbf{Single-meme diffusion.} Numerous studies have addressed  single-meme diffusion,  where the notion of meme is a generic term for the information propagating on the network. For the diffusion model of ``flu-like'' Susceptible-Infected-Susceptible (i.e., $SIS$), Wang \emph{et al.} \cite{wang2003epidemic}
presented an approximate analysis and showed
that the
epidemic threshold in a network equals the reciprocal of
  the largest eigenvalue of the network's adjacency matrix. This result was formally proved by Van Mieghem \emph{et al.} \cite{van2009virus}. Van Mieghem \cite{van2012epidemic} further derived the expression for the steady-state fraction of infected
nodes. We refer interested readers to a survey \cite{guille2013information} for more related work on single-meme diffusion.

\textbf{Competitive diffusion.}
Prakash \emph{et al.} \cite{prakash2012winner} studied the $\textrm{Susceptible-}  \textrm{Infected}_1  \textrm{-Infected}_2  \textrm{-Susceptible}$ (i.e., $S{I_1}{I_2}S$) diffusion problem for   a pair of competing products and show that ``winner takes all'' (more accurately, the weaker product will soon become
extinct). Wei \emph{et al.} \cite{6517109} observed an analogous result in the setting where two products propagate on two different networks defined on the same set of users.  Fazeli~\emph{et~al.}~\cite{7451232} considered two companies competing to maximize the consumption of their products by users which decide   consumption  based on best response dynamics. The goal of~\cite{7451232} was to investigate whether a firm  with a limited budget should spend on the quality of the product or on initial seeding in
the network.
Bimpikis \textit{et al.} \cite{bimpikis2016competitive} also considered firms' optimal strategies under competitive diffusion and investigated their connections with  the underlying social
network structure. Alon~\cite{alon2010note} used a game to model the competition between many products (i.e., possibly more than two) for diffusion on  a social network, and obtained the  relation between the existence of pure Nash equilibria  and  the diameter of the network. Beutel \emph{et al.} \cite{beutel2012interacting} modified the $S{I_1}{I_2}S$ model to allow a user to be in both $I_1$ and $I_2$ states (note that we use the standard $S{I_1}{I_2}S$ model where a user can be in only one state), and showed that competing products can co-exist in the equilibrium. However, this nice work \cite{beutel2012interacting} had the following limitations: 1) its theoretical result was  only for  a complete graph (i.e.,  a full clique) although their experiments consider more general topologies; and 2) \cite{beutel2012interacting} lacked   a proof to show that the  equilibrium is   stable (proving the stability of an equilibrium is often much more challenging than finding an equilibrium~\cite{prakash2012winner}). {Compared with our work}, all the above reference \cite{prakash2012winner,6517109,7451232,bimpikis2016competitive,alon2010note} did not take users' privacy into consideration.

\textbf{Few recent studies on \mbox{privacy-aware} diffusion.} So far, there has been only few  work \cite{giakkoupis2015privacy,harranetoward,zhu2013prince} recently on \mbox{privacy-aware} diffusion. Giakkoupis \textit{et al.}
\cite{giakkoupis2015privacy} presented a distributed algorithm for \mbox{privacy-aware} information diffusion
in social networks, where each user's privacy means her opinion on the information and is modeled by  how likely she forwards the information (the information itself is not perturbed). If the user favors the information, she forwards the information with higher probability compared with the case where she dislikes the information. Giakkoupis \textit{et al.}
\cite{giakkoupis2015privacy} showed that the information spreads to a constant fraction of the network if it appeals to a constant fraction of users, and dies out if few users likes it. Harrane
\textit{et al.}
\cite{harranetoward} recently used \mbox{privacy-aware}   diffusion to solve distributed inference problems, where privacy means that each user's measurement is perturbed before being propagated to its neighbors. Zhu~\textit{et~al.}~\cite{zhu2013prince} considered private link exchange over social networks, where a user's private information is her friend list and she obfuscates the list before sending it to her neighbors to preserve privacy. Different from our work here, these recent studies \cite{giakkoupis2015privacy,harranetoward,zhu2013prince} did not consider \textit{competition} between different information in diffusion and their privacy signals were different from our privacy signal that captures which product the user adopts.


Notably, Krishnamurthy
 and Wills~\cite{krishnamurthy2009privacy} reported a longitudinal study of diffusion of users' private data.
Avgerou and Stamatiou \cite{avgerou2015privacy} developed a game theory framework to show that social networks help spread privacy consciousness to large populations.
Akcora \textit{et al.} \cite{akcora2012privacy} proposed a   measure to quantify the risk of  disclosing private information in social networks. Heatherly \textit{et al.} \cite{heatherly2013preventing} investigated  how to utilize released social networking data to infer undisclosed private
information about users. Baden \textit{et al.} \cite{baden2009persona} presented an online social network called Persona where users' privacy can be protected via cryptographic techniques.

\section{System Model} \label{section:SystemModel}

%
In what follows, we present the system model  in detail. In Section \ref{sec-Traditional-Competitive-Diffusion}, we review traditional competitive diffusion, where user privacy is not considered. In Section \ref{sec-Privacy-Competitive-Diffusion}, we incorporate user privacy into competitive diffusion to introduce the novel problem of \emph{\mbox{privacy-aware} competitive diffusion}. Since one may tempt to convert \mbox{privacy-aware} competitive diffusion into traditional (\mbox{privacy-oblivious}) competitive diffusion, we will elaborate on why this approach does not work at the end of this section.

%

\subsection{Traditional (Privacy-Oblivious) Competitive Diffusion} \label{sec-Traditional-Competitive-Diffusion}

We review traditional  competitive diffusion that does not take into consideration of user privacy. In particular, we look at the scenario where two competing products spread on
a social network according to the following diffusion model. The diffusion model $S{I_1}{I_2}S$ extends the widely studied ``flu-like'' $SIS$ model from one product to the case of two products. Each user in the network can be in one of the following three
states: Susceptible (healthy), $I_1$ (adopting product 1), or $I_2$
(adopting product 2).

We now describe several parameters of the $S{I_1}{I_2}S$ model. For convenience, we still use the words ``infection'' and ``healing'' as in the case of disease diffusion, although the diffused information is product adoption.

\textbf{Infection rates: $\beta_1$ and $\beta_2 $.} A healthy user
gets infected by  her infected neighbors, and the infection rate of product 1 (resp., product 2)
is denoted by $\beta_1$ (resp., $\beta_2 $). Specifically, an infected
user transmits her adoption of product 1 (resp., product 2) to each of its neighbors {independently} at rate $\beta_1$ (resp., $\beta_2 $). In other words, the time taken for each
infected user to spread product 1 (resp., product 2) to a neighbor is exponentially distributed with parameter $\beta_1$ (resp., $\beta_2 $).

\textbf{Healing rates: $\delta_1$ and $\delta_2 $. }If a user is in state
$I_1$ (resp., $I_2 $), she recovers on her own with rate $\delta_1$ (resp., $\delta_2 $). In other words, the time taken for a  user infected by product 1 (resp., product 2) to heal
is exponentially distributed with parameter $\delta_1$ (resp., $\delta_2 $). After a user is healed, it can get infected by product 1 or product 2 again.

Note that the infection rates $\beta_1$ and $\beta_2 $, and the healing rates $\delta_1$ and $\delta_2$ are common across all users in the network; i.e., all rates are homogeneous with respect to users. This is an assumption widely made in many diffusion studies   \cite{prakash2012winner,6517109,7451232,bimpikis2016competitive,beutel2012interacting} to have tractable analyses. Removing this assumption to consider the heterogeneity of the infection and healing rates across users will be an interesting direction, yet the analysis may become intractable.

Dividing the infection rate by the healing rate, we obtain the spreading strengths of products 1 and 2 given by
$\sigma_1  \da \frac{\beta_1}{\delta_1}$ and $\sigma_2  \da \frac{\beta_2}{\delta_2}$, respectively. In the network, each product competes with the other product
for healthy victims. We also assume full
mutual immunity: a user can not be infected by two products at the same time.

\subsection{Privacy-Aware Competitive Diffusion} \label{sec-Privacy-Competitive-Diffusion}

We now present our model for \emph{\mbox{privacy-aware} competitive diffusion}. Two products  compete for  adoption by users
  in a connected social network. A user's private data indicates which product she is using. Users are \mbox{privacy-aware} in the sense that an user using a product may pretend adopting another product and spread the latter product to her social friends. We thus incorporate privacy into the above $S{I_1}{I_2}S$ diffusion problem as follows.

The technique here to preserve users' privacy is   conceptually similar to the
randomized response technique \cite{warner1965randomized},  introduced first
for survey interviews.
For each user  adopting product $1$ (i.e., in state $I_1$), her action has the following three possibilities: (i) she spreads product $1$ to her neighbors with probability $r_{11}$;  (ii) she pretends that she is using product $2$ and diffuses product $2$ to her neighbors with probability $r_{12}$; (iii) she does not spread any product with probability $1-r_{11}-r_{12}$. Similarly, for a user adopting product $2$ (i.e., in state $I_2$), her action has the following three possibilities: (i) she spreads product $2$ to her neighbors with probability $r_{22}$; (ii) she pretends that she is using product $1$ and disseminates product $1$ to her neighbors with probability $r_{21}$; (iii) she does not spread any product with probability $1-r_{21}-r_{22}$. Finally, for a user that has not yet adopted any product, she does not   advocate any product.

We   represent the privacy parameters by a matrix $
\begin{bmatrix}
    r_{11} & r_{12} & 1 - r_{11} -  r_{12} \\     r_{21} & r_{22}&  1 - r_{21} -  r_{22}
     \end{bmatrix}
$, and refer to this matrix as \vspace{1pt} the \emph{privacy scheme}. When $r_{11} = 1$, $r_{12} = 0$, $r_{21} = 0$, and $r_{22} = 1$, the problem reduces to the traditional (i.e., \mbox{privacy-oblivious}) competitive diffusion problem $S{I_1}{I_2}S$ studied by Prakash~\emph{et~al.}~\cite{prakash2012winner}. In the rest of paper, we focus on \mbox{privacy-aware} competitive diffusion and thus assume that $r_{11}$, $r_{12}$, $r_{21}$, and $r_{22}$ are strictly positive. More formally, we always enforce the condition (\ref{rconditions}) given on Page \pageref{rconditions}; i.e., \mbox{$r_{11},r_{12},r_{21},r_{22}\in(0,1)$} and  $r_{11} + r_{12} \leq 1$,~$  r_{21} + r_{22} \leq 1$.

In the literature, a widely accepted privacy definition is the renowned notion of \emph{differential privacy} \cite{Dwork2006,dwork2014algorithmic,bun2017make,ddp}. Specifically, $\epsilon$-differential privacy for a mechanism that provides a randomized answer for a query on a database means that if a single record changes in the database, then the probability that the same answer is given differs by at most a multiplicative factor of $e^{\epsilon}$ (smaller $\epsilon$ means better privacy). In our privacy-aware competitive diffusion problem, each user's private signal is her state (which is one of the three states: ${I_1}$, ${I_2}$, and $S$) and the randomized output is the state that corresponds to the product that she advocates.
More relevant to our setting is the local model of differential
privacy \cite{duchi2013local,wang2016value}, where each individual maintains its
own data (a database of size 1) and  answers only the question about the data
in a differentially private manner.


Built on the above,  the \emph{privacy-preserving mechanism} $
\begin{bmatrix}
    r_{11} & r_{12} & 1 - r_{11} -  r_{12} \\     r_{21} & r_{22}&  1 - r_{21} -  r_{22}
     \end{bmatrix}
$ in competitive diffusion achieves $\epsilon$-(local) differential privacy for $\epsilon$   given by
\begin{align}
\epsilon : = \ln \max \left\{ \begin{array}{l}
	\mathlarger{\frac{r_{11}}{r_{21}},~ \frac{r_{21}}{r_{11}}, ~\frac{r_{12}}{r_{22}}, ~ \frac{r_{22}}{r_{12}}},\\[6pt] \mathlarger{\frac{1 - r_{11} -  r_{12}}{1 - r_{21} -  r_{22}},~\frac{1 - r_{21} -  r_{22}}{1 - r_{11} -  r_{12}}}
\end{array} \right\} .\nonumber
\end{align}
If $r_{11} = r_{21}$ and $r_{12} = r_{22}$, then $\epsilon = 0$; i.e., the scheme has \emph{perfect privacy} if the first row and the second row of the privacy matrix are the same. In such case, user's spreading behavior is   independent of her private state.
 %

\textbf{Problem Statement.} We now state the research problem of \mbox{privacy-aware} competitive diffusion. Given a connected social network and  the diffusion parameters (the infection rate $\beta_1$ and the healing rate $\delta_1$ for product 1, and the infection rate $\beta_2$ and the healing rate $\delta_2$ for product 2), our goal is to find a privacy scheme $
\begin{bmatrix}
    r_{11} & r_{12} & 1 - r_{11} -  r_{12} \\     r_{21} & r_{22}&  1 - r_{21} -  r_{22}
     \end{bmatrix}
$ \vspace{1pt} such that competing products co-exist in the stable equilibrium.

\textbf{Answer.} Our answer to the above problem is different for different network topologies. Specifically, for a connected social network with an \textit{arbitrary} topology, \vspace{2pt} we consider \textit{perfect privacy} schemes $
\begin{bmatrix}
    \gamma_{1} & \gamma_{2} & 1 - \gamma_{1} -  \gamma_{2} \\     \gamma_{1} & \gamma_{2}&  1 - \gamma_{1} -  \gamma_{2}
     \end{bmatrix}
$; \vspace{1.5pt} for a social network with a  \textit{complete graph} topology, we consider \textit{general} privacy schemes $
\begin{bmatrix}
    r_{11} & r_{12} & 1 - r_{11} -  r_{12} \\     r_{21} & r_{22}&  1 - r_{21} -  r_{22}
     \end{bmatrix}
$.

\textbf{Privacy-aware competitive diffusion cannot be converted into traditional (\mbox{privacy-oblivious}) competitive diffusion.}
One may tempt to convert \mbox{privacy-aware} competitive diffusion into traditional (\mbox{privacy-oblivious}) competitive diffusion. For example, one may consider a \mbox{privacy-oblivious} diffusion problem $S{I_1}{I_2}S$, where product 1 has an infection rate of $\beta_1 r_{11} + \beta_2 r_{21}$ and a healing rate of $\delta_1$, and product 2 has an infection rate of $\beta_1 r_{12} + \beta_2 r_{22}$ and a healing rate of $\delta_2$, and treat this problem as an equivalent of our \mbox{privacy-aware}  diffusion problem $S{I_1}{I_2}S$. This approach does not work as explained below. In the \mbox{privacy-oblivious} diffusion problem, if at some point a product dies out, then the product becomes extinct forever and cannot be reborn again. However, in our \mbox{privacy-aware}  diffusion problem, even if at some point an product is wiped out,   the product may revive again, because some users adopting the other product may pretend using this product and spread this product to its neighbors, which can resurrect the product.



\section{Main Results}\label{section:Results}





%
%


We present the main results in this section.
We present conditions for a stable co-existence equilibrium in Section \ref{sec-co-exist}, and conditions for no stable co-existence equilibrium in Section \ref{sec-co-exist-no}. The notion of a \textit{{stable co-existence equilibrium}} means that competing products co-exist in a equilibrium which is  also stable (i.e., attracting), and the co-existence is in the strong sense: the probability of any user adopting any product is positive (i.e., greater than $0$) in the stable equilibrium.

\subsection{Conditions for a Stable {Co-Existence Equilibrium}} \label{sec-co-exist}

To have a stable co-existence equilibrium, we discuss in Theorem \ref{thm:main:generalgraph-aux} privacy schemes on a connected social network with an \emph{arbitrary} topology, and present in Theorem \ref{thm:main:compete-graph} general privacy schemes on a social network with a \emph{complete graph} topology.

\begin{thm}[\textbf{Perfect Privacy Schemes on \textit{Arbitrary} Network Topologies}] \label{thm:main:generalgraph-aux}
Consider a connected social network with an \textbf{\mbox{arbitrary}} topology, with $\lambda$ denoting the largest eigenvalue of the adjacency matrix. For the \mbox{privacy-aware} $SI_1I_2S$ problem with a perfect privacy \text{\rm scheme} $
\begin{bmatrix}
    \gamma_{1} & \gamma_{2} & 1 - \gamma_{1} -  \gamma_{2} \\     \gamma_{1} & \gamma_{2}&  1 - \gamma_{1} -  \gamma_{2}
     \end{bmatrix}
$ \vspace{1pt} \text{\rm satisfying} $\gamma_1,\gamma_2\in(0,1)$ and $\gamma_{1} + \gamma_{2} \leq 1$,
\text{\rm if}
\begin{align}
\sigma_1 \gamma_1+\sigma_2 \gamma_2>1/\lambda, \label{betastardeltastarlambda2-thmeq}
\end{align}
then the system will reach a stable {co-existence equilibrium}.
\end{thm}

\begin{rem}  \label{thm:main:generalgraph-aux-condv2}
In Remark \ref{rem-thm3-condv2} of Theorem \ref{thm:main:generalgraph-aux-partii} later, we will show that if the sign ``$>$'' in the condition (\ref{betastardeltastarlambda2-thmeq}) is replaced by ``$<$'', the system will reach a stable equilibrium where both products die out.
\end{rem}

We explain the basic ideas and the corresponding challenges to establish Theorem \ref{thm:main:generalgraph-aux} in Section \ref{sec-ideas-thm:main:generalgraph-aux} on  Page \pageref{sec-ideas-thm:main:generalgraph-aux}. The proof details are given in Appendix \ref{sec:Establishing:thm:main:generalgraphC-aux} on Page \pageref{sec:Establishing:thm:main:generalgraphC-aux}.

\begin{thm}[\textbf{\textit{General} Privacy Schemes on Complete Network Topologies}] \label{thm:main:compete-graph}
Consider a social network of $n$ nodes with a complete graph topology. For the \mbox{privacy-aware} $SI_1I_2S$ problem with a \textbf{general} privacy scheme $
\begin{bmatrix}
    r_{11} & \hspace{-3pt} r_{12} & \hspace{-3pt}1 - r_{11} -  r_{12} \\     r_{21} & \hspace{-3pt}r_{22}& \hspace{-3pt} 1 - r_{21} -  r_{22}
     \end{bmatrix}
$ satisfying \mbox{$r_{11},r_{12},r_{21},r_{22}\in(0,1)$}\\and  $r_{11} + r_{12} \leq 1$,~$  r_{21} + r_{22} \leq 1$, if  \begin{align} \frac{1}{2}\left[\begin{array}{l}
	\sigma_1 r_{11} + \sigma_2 r_{22} \\[2pt] + \sqrt{(\sigma_1 r_{11}-\sigma_2 r_{22})^2 + 4 \sigma_1 \sigma_2 r_{12} r_{21}}
\end{array}\right]
  >  \frac{1}{n-1},  \label{completegraph-equi-condition-lambda-repeat-thmeq}
\end{align}
then the system will reach a stable {co-existence equilibrium}.
\end{thm}

\begin{rem}  \label{thm:main:compete-graph-condv2}
In Theorem \ref{thm:main:generalgraph-aux-partii} later, we will show that if the sign ``$>$'' in the condition (\ref{completegraph-equi-condition-lambda-repeat-thmeq}) is replaced by ``$<$'', the system will reach a stable equilibrium where both products die out.
\end{rem}

We explain the basic ideas and the corresponding challenges to establish Theorem \ref{thm:main:compete-graph} in Section \ref{sec-ideas-thm:main:generalgraph-aux} on  Page \pageref{sec-ideas-thm:main:generalgraph-aux}. The proof details are given in Appendix D of the full version \cite{fullversion}.


\textbf{Comparing Theorems \ref{thm:main:generalgraph-aux} and \ref{thm:main:compete-graph}.} Although Theorem \ref{thm:main:generalgraph-aux} considers only perfect privacy schemes, but it applies to arbitrary network topologies. In contrast, Theorem  \ref{thm:main:compete-graph} considers general privacy schemes at the sacrifice of requiring network topologies to be complete graphs. The special case addressed by both Theorems \ref{thm:main:generalgraph-aux} and \ref{thm:main:compete-graph} is perfect privacy schemes on complete graphs. For this special case, Theorems \ref{thm:main:generalgraph-aux} and \ref{thm:main:compete-graph} are consistent, as explained below.

We first apply Theorem  \ref{thm:main:generalgraph-aux} to complete graphs.
For a complete graph of $n$ nodes, the largest eigenvalue of the adjacency matrix is $n-1$, as explained in the footnote here\footnote{With $\boldsymbol{I}_n$ denoting the $n \times n$ identity matrix (i.e., unit matrix) and $\boldsymbol{J}_n$ denoting the $n \times n$ matrix whose elements are all $1$, then the adjacency matrix for a complete graph of $n$ nodes is $\boldsymbol{J}_n-\boldsymbol{I}_n$, and it is straightforward to check its eigenvalues are $n-1$ and  $−1$ (of multiplicity $n-1$).\label{footnotelambdan1}}.  Hence, if we apply Theorem  \ref{thm:main:generalgraph-aux} to complete graphs, $\lambda$ in the condition
(\ref{betastardeltastarlambda2-thmeq}) becomes $n-1$ so that (\ref{betastardeltastarlambda2-thmeq}) becomes $\sigma_1 \gamma_1+\sigma_2 \gamma_2>\frac{1}{n-1}$.

We now apply Theorem \ref{thm:main:compete-graph} to perfect privacy schemes. Under a perfect privacy \text{\rm scheme} $
\begin{bmatrix}
    \gamma_{1} & \gamma_{2} & 1 - \gamma_{1} -  \gamma_{2} \\     \gamma_{1} & \gamma_{2}&  1 - \gamma_{1} -  \gamma_{2}
     \end{bmatrix}
$, we substitute
$r_{11}=r_{21} = \gamma_1$ and $r_{12}=r_{22} = \gamma_2$ into (\ref{completegraph-equi-condition-lambda-repeat-thmeq}) so that \begin{align}
&\textstyle{\frac{1}{2}\big[\sigma_1 r_{11} + \sigma_2 r_{22} + \sqrt{(\sigma_1 r_{11}-\sigma_2 r_{22})^2 + 4 \sigma_1 \sigma_2 r_{12} r_{21}}\hspace{1.3pt}\big] }\nonumber \\ & =\textstyle{\frac{1}{2}\big[\sigma_1 \gamma_1 + \sigma_2 \gamma_2 + \sqrt{(\sigma_1 \gamma_1-\sigma_2 \gamma_2)^2 + 4 \sigma_1 \sigma_2 \gamma_2 \gamma_1}\hspace{1.3pt}\big]\nonumber} \\ & =\sigma_1 \gamma_1 + \sigma_2 \gamma_2,\label{conditions-transb}
\end{align}
which converts (\ref{completegraph-equi-condition-lambda-repeat-thmeq}) into $\sigma_1 \gamma_1+\sigma_2 \gamma_2>\frac{1}{n-1}$.

From the above, applying Theorem  \ref{thm:main:generalgraph-aux} (with perfect privacy schemes) to complete graphs and applying Theorem \ref{thm:main:compete-graph} (with complete graphs) to perfect privacy schemes give consistent results. In addition, the case addressed by Theorem  \ref{thm:main:generalgraph-aux} but not Theorem \ref{thm:main:compete-graph} is incomplete graphs with perfect privacy schemes, and the case addressed by Theorem \ref{thm:main:compete-graph} but not Theorem  \ref{thm:main:generalgraph-aux} is complete graphs with imperfect privacy schemes.





We further obtain Remark \ref{rem-thm1-cond}  from Theorem \ref{thm:main:generalgraph-aux} by discussing the relationships between the spreading strengths $\sigma_1,\sigma_2$ and the threshold $1/\lambda$. Remark \ref{rem-thm1-cond} shows that a stable co-existence equilibrium can be achieved under suitable privacy schemes if at least one product's strength is above the threshold $1/\lambda$.

\begin{rem}[\textbf{Discussing the conditions of $\sigma_1$ and $\sigma_2$ in Theorem \ref{thm:main:generalgraph-aux}}] \label{rem-thm1-cond}
In Theorem \ref{thm:main:generalgraph-aux}, the conditions of $\sigma_1$ and $\sigma_2$ for a stable co-existence equilibrium are given by
\begin{align}
\begin{Bmatrix}
\sigma_1 \gamma_1+\sigma_2 \gamma_2>1/\lambda, \\ 0 < \gamma_{1} < 1,~~~~~~~   0 < \gamma_{2} < 1 ,~~~~~~~    \gamma_{1} + \gamma_{2} \leq 1.
\end{Bmatrix}, \label{rconditions-perfect-newcond}
\end{align}

To have a feasible $(\gamma_1,\gamma_2)$ satisfying (\ref{rconditions-perfect-newcond}), at least one of $\sigma_1$ and $\sigma_2$ should be greater than $1/\lambda$. Then we have the following three cases. \\
(a)
Under $\sigma_1>1/\lambda$ and $\sigma_2>1/\lambda$, (\ref{rconditions-perfect-newcond}) is equivalent to
\begin{align}
\begin{Bmatrix}0<\gamma_{1}<1,~~ \max\big\{0,\frac{\lambda^{-1}-\sigma_1\gamma_{1}}{\sigma_2}\big\}<\gamma_{2}\leq 1 - \gamma_{1}.
\end{Bmatrix} \nonumber
\end{align}
(b)
Under $\sigma_1 >1/\lambda \geq \sigma_2$, (\ref{rconditions-perfect-newcond}) is equivalent to
\begin{align}
\begin{Bmatrix} \frac{\lambda^{-1}-\sigma_2}{\sigma_1-\sigma_2}<\gamma_{1}<1,~~ \max\big\{0,\frac{\lambda^{-1}-\sigma_1\gamma_{1}}{\sigma_2}\big\}<\gamma_{2}\leq 1 - \gamma_{1}.
\end{Bmatrix} \nonumber
\end{align}
(c)
Under $\sigma_2 >1/\lambda \geq \sigma_1$, (\ref{rconditions-perfect-newcond}) is equivalent to
\begin{align}
\begin{Bmatrix} 0<\gamma_{1}<\frac{\sigma_2-\lambda^{-1}}{\sigma_2-\sigma_1},~~ \frac{\lambda^{-1}-\sigma_1\gamma_{1}}{\sigma_2}<\gamma_{2}\leq 1 - \gamma_{1}.
\end{Bmatrix} \nonumber
\end{align}
\end{rem}
The above three cases of Remark \ref{rem-thm1-cond} are illustrated by Figure~\ref{gamma1gamma2} and will be proved in Appendix \ref{sec:Establishing:thm:main:generalgraphC} on Page \pageref{sec:Establishing:thm:main:generalgraphC}. In addition, in the scenario of $\sigma_1<1/\lambda$ and  $\sigma_2<1/\lambda$, which the above three cases do not cover, we will discuss in Remark \ref{rem-thm3-cond} after Theorem~\ref{thm:main:generalgraph-aux-partii} later that both products always die out under any privacy scheme.

\begin{figure}[!t]
\vspace{0mm} \centerline{\includegraphics[width=0.5\textwidth]{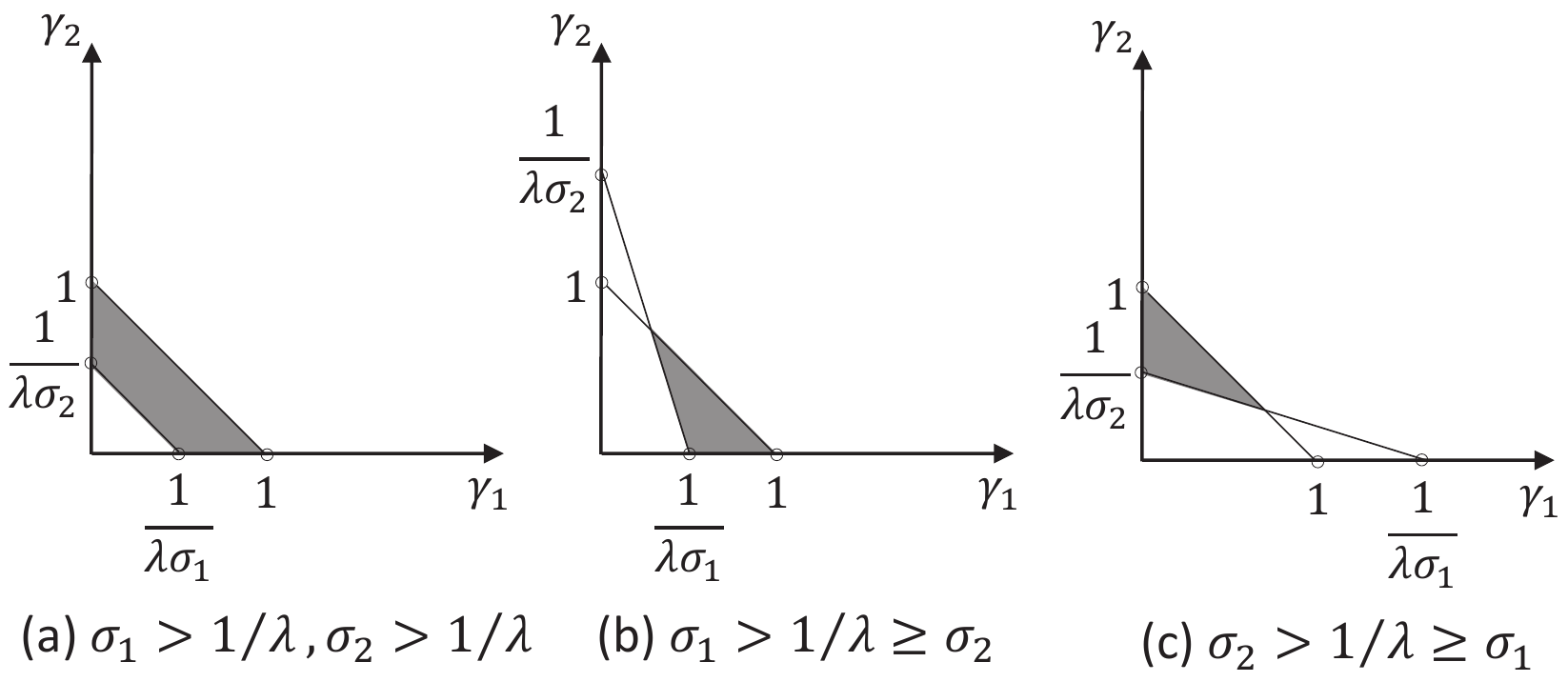}}
\caption{The shaded area in each subfigure presents the set of $(\gamma_1,\gamma_2)$ considered in Theorem \ref{thm:main:generalgraph-aux}.}
\label{gamma1gamma2}
\end{figure}

\subsection{Conditions for No Stable {Co-Existence Equilibrium}} \label{sec-co-exist-no}


\begin{thm} \label{thm:main:generalgraph-aux-partii}
Consider a connected social network with an \textbf{arbitrary} topology, with $\lambda$ denoting the largest eigenvalue of the adjacency matrix. For the \mbox{privacy-aware} $SI_1I_2S$ \vspace{1pt} problem with a privacy scheme $
\begin{bmatrix}
    r_{11} & r_{12} & 1 - r_{11} -  r_{12} \\     r_{21} & r_{22}&  1 - r_{21} -  r_{22}
     \end{bmatrix}
$, if  \begin{align}
\frac{1}{2}\left[\begin{array}{l}
	\sigma_1 r_{11} + \sigma_2 r_{22} \\[2pt] + \sqrt{(\sigma_1 r_{11}-\sigma_2 r_{22})^2 + 4 \sigma_1 \sigma_2 r_{12} r_{21}}
\end{array}\right]< \frac{1}{\lambda}  ,\label{equi-condition-lambda-repeat2}
\end{align}
the system will reach a stable equilibrium where both products die out (more specifically, the probability of any user adopting any product is zero in the stable equilibrium).
\end{thm}

We explain the basic ideas and the corresponding challenges to establish Theorem \ref{thm:main:generalgraph-aux-partii} in Section \ref{sec-ideas-thm:main:generalgraph-aux-partii} on  Page \pageref{sec-ideas-thm:main:generalgraph-aux-partii}. The proof details are given in Appendix  \ref{sec:Establishing:thm:main:generalgraphB-partII} on Page \pageref{sec:Establishing:thm:main:generalgraphB-partII}. Theorem \ref{thm:main:generalgraph-aux-partii} further implies Remark \ref{rem-thm3-cond} below.
\begin{rem}  \label{rem-thm3-cond}
If $\sigma_1<1/\lambda$ and  $\sigma_2<1/\lambda$, any privacy scheme $
\begin{bmatrix}
    r_{11} & r_{12} & 1 - r_{11} -  r_{12} \\     r_{21} & r_{22}&  1 - r_{21} -  r_{22}
     \end{bmatrix}
$ \vspace{1pt} satisfies (\ref{equi-condition-lambda-repeat2}). Then Theorem \ref{thm:main:generalgraph-aux-partii} implies  that if $\sigma_1<1/\lambda$ and  $\sigma_2<1/\lambda$, both products always die out under any privacy scheme.
\end{rem}
Remark \ref{rem-thm3-cond} will be proved in Appendix \ref{sec:Establishing:thm:main:generalgraphB-partI}.
\begin{rem}  \label{rem-thm3-condv2}
Given (\ref{conditions-transb}), we use Theorem \ref{thm:main:generalgraph-aux-partii} to obtain Remark \ref{thm:main:generalgraph-aux-condv2}; i.e., under a perfect privacy scheme \vspace{1pt} $
\begin{bmatrix}
    \gamma_{1} & \gamma_{2} & 1 - \gamma_{1} -  \gamma_{2} \\     \gamma_{1} & \gamma_{2}&  1 - \gamma_{1} -  \gamma_{2}
     \end{bmatrix}
$, if $\sigma_1 \gamma_1+\sigma_2 \gamma_2<1/\lambda$,
 the system will reach a stable equilibrium where both products die out.
\end{rem}

\section{Experiments}\label{section:Experiments}

To confirm our analytical results in Section \ref{section:Results}, we perform experiments on social networks and plot a few figures.

\afterpage{
\begin{figure}
\vspace{-10pt}
 \addtolength{\subfigcapskip}{-4pt}
\centering
\subfigure[]{\label{diffusion1a}\includegraphics[width=.23\textwidth]{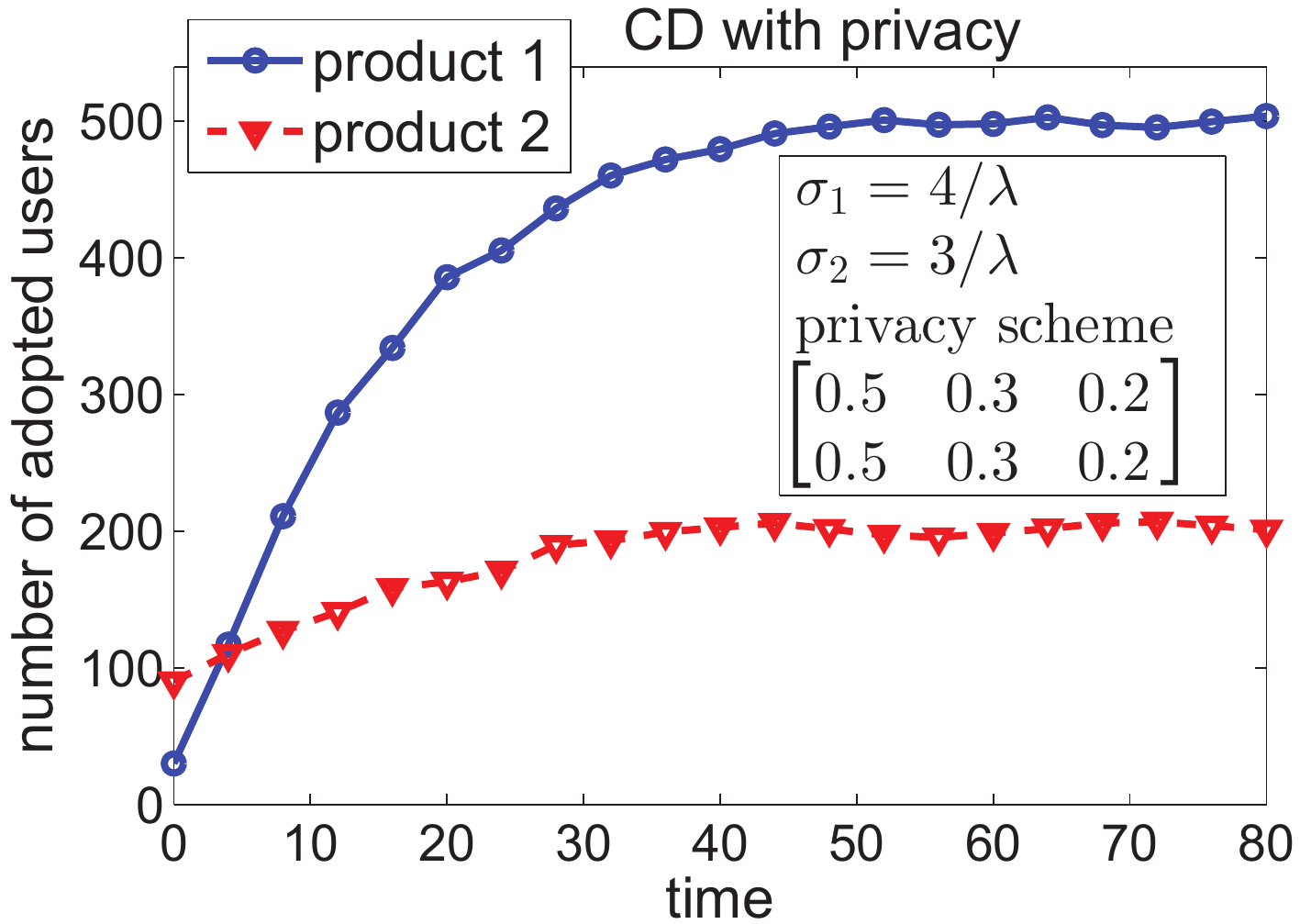}}
\hspace{5pt}\subfigure[]{\label{diffusion1b}\includegraphics[width=.23\textwidth]{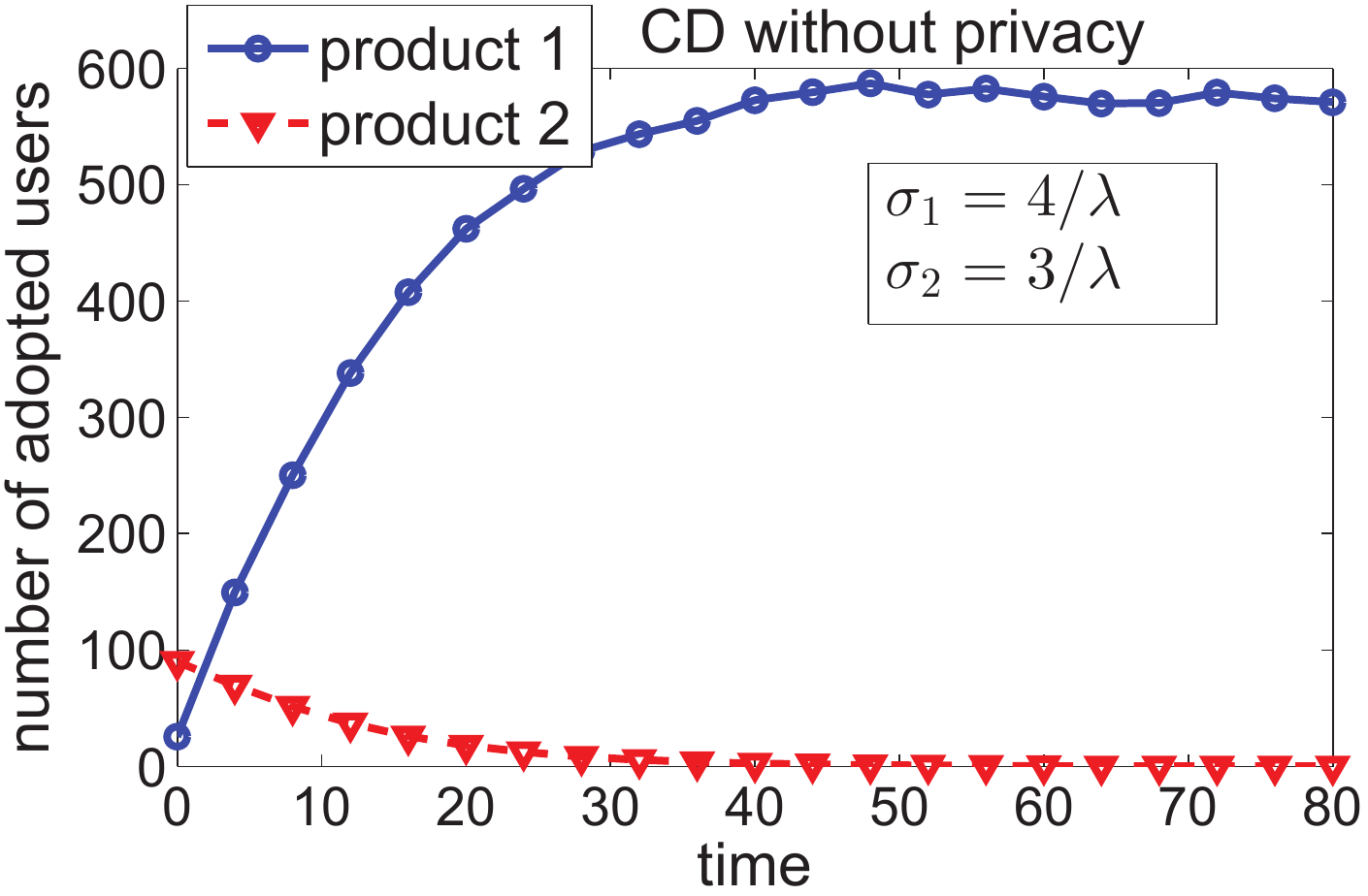}}
\\[-5pt] \hspace{\fill}\line(1,0){240}\hspace{\fill}\subfigure[]{\label{diffusion1c}\includegraphics[width=.23\textwidth]{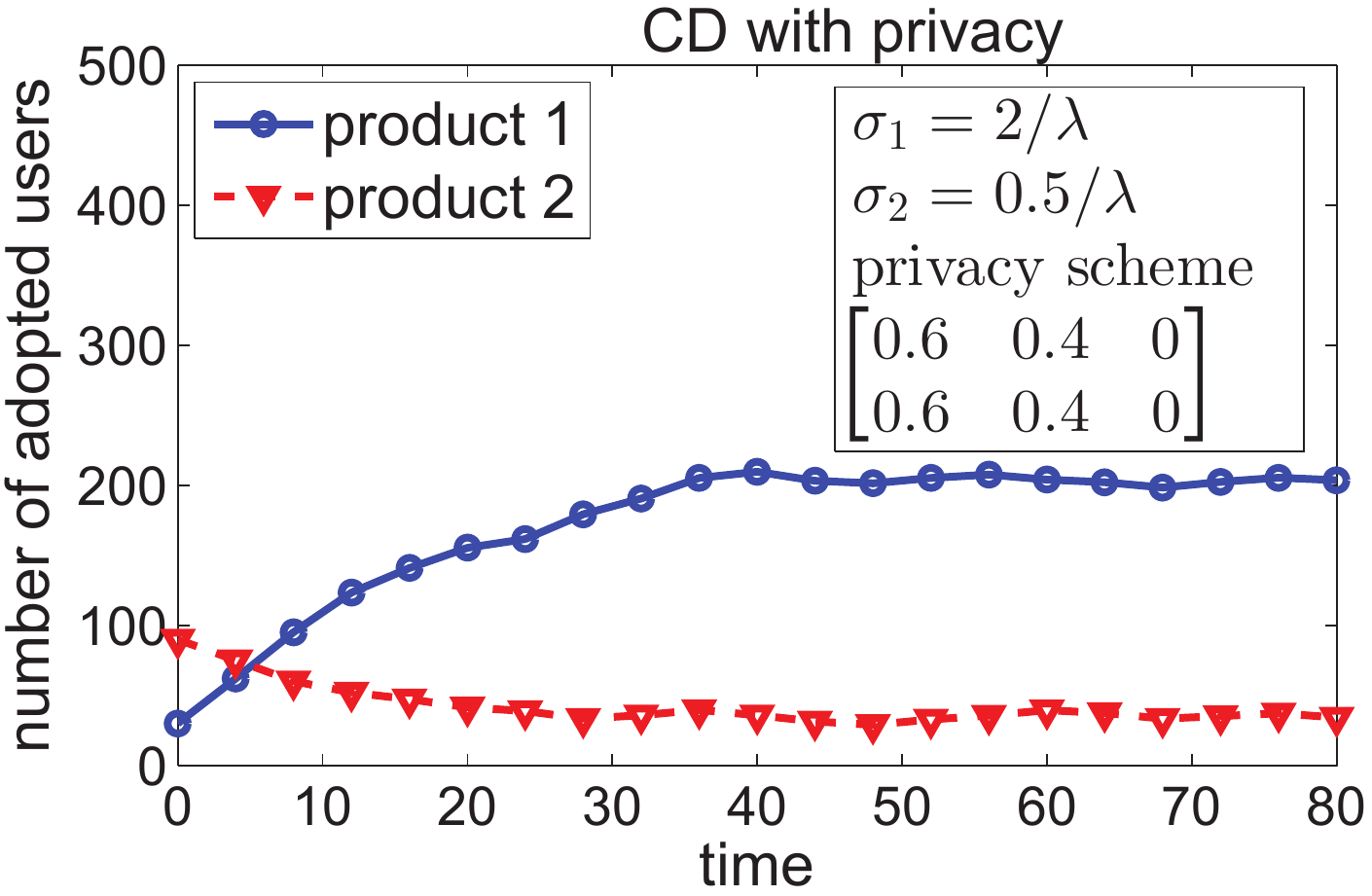}}
\hspace{5pt}\subfigure[]{\label{diffusion1d}\includegraphics[width=.23\textwidth]{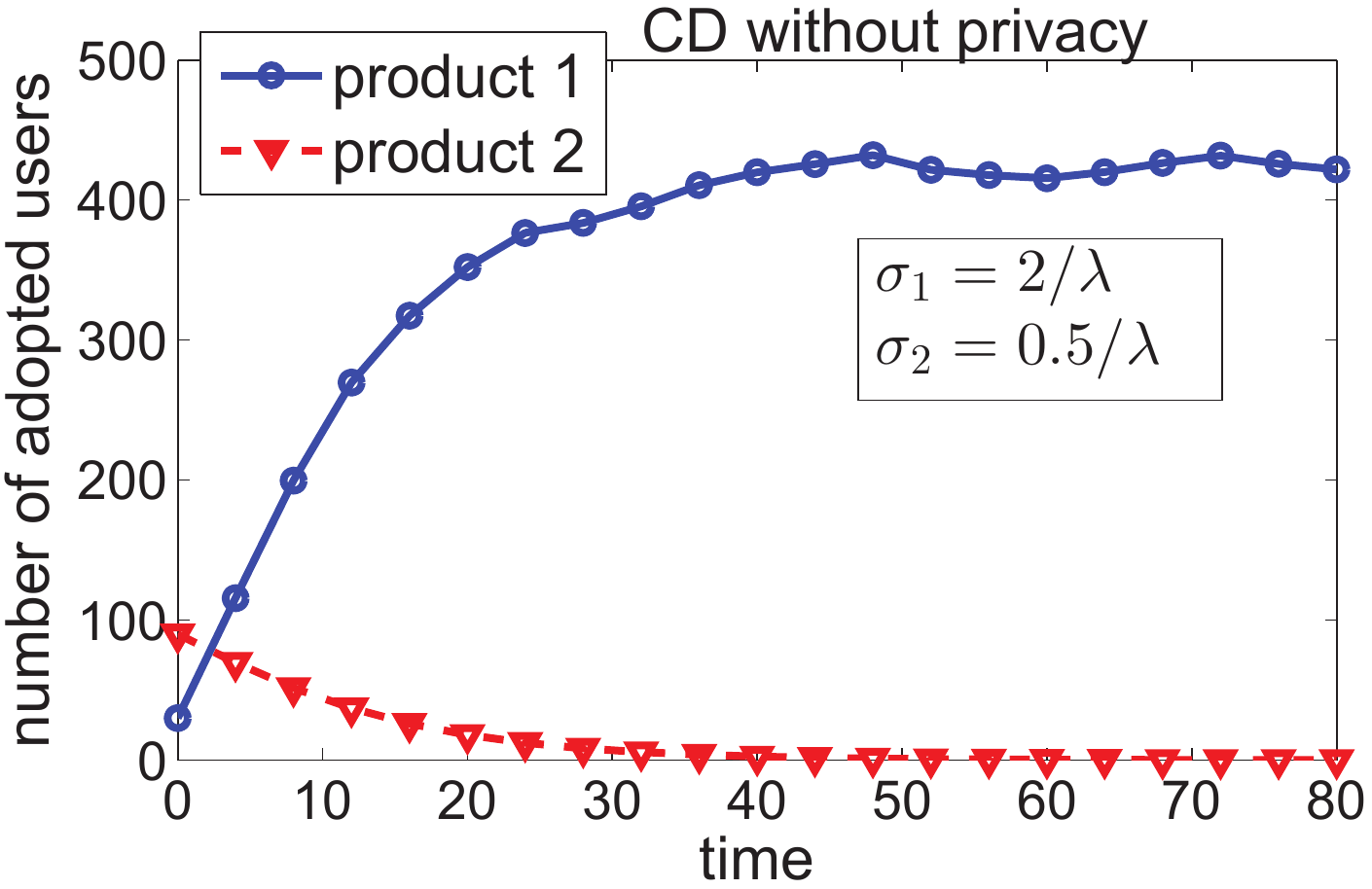}}
\vspace{-9pt}\caption{Experiments on the \textit{HumanSocial} network, where ``CD'' is short for competitive diffusion. In  Figure \ref{diffusion1a} (resp., \ref{diffusion1b}),  we consider \mbox{privacy-aware} (resp., \mbox{privacy-oblivious}) competitive diffusion, and set $\sigma_1$ and $\sigma_2$ both greater than $1/\lambda$. In  Figure \ref{diffusion1c} (resp., \ref{diffusion1d}),  we look at \mbox{privacy-aware} (resp., \mbox{privacy-oblivious}) competitive diffusion and consider $\sigma_1>1/\lambda>\sigma_2$. \protect\\[-5pt] \hspace{\fill}\line(1,0){250}\hspace{\fill}\protect\\[-7pt] \hspace{\fill}\line(1,0){250}\hspace{\fill}\vspace{-10pt}} \label{diffusion1}
\end{figure}

 \begin{figure}
\vspace{3pt}
 \addtolength{\subfigcapskip}{-4pt}
\centering
\subfigure[]{\label{diffusion2a}\includegraphics[width=.23\textwidth]{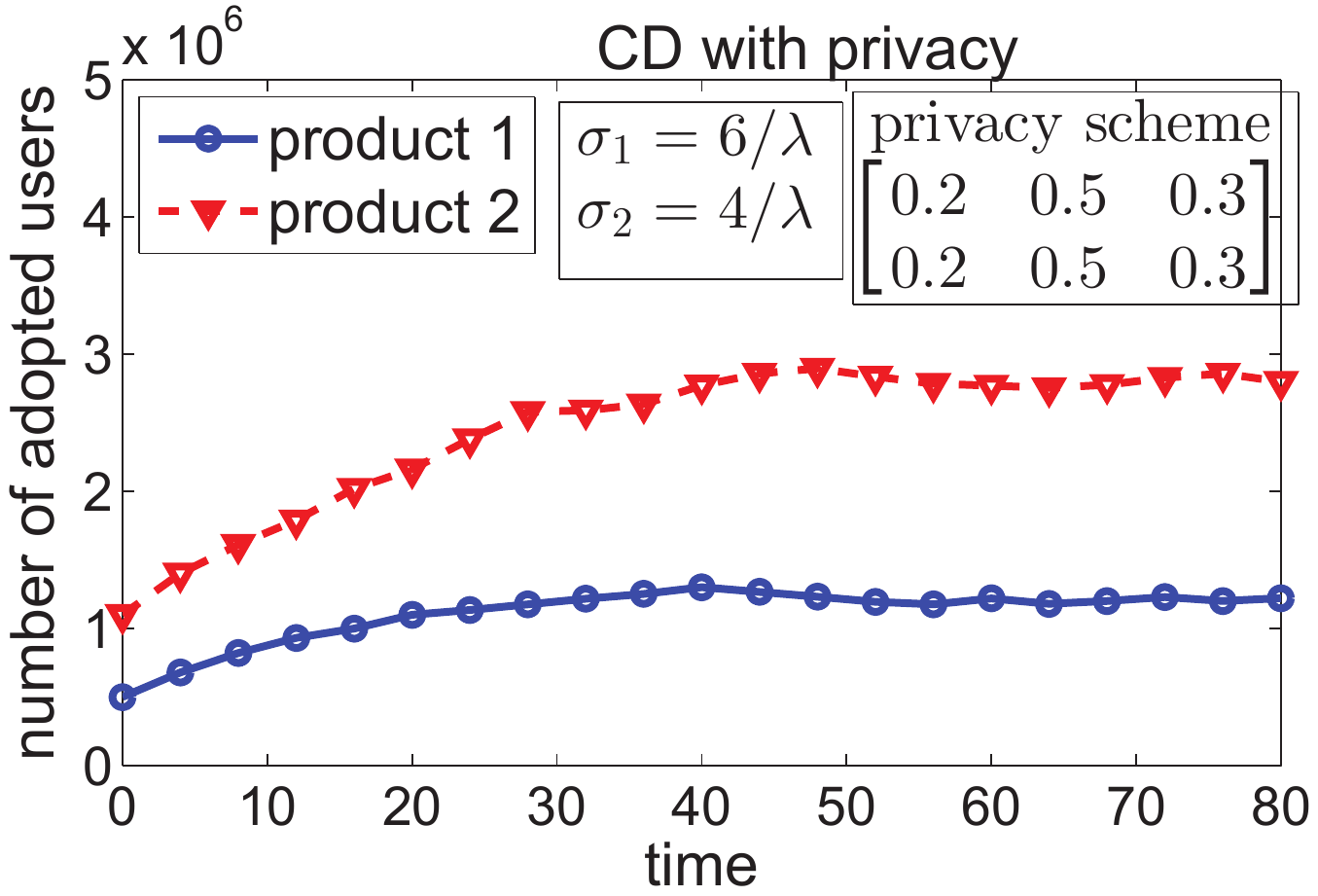}}
\hspace{5pt}\subfigure[]{\label{diffusion2b}\includegraphics[width=.23\textwidth]{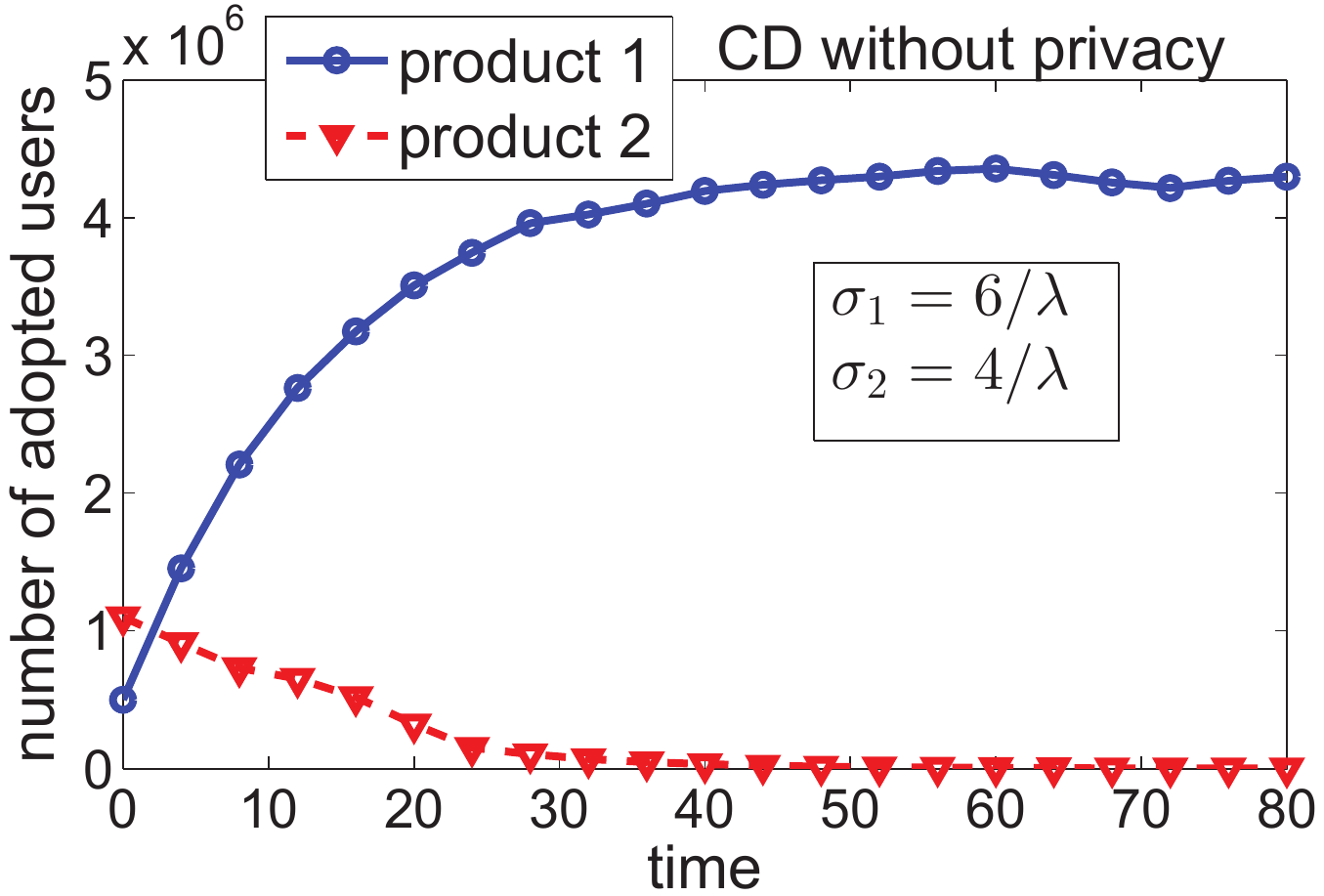}}
\\[-5pt] \hspace{\fill}\line(1,0){240}\hspace{\fill}\subfigure[]{\label{diffusion2c}\includegraphics[width=.23\textwidth]{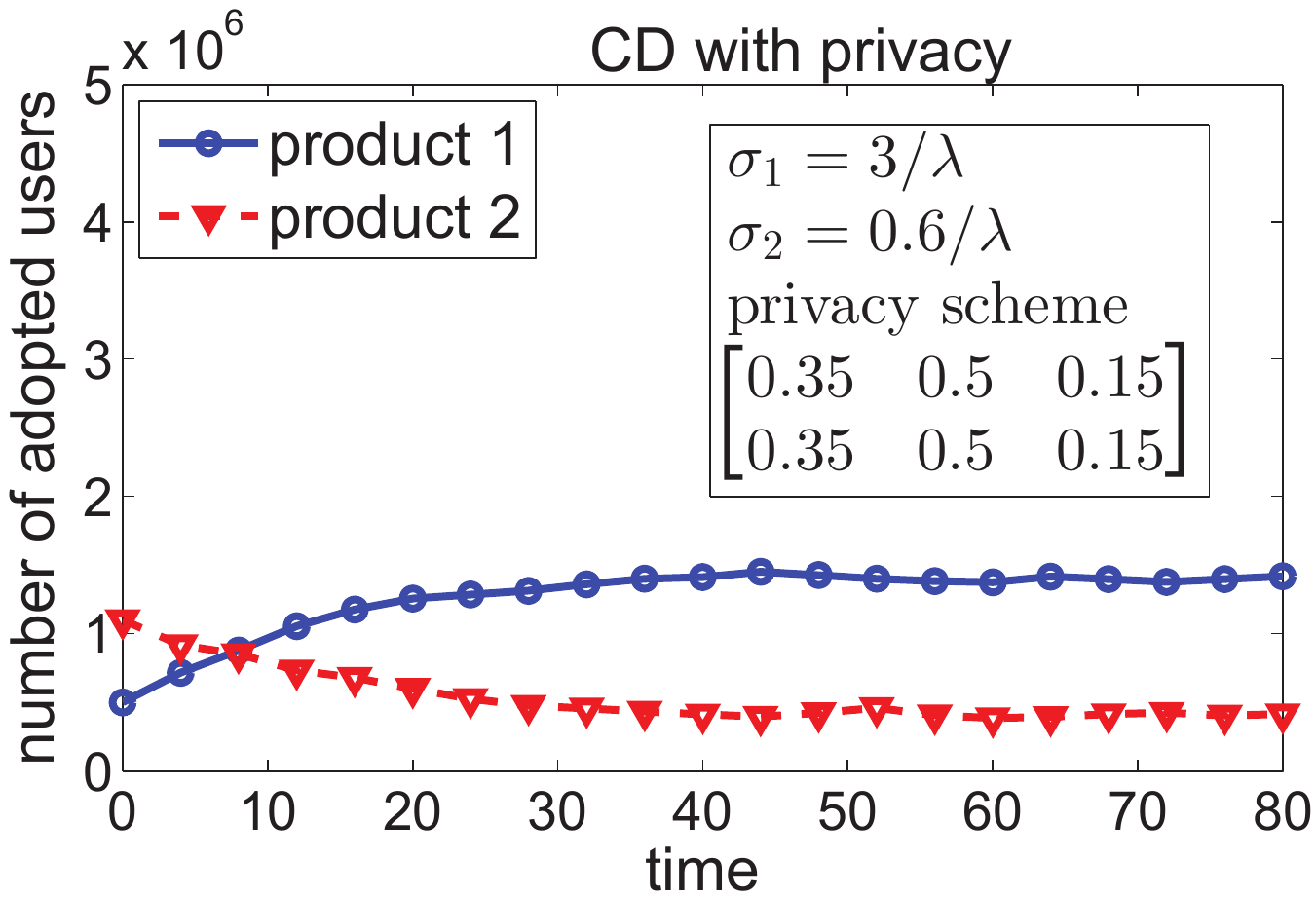}}
\hspace{5pt}\subfigure[]{\label{diffusion2d}\includegraphics[width=.23\textwidth]{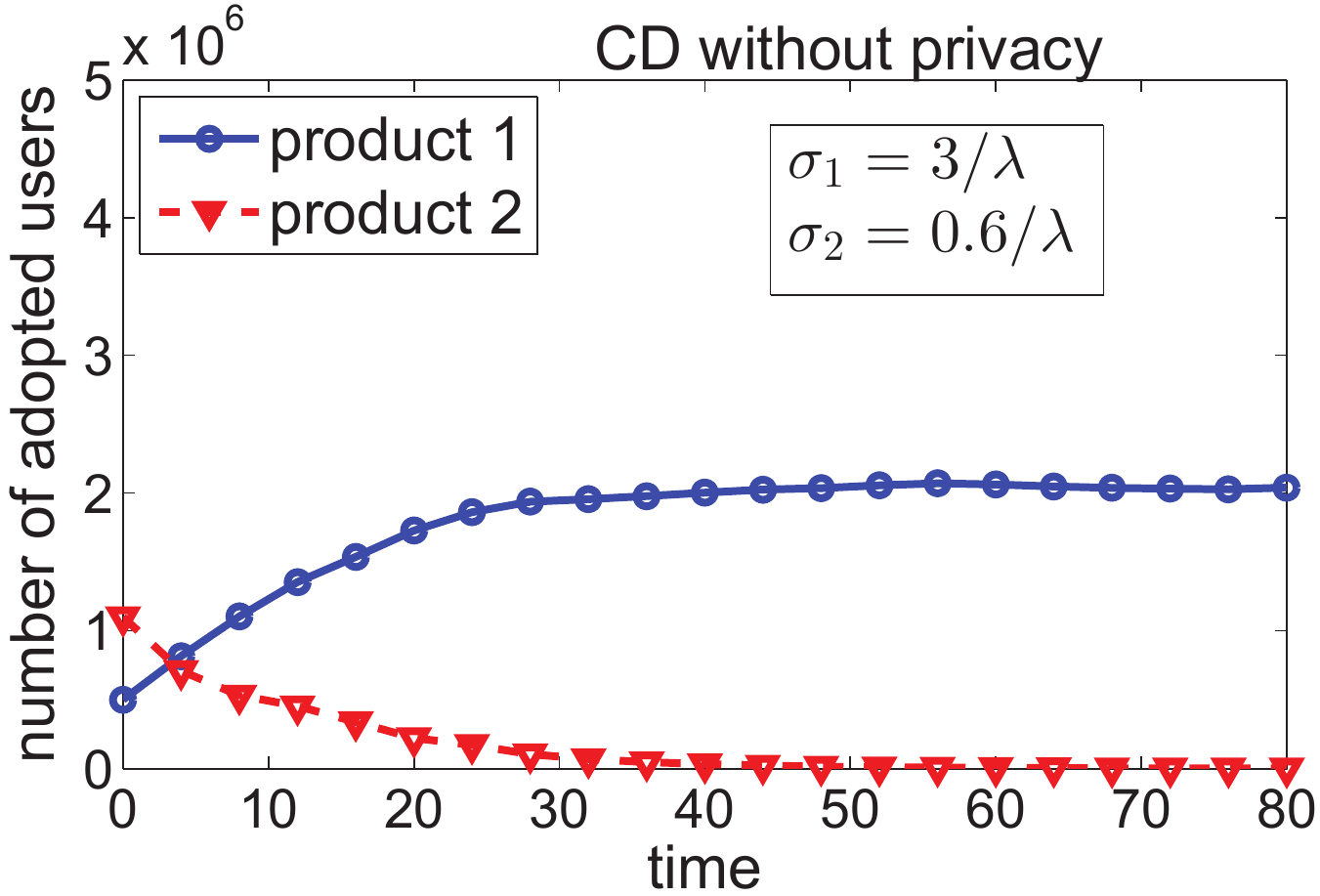}}
\vspace{-9pt}\caption{Experiments on the Google+ network, where ``CD'' is short for competitive diffusion. In  Figure \ref{diffusion2a} (resp., \ref{diffusion2b}),  we consider \mbox{privacy-aware} (resp., \mbox{privacy-oblivious}) competitive diffusion under  $\sigma_1>\sigma_2>1/\lambda$. In  Figure \ref{diffusion2c} (resp., \ref{diffusion2d}),  we look at \mbox{privacy-aware} (resp., \mbox{privacy-oblivious}) competitive diffusion under $\sigma_1>1/\lambda>\sigma_2$.\protect\\[-5pt] \hspace{\fill}\line(1,0){250}\hspace{\fill}\protect\\[-7pt] \hspace{\fill}\line(1,0){250}\hspace{\fill}\vspace{-10pt}} \label{diffusion2}
\end{figure}

 \begin{figure}[!b]
\vspace{3pt}
 \addtolength{\subfigcapskip}{-4pt}
\centering
\subfigure[]{\label{PhysicalContact-coexist1}\includegraphics[width=.23\textwidth]{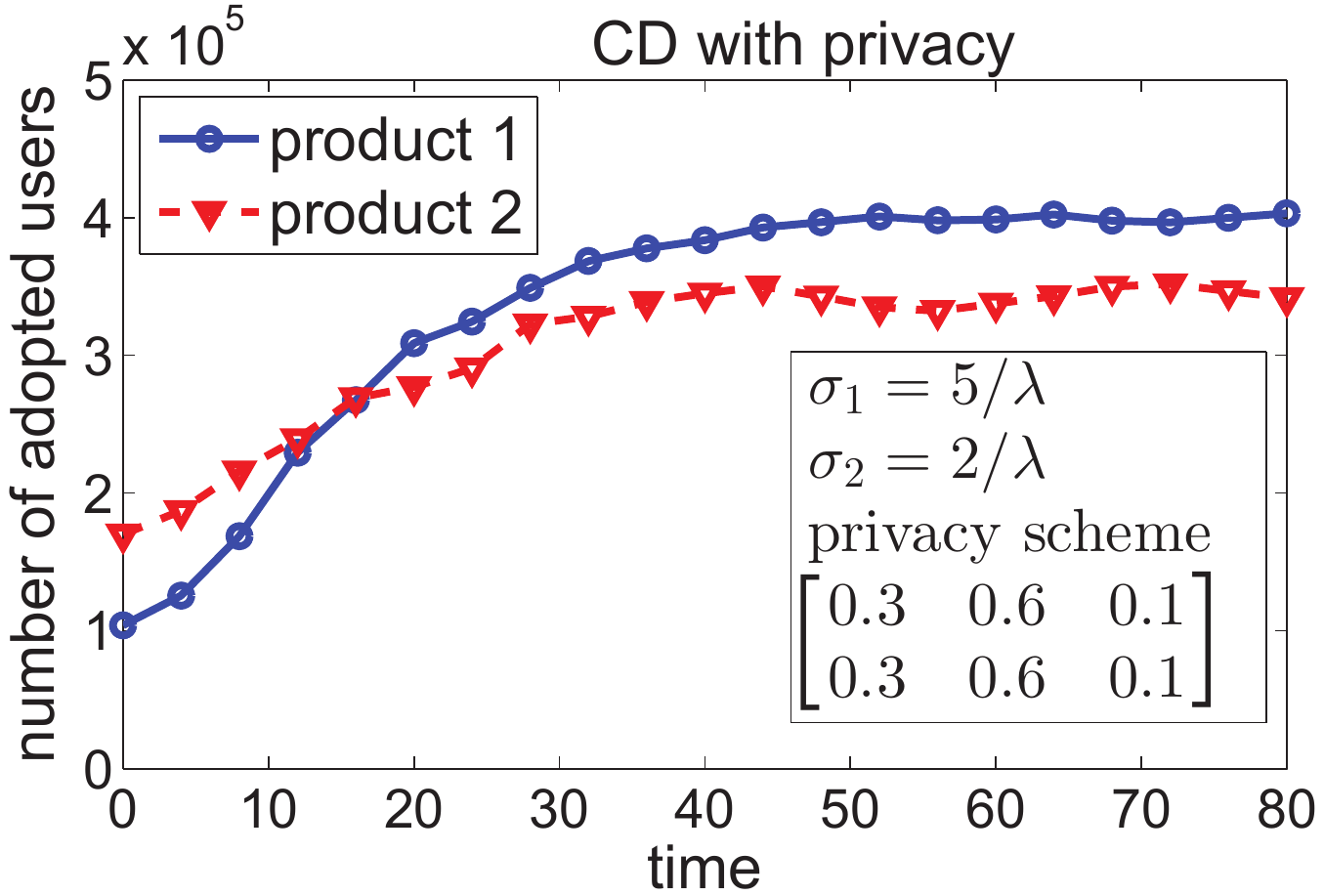}}
\hspace{5pt}\subfigure[]{\label{PhysicalContact-coexist2}\includegraphics[width=.23\textwidth]{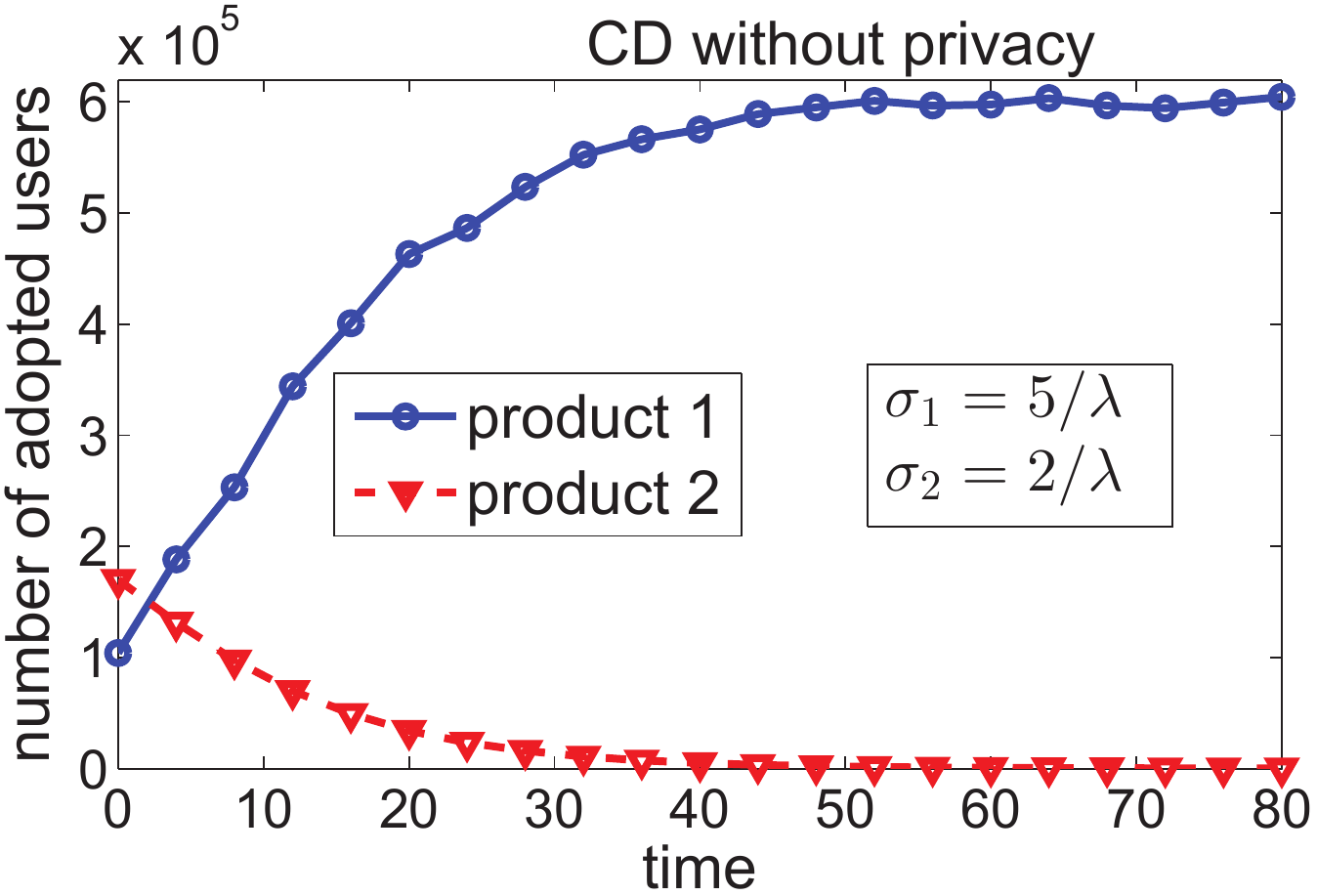}}
\\[-5pt] \hspace{\fill}\line(1,0){240}\hspace{\fill}\subfigure[]{\label{PhysicalContact-coexist3}\includegraphics[width=.23\textwidth]{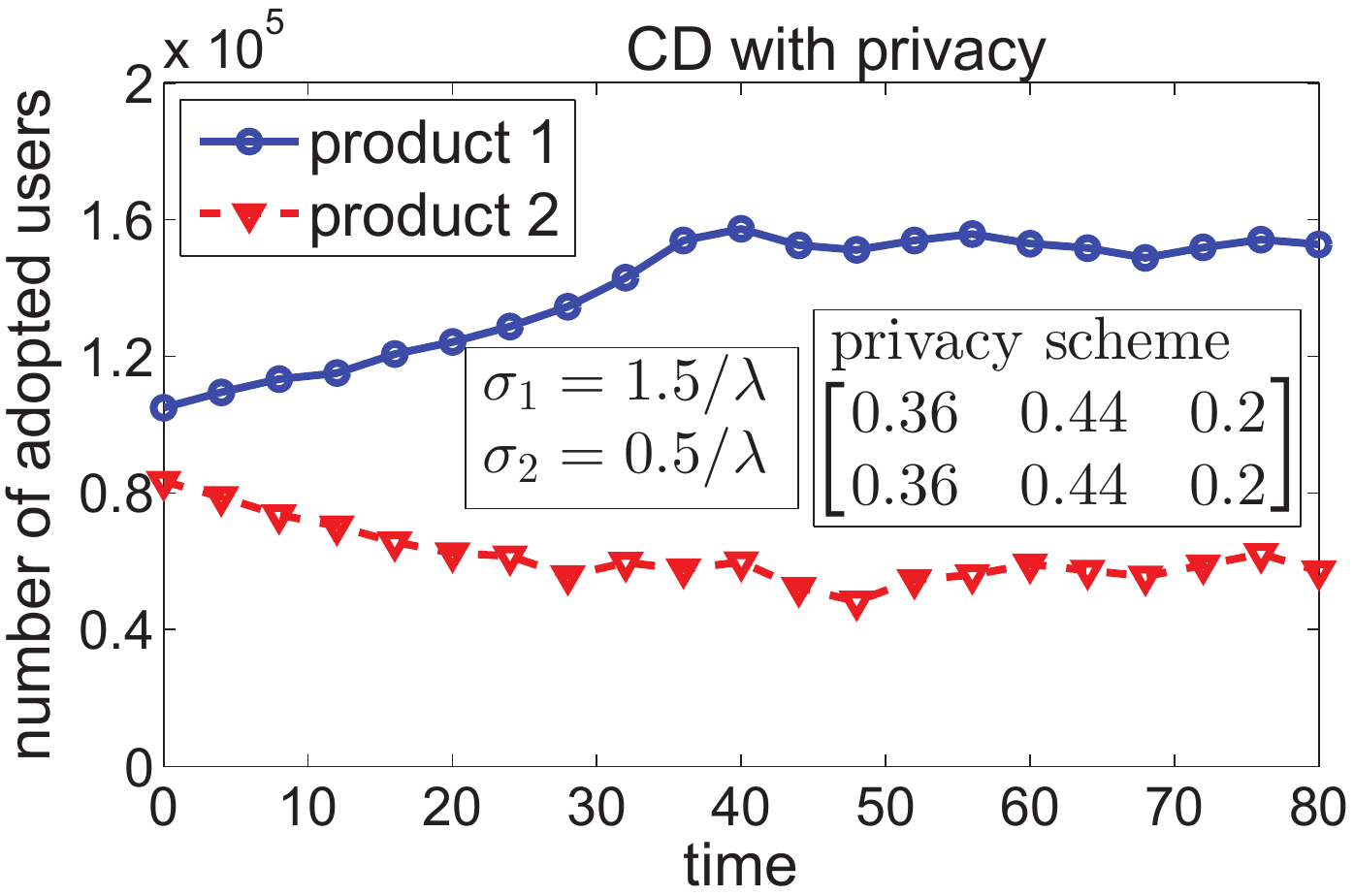}}
\hspace{5pt}\subfigure[]{\label{PhysicalContact-coexist4}\includegraphics[width=.23\textwidth]{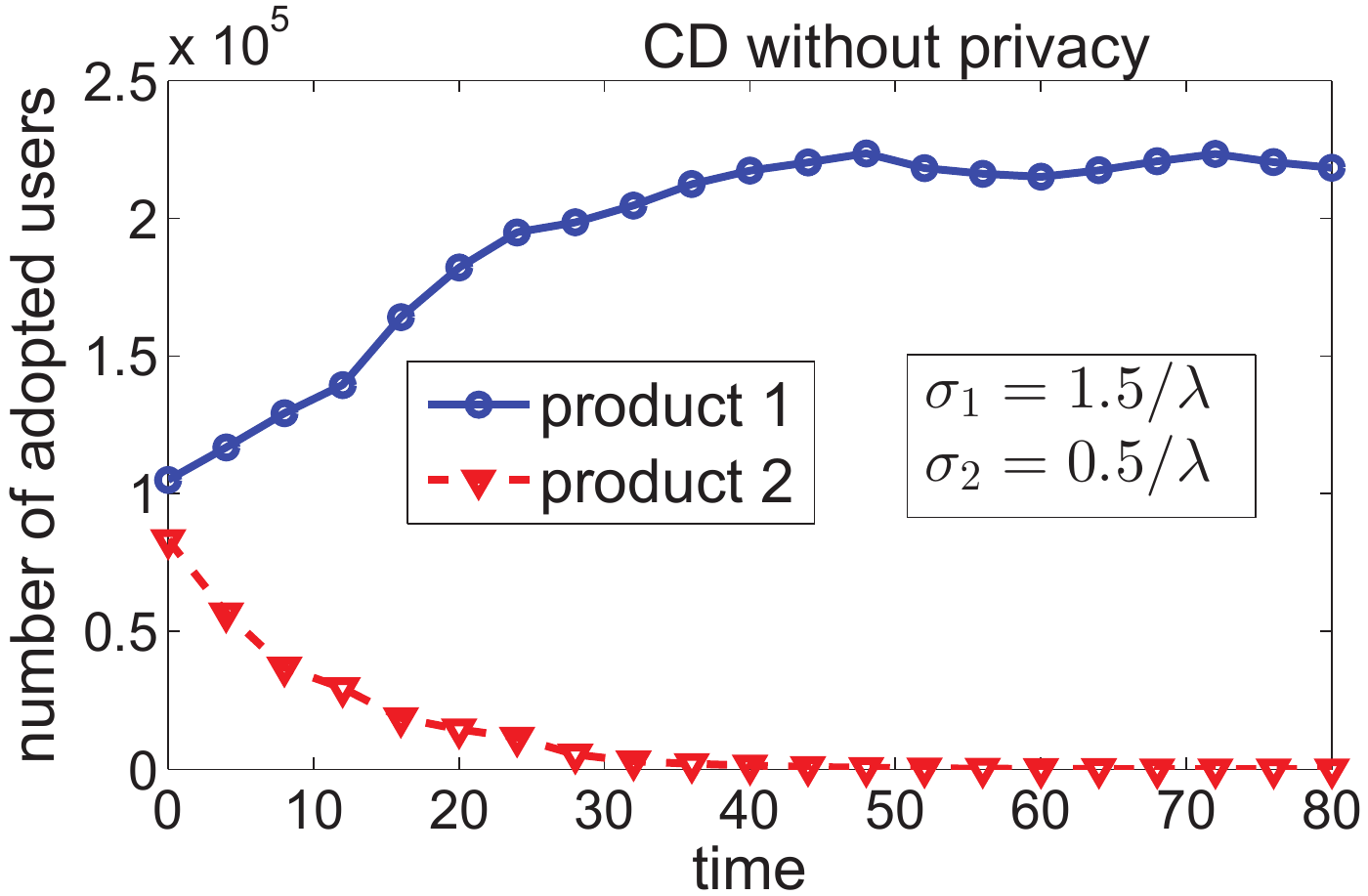}}
\vspace{-9pt}\caption{Experiments on the \textit{PhysicalContact} network, where ``CD'' is short for competitive diffusion. In  Figure \ref{PhysicalContact-coexist1} (resp., \ref{PhysicalContact-coexist2}),  we consider \mbox{privacy-aware} (resp., \mbox{privacy-oblivious}) competitive diffusion under  $\sigma_1>\sigma_2>1/\lambda$. In  Figure \ref{PhysicalContact-coexist3} (resp., \ref{PhysicalContact-coexist4}),  we consider \mbox{privacy-aware} (resp., \mbox{privacy-oblivious}) competitive diffusion under $\sigma_1>1/\lambda>\sigma_2$.} \label{PhysicalContact}
\end{figure}

\begin{figure}
\vspace{-10pt}
 \addtolength{\subfigcapskip}{-4pt}
\centering
\subfigure[]{\label{diffusion5a}\includegraphics[width=.24\textwidth]{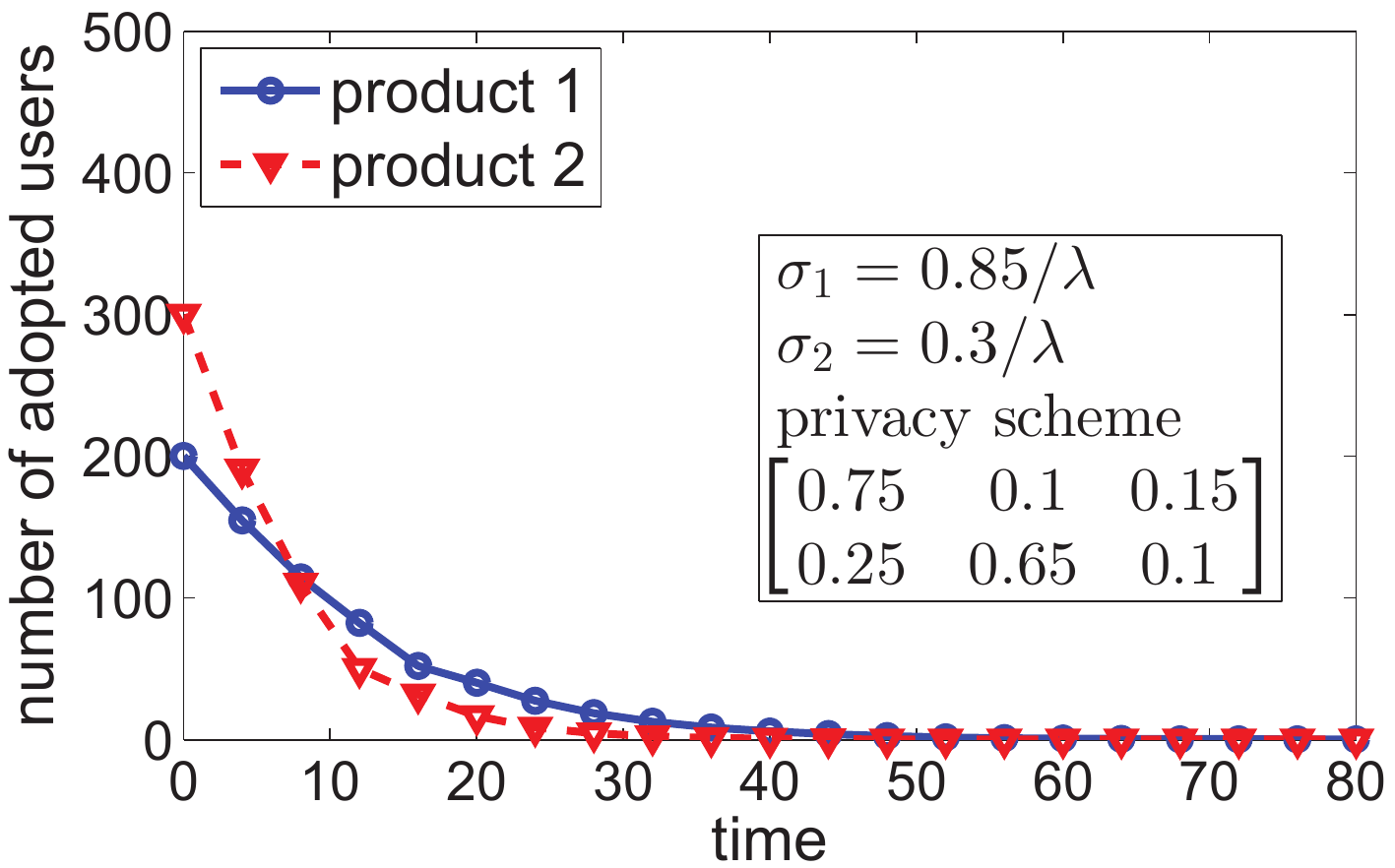}}
\hspace{-2pt}\subfigure[]{\label{diffusion5b}\includegraphics[width=.24\textwidth]{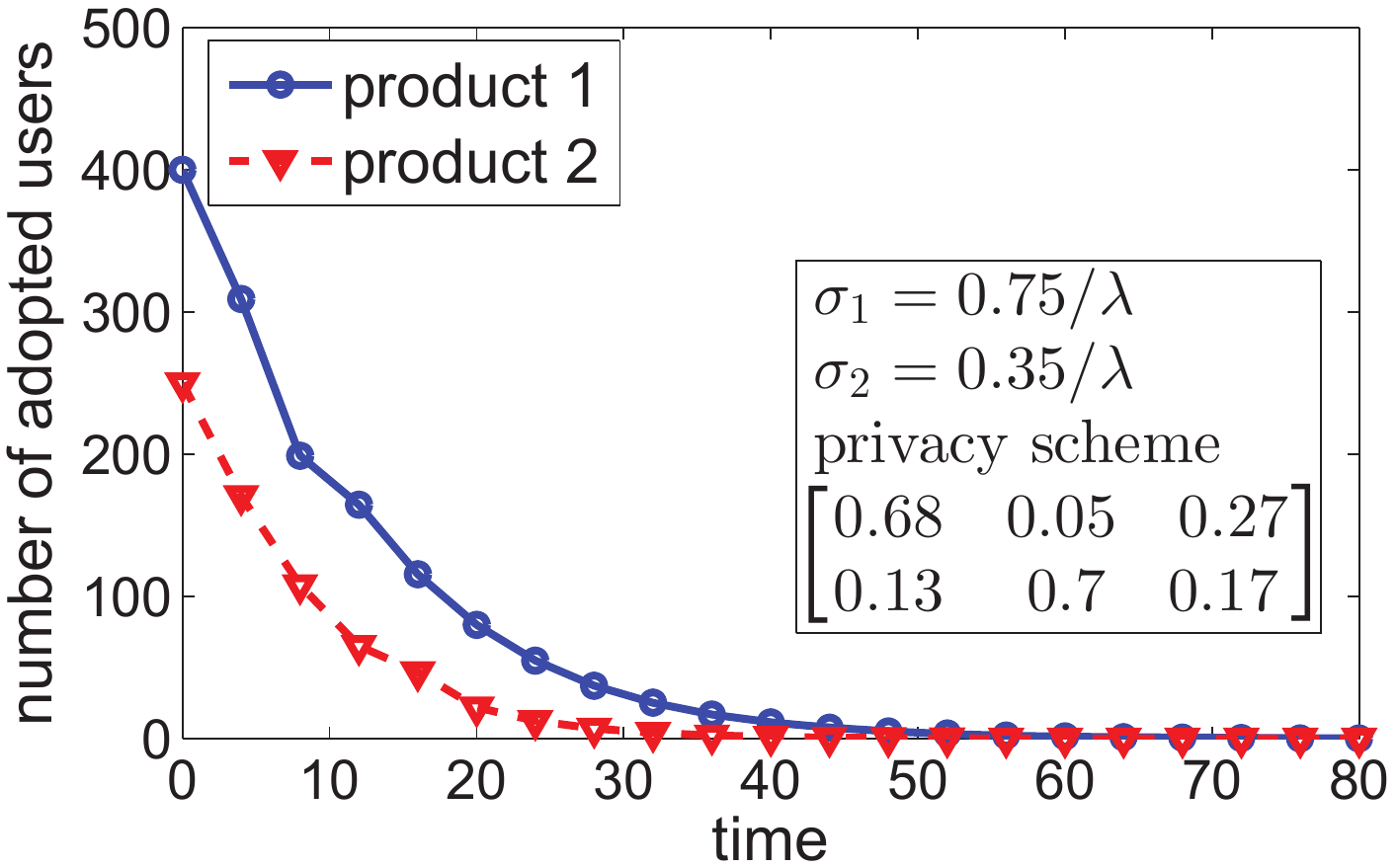}}
\\[-5pt] \hspace{\fill}\line(1,0){240}\hspace{\fill}\subfigure[]{\label{diffusion5c}\includegraphics[width=.23\textwidth]{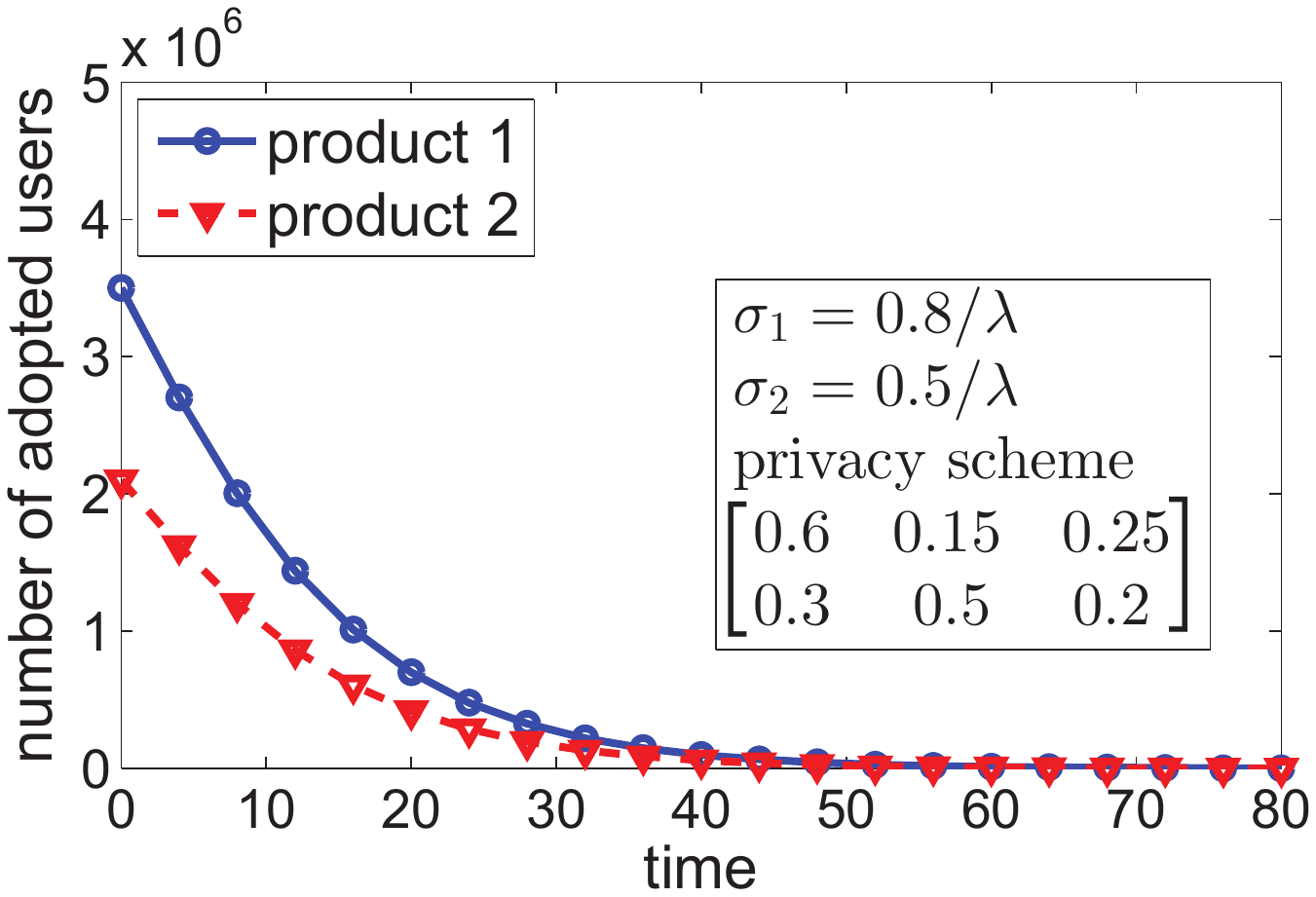}}
\hspace{5pt}\subfigure[]{\label{diffusion5d}\includegraphics[width=.23\textwidth]{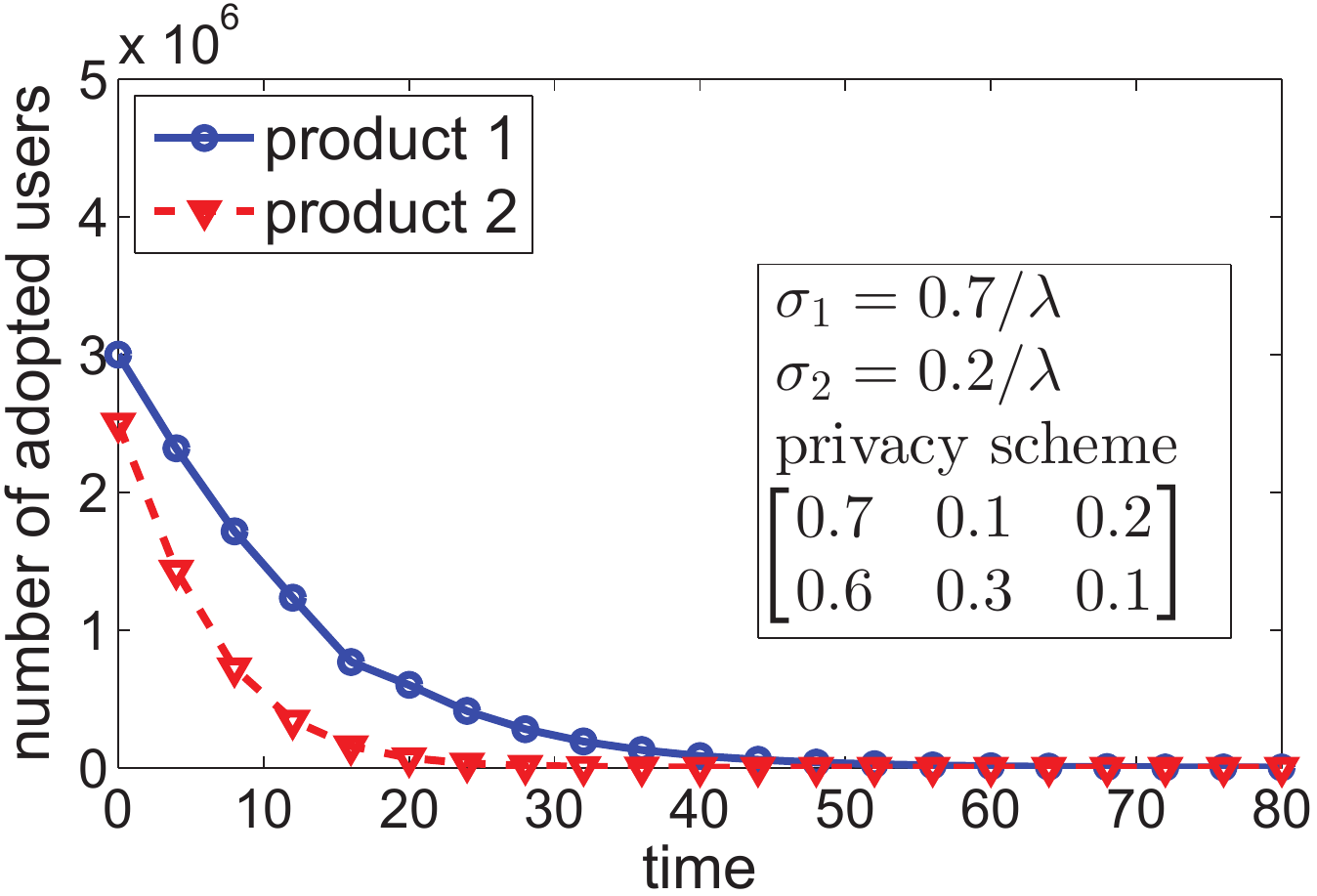}}\\[-5pt] \hspace{\fill}\line(1,0){240}\hspace{\fill}\subfigure[]{\label{diffusion5e}\includegraphics[width=.23\textwidth]{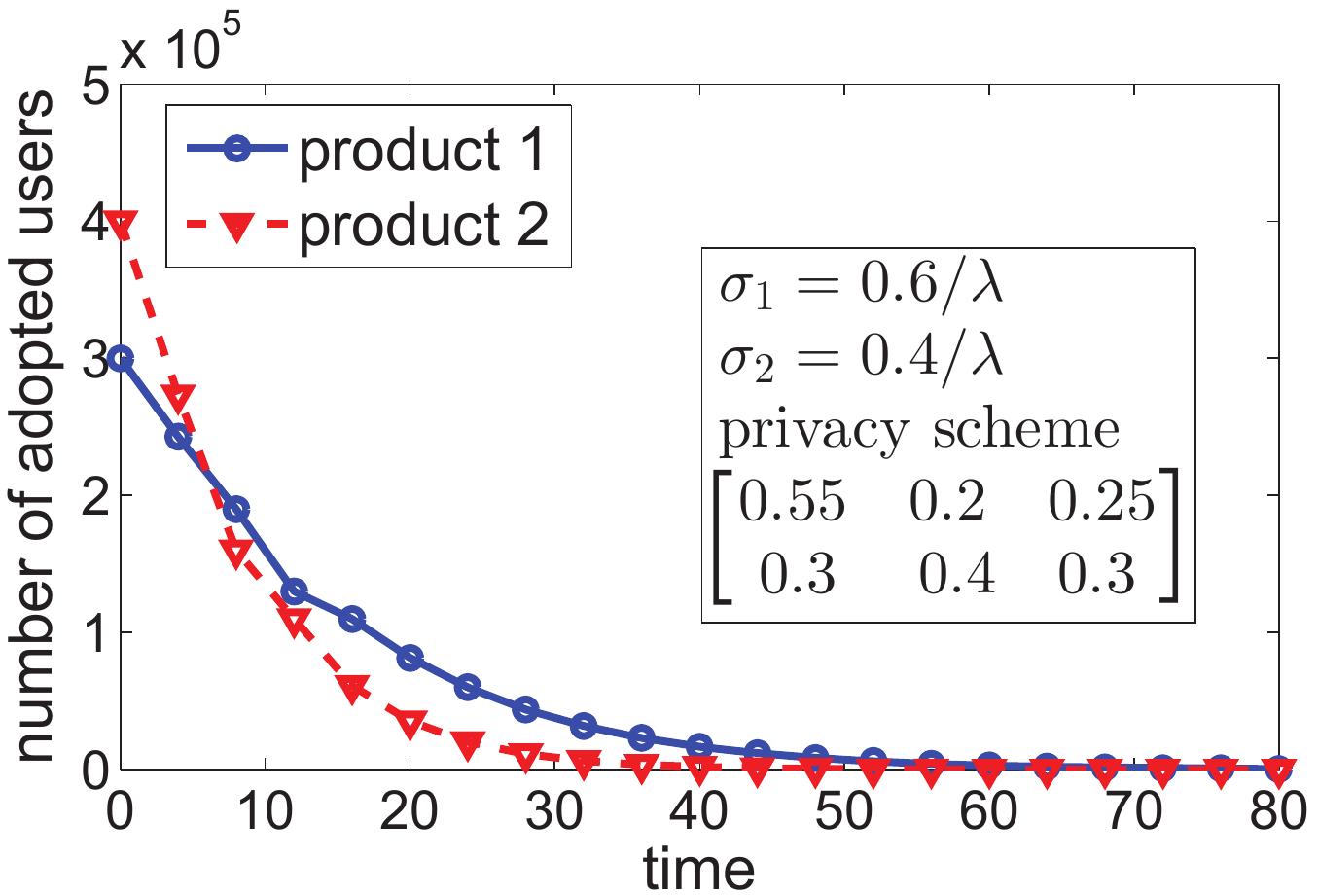}}
\hspace{5pt}\subfigure[]{\label{diffusion5f}\includegraphics[width=.23\textwidth]{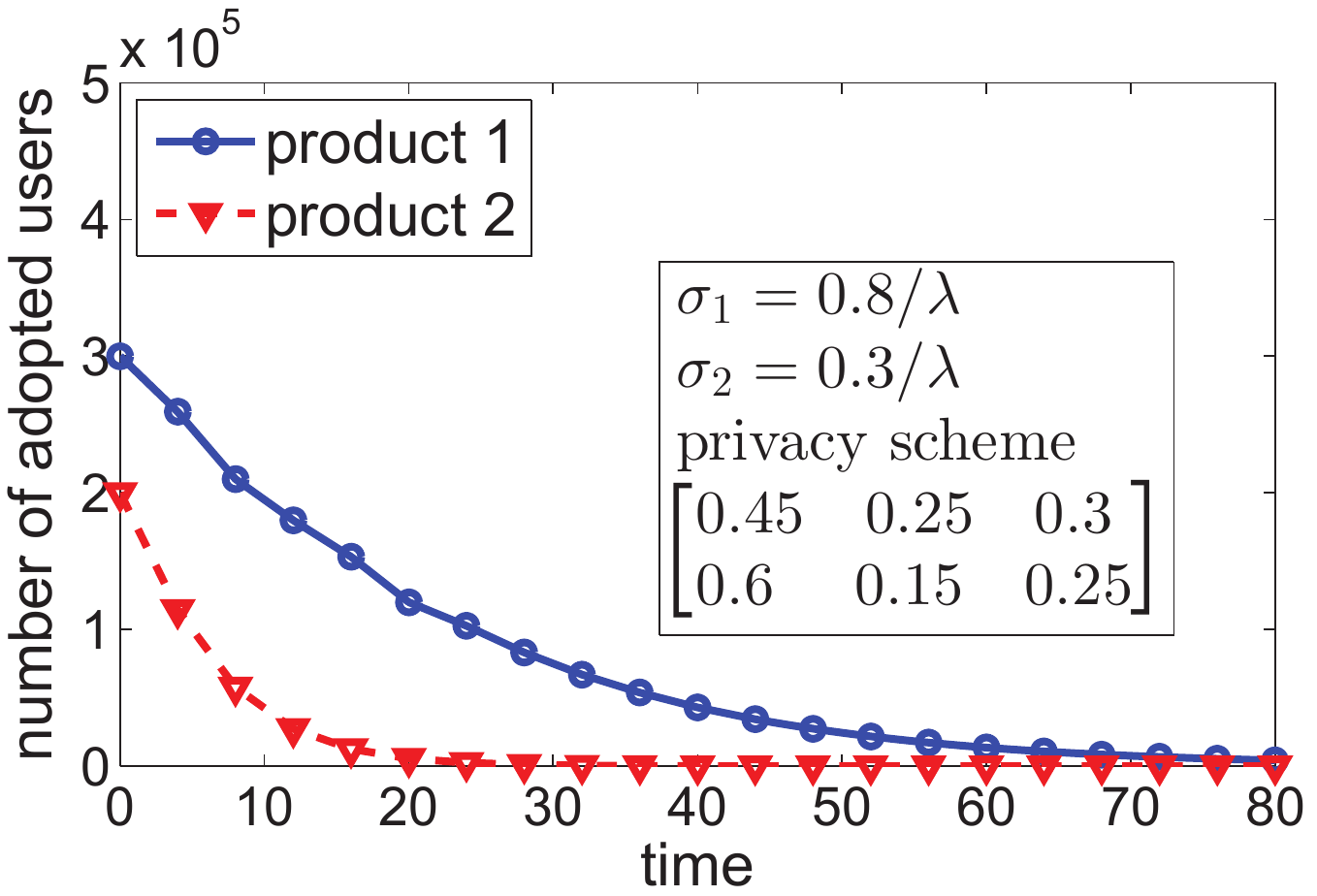}}
\vspace{-9pt}\caption{Experiments of \mbox{privacy-aware} competitive diffusion when $\sigma_1<1/\lambda$ and  $\sigma_2<1/\lambda$ (Figures \ref{diffusion5a} and \ref{diffusion5b} on the \textit{HumanSocial} network, Figures \ref{diffusion5c} and \ref{diffusion5d} on the Google+ network, Figures \ref{diffusion5e} and \ref{diffusion5f} on the \textit{PhysicalContact} network). \protect\\[-5pt] \hspace{\fill}\line(1,0){250}\hspace{\fill}\protect\\[-7pt] \hspace{\fill}\line(1,0){250}\hspace{\fill}\vspace{-10pt}} \label{diffusion5}
\end{figure}

\begin{figure}
\vspace{0pt}
 \addtolength{\subfigcapskip}{-4pt}
\centering
\subfigure[]{\label{completegraph-coexist1}\includegraphics[width=.24\textwidth]{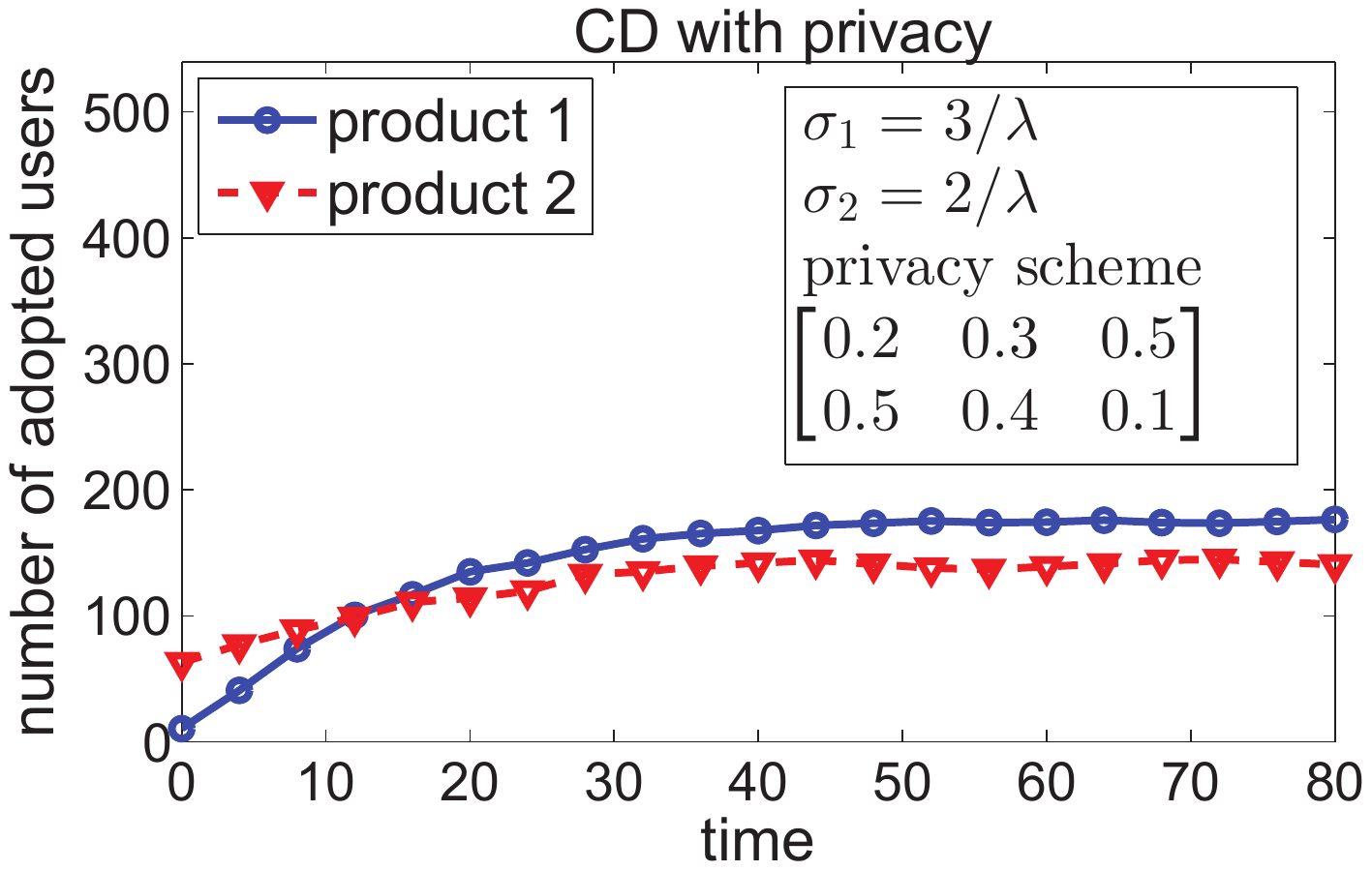}}
\hspace{0pt}\subfigure[]{\label{completegraph-coexist2}\includegraphics[width=.24\textwidth]{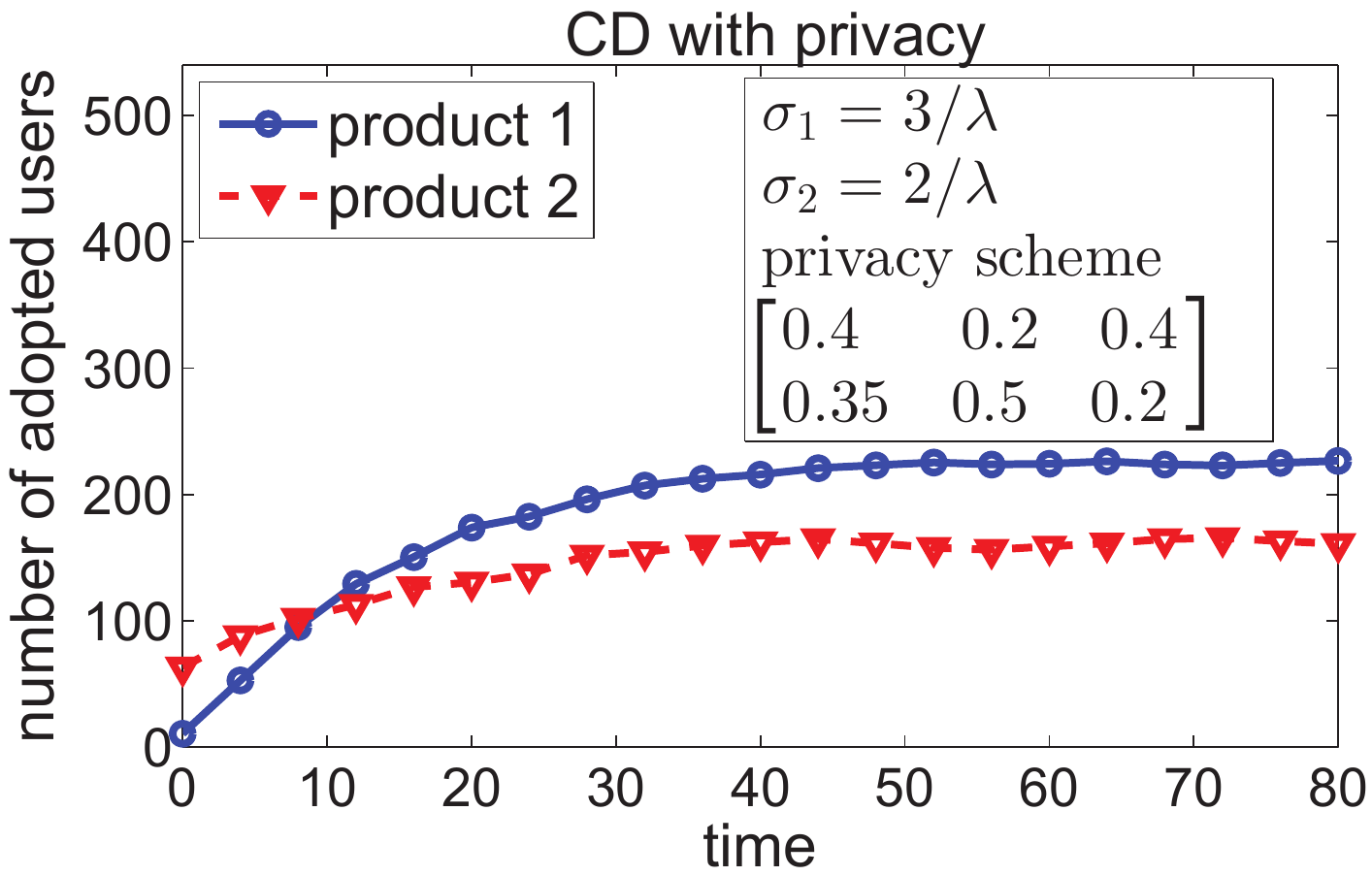}}
\\[-5pt] \hspace{\fill}\line(1,0){240}\hspace{\fill}\subfigure[]{\label{completegraph-coexist3}\includegraphics[width=.24\textwidth]{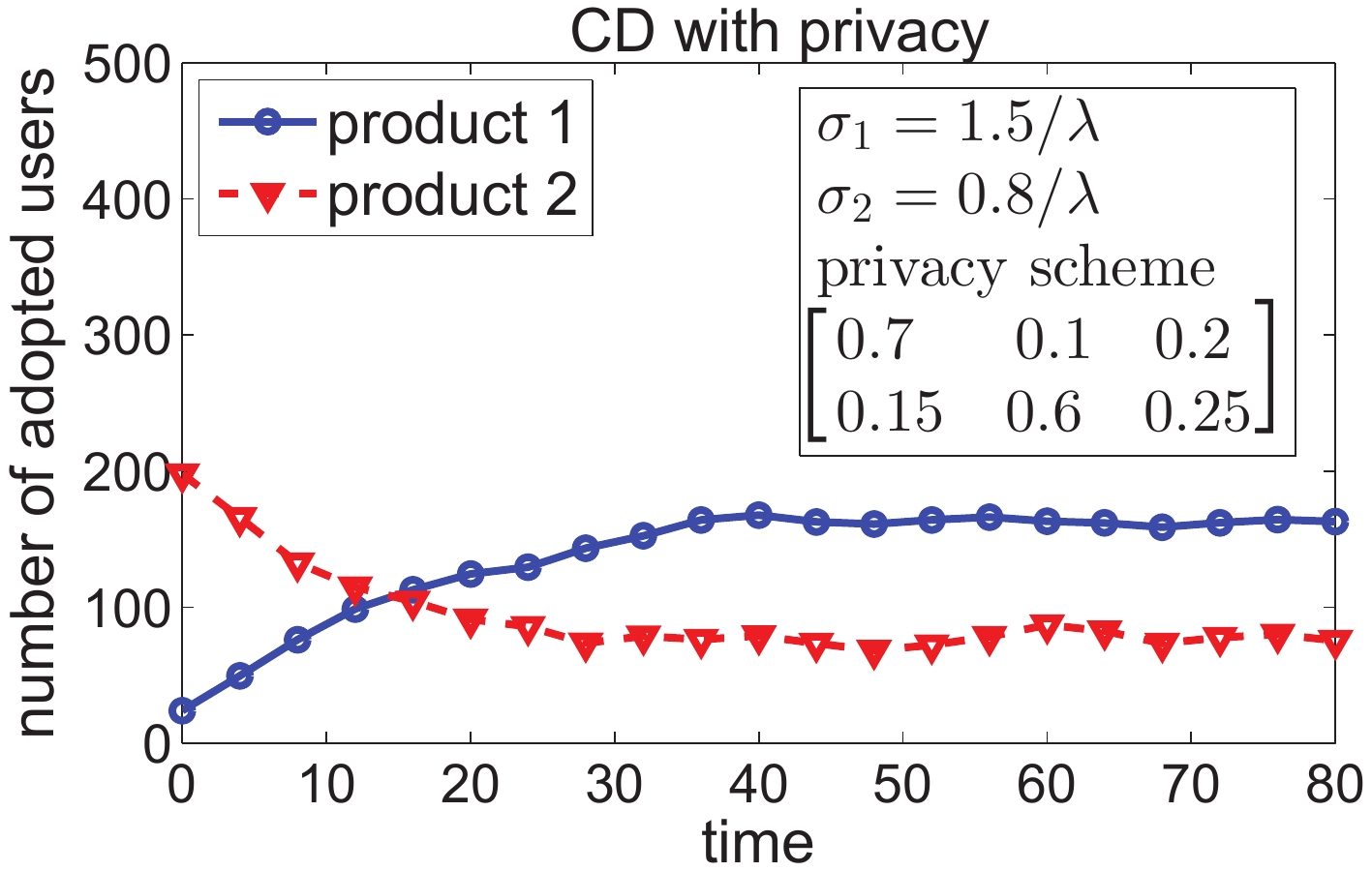}}
\hspace{0pt}\subfigure[]{\label{completegraph-coexist4}\includegraphics[width=.24\textwidth]{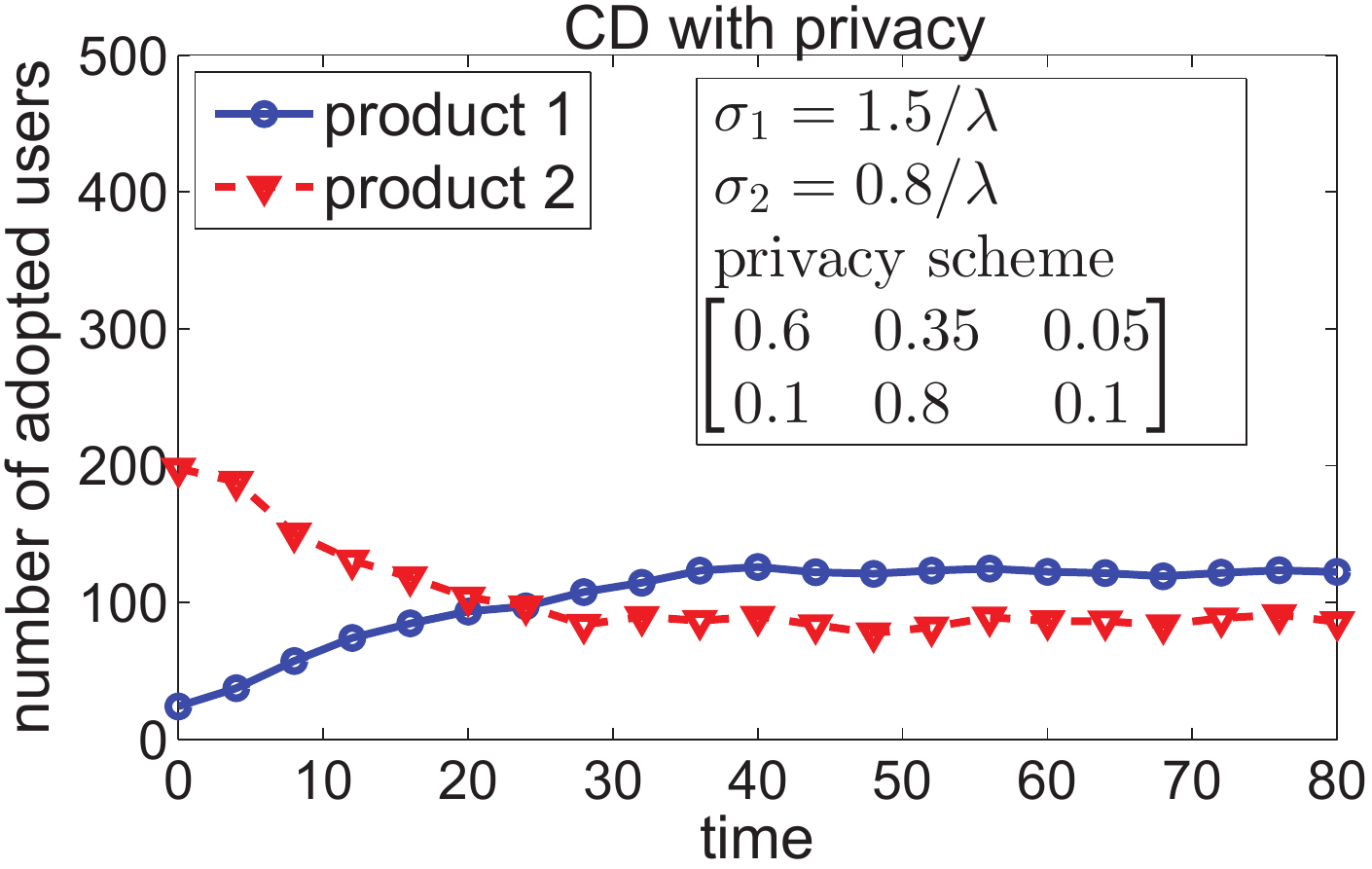}}
\vspace{-9pt}\caption{Experiments of \mbox{privacy-aware} competitive diffusion on a complete network of $2000$ nodes. In  Figure \ref{completegraph-coexist1} (resp., \ref{completegraph-coexist2}),  we consider \mbox{privacy-aware} (resp., \mbox{privacy-oblivious}) competitive diffusion, and set $\sigma_1$ and $\sigma_2$ both greater than $1/\lambda$. In  Figure \ref{completegraph-coexist3} (resp., \ref{completegraph-coexist4}),  we look at \mbox{privacy-aware} (resp., \mbox{privacy-oblivious}) competitive diffusion and consider $\sigma_1>1/\lambda>\sigma_2$. \protect\\[-5pt] \hspace{\fill}\line(1,0){250}\hspace{\fill}\protect\\[-7pt] \hspace{\fill}\line(1,0){250}\hspace{\fill}\vspace{-10pt}} \label{completegraph-coexist}
\end{figure}

\begin{figure}
\vspace{0pt}
 \addtolength{\subfigcapskip}{-4pt}
\centering
\subfigure[]{\label{completegraph-no-coexist1}\includegraphics[width=.24\textwidth]{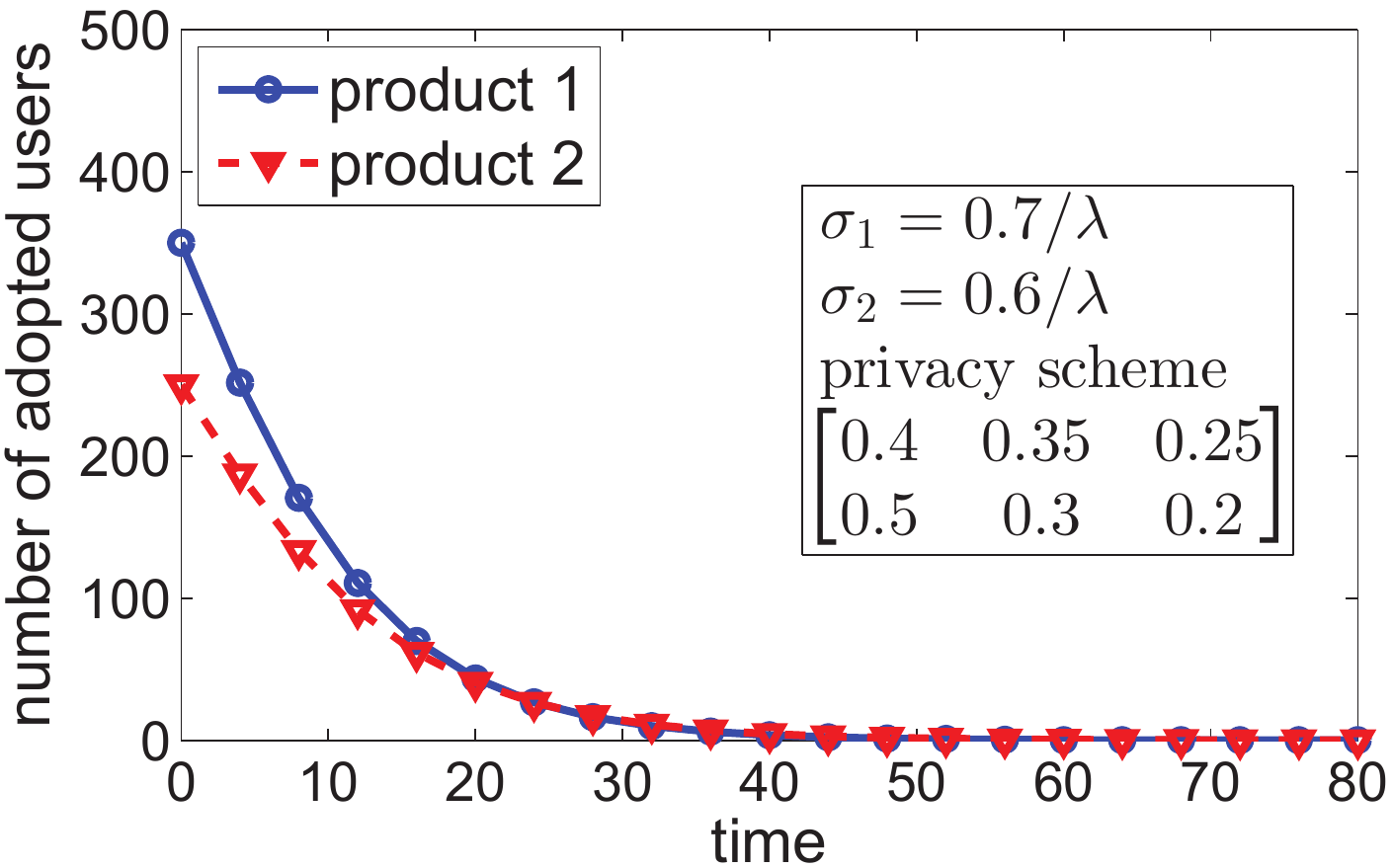}}
\hspace{-2pt}\subfigure[]{\label{completegraph-no-coexist2}\includegraphics[width=.24\textwidth]{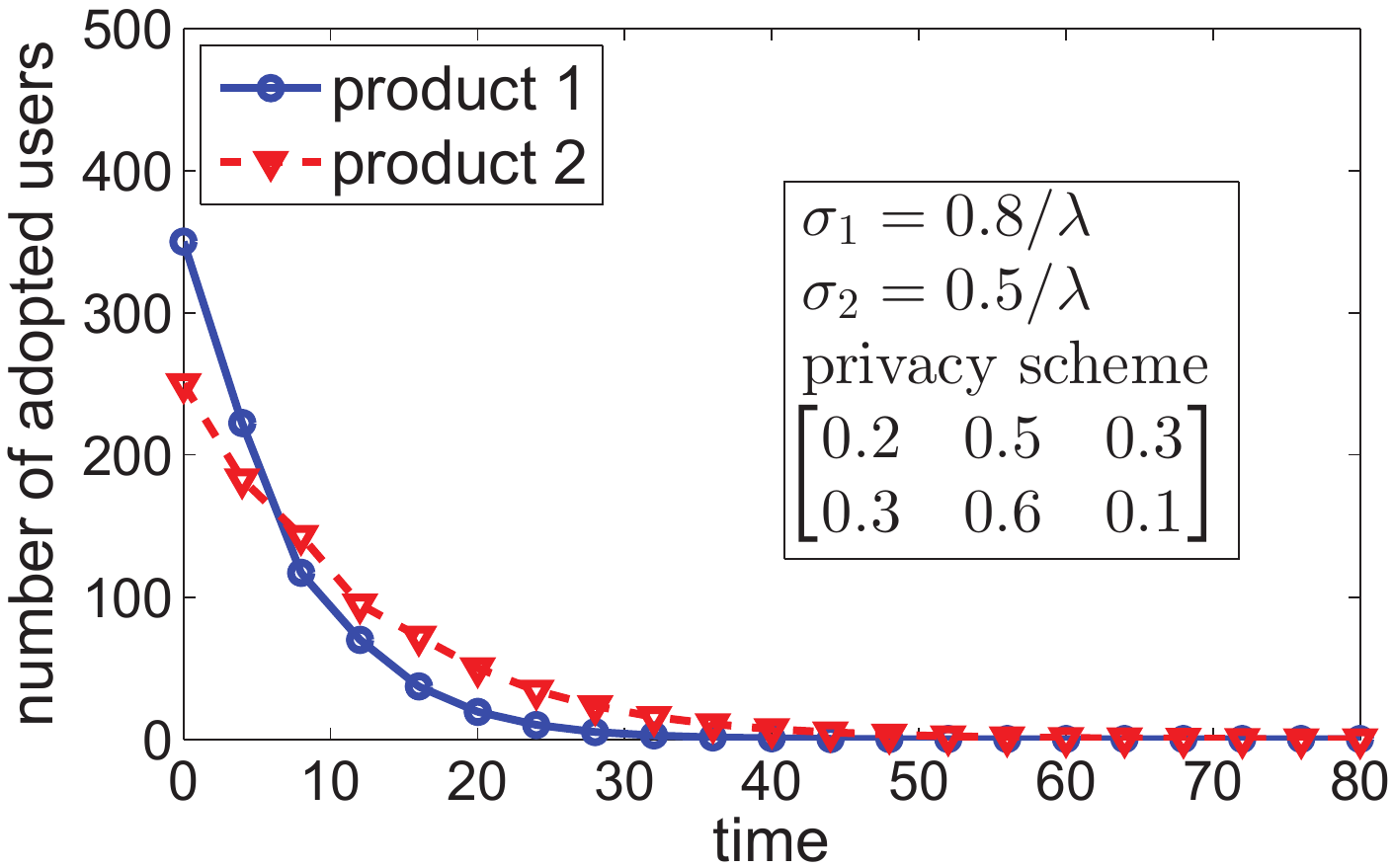}}
\vspace{-9pt}\caption{Experiments of \mbox{privacy-aware} competitive diffusion on a complete network of $2000$ nodes. We   set $\sigma_1$ and $\sigma_2$ both smaller than $1/\lambda$.} \label{completegraph-no-coexist}
\end{figure}

\clearpage
}

For a comprehensive study, we consider networks of different scales: a small-scale social network called  \textit{HumanSocial}, a large-scale physical contact network called  \textit{PhysicalContact}, and a vast-scale Google+ network \cite{gong2012evolution}. The \textit{HumanSocial} and  \textit{PhysicalContact} networks are ``physical" social networks, while the Google+ network is an online social network. The \textit{HumanSocial} network  from the Koblenz Network Collection\footnote{\url{http://konect.uni-koblenz.de/networks/moreno_health}} has 2,539 nodes and 12,969 edges, where nodes represent students, and an edge between two nodes means that one node is among the other node's 5 best female friends or 5 best male friends (we remove directions in edges of the original network). The \textit{PhysicalContact} network~\cite{NDSSL} represents a synthetic population of
the city of Portland, Oregon, USA, and contains about 31 millions links
(interactions) among about 1.6 millions nodes (people). The Google+ network \cite{gong2012evolution},
crawled from July 2011 to October 2011,  consists of
28,942,911 users and 947,776,172 edges.

In Figures \ref{diffusion1} (a)--(d), we perform experiments on the \textit{HumanSocial} network. The abbreviation ``CD'' on the top of figures is short for competitive diffusion, and $\lambda$ denotes the largest eigenvalue of the adjacency matrix of the network. In  Figure \ref{diffusion1a} (resp., \ref{diffusion1b}),  we consider \mbox{privacy-aware} (resp., \mbox{privacy-oblivious}) competitive diffusion, and set $\sigma_1$ and $\sigma_2$ both greater than $1/\lambda$ (i.e., the
epidemic threshold for single-product diffusion). In  Figure \ref{diffusion1c} (resp., \ref{diffusion1d}),  we look at \mbox{privacy-aware} (resp., \mbox{privacy-oblivious}) competitive diffusion and consider $\sigma_1>1/\lambda>\sigma_2$. For \mbox{privacy-aware} competitive diffusion in Figures \ref{diffusion1a} and \ref{diffusion1c}, we   observe the co-existence of products in the equilibrium where the number of adopted users for each product is positive. For \mbox{privacy-oblivious} competitive diffusion in Figures \ref{diffusion1b} and \ref{diffusion1d}, we   see that the product with weaker strength will die out. We also note that for the initial numbers of adopted users, we select the same combinations across some figures and different combinations across some other figures for a comprehensive comparison.
 In Figures \ref{diffusion2} (a)--(d) and \ref{PhysicalContact} (a)--(d), we perform experiments on the large-scale Google+ network and \textit{PhysicalContact} network. The explanations of  Figures \ref{diffusion2} (a)--(d) and \ref{PhysicalContact} (a)--(d) are similar to those of Figures \ref{diffusion1} (a)--(d), and we do not repeat them here.

 Figures \ref{diffusion1}--\ref{PhysicalContact} (a)--(d) support Theorem \ref{thm:main:generalgraph-aux} and its Remark \ref{rem-thm1-cond}. We further use  Figures \ref{diffusion5} (a)--(f) to confirm Theorem \ref{thm:main:generalgraph-aux-partii} and its Remark \ref{rem-thm3-cond}, where we show that if $\sigma_1<1/\lambda$ and  $\sigma_2<1/\lambda$, then no privacy scheme $
\begin{bmatrix}
    r_{11} & r_{12} & 1 - r_{11} -  r_{12} \\     r_{21} & r_{22}&  1 - r_{21} -  r_{22}
     \end{bmatrix}
$ induces the co-existence of products in the equilibrium. We consider \mbox{privacy-aware} competitive diffusion in all of Figures \ref{diffusion5} (a)--(f). Figures \ref{diffusion5a} and \ref{diffusion5b} are for the \textit{HumanSocial} network; Figures \ref{diffusion5c} and \ref{diffusion5d} are for the Google+ network; Figures \ref{diffusion5e} and \ref{diffusion5f} are for the \textit{PhysicalContact} network. As we can see, both products become extinct in Figures \ref{diffusion5} (a)--(d). This is in consistence with Theorem \ref{thm:main:generalgraph-aux-partii} and its Remark \ref{rem-thm3-cond}.

Theorem \ref{thm:main:compete-graph} considers complete graphs, so we plot Figures \ref{completegraph-coexist} and \ref{completegraph-no-coexist} for complete graphs. We consider the case of $\sigma_1>1/\lambda$ and  $\sigma_2>1/\lambda$ in Figures \ref{completegraph-coexist} (a) and (b), consider $\sigma_1>1/\lambda>\sigma_2$ in Figures \ref{completegraph-coexist} (c) and (d), and consider $\sigma_1<1/\lambda$ and  $\sigma_2<1/\lambda$ in Figure \ref{completegraph-no-coexist} ($\lambda$ equals $n-1$ for complete graphs of $n$ nodes, as explained in Footnote \ref{footnotelambdan1}). For \mbox{privacy-aware} competitive diffusion in Figures \ref{completegraph-coexist1} and \ref{completegraph-coexist3}, we   observe the co-existence of products in the equilibrium where the number of adopted users for each product is positive. For \mbox{privacy-oblivious} competitive diffusion in Figures \ref{completegraph-coexist2} and \ref{completegraph-coexist4}, the product with weaker strength will die out. Hence, Figures \ref{completegraph-coexist} (a)--(d) and \ref{completegraph-no-coexist} (a)--(b) all confirm Theorem \ref{thm:main:compete-graph}.


Summarizing the above, the experiments have confirmed our theoretical results in Section \ref{section:Results}.

\section{Proof Sketches of Theorem \ref{thm:main:generalgraph-aux}--\ref{thm:main:generalgraph-aux-partii}} \label{section:Preliminaries}

In this section, we sketch the proofs for establishing Theorem \ref{thm:main:generalgraph-aux}--\ref{thm:main:generalgraph-aux-partii} of Section \ref{section:Results}.

\subsection{Basic Proof Ideas} \label{subsection:basic-ideas}

As discussed in the ``Technical approach'' paragraph on Page \pageref{para-approach}, the proofs consist of the following steps:
\begin{itemize}
\item \textbf{Dynamical System.} We build a   dynamical
system of differential equations to  describe the \mbox{privacy-aware} competitive diffusion problem.
\item \textbf{Finding  Equilibria.} Based on the dynamical system, we find all possible equilibria.
\item \textbf{Stability of Equilibria.} We analyze whether each equilibrium is stable (i.e., attracting).
\end{itemize}
We now elaborate the above three steps respectively.

\textbf{Dynamical System.} We construct a   dynamical
system of differential equations to  describe the \mbox{privacy-aware} competitive diffusion \vspace{1pt} problem $S{I_1}{I_2}S$ with a privacy scheme $
\begin{bmatrix}
    r_{11} & r_{12} & 1 - r_{11} -  r_{12} \\     r_{21} & r_{22}&  1 - r_{21} -  r_{22}
     \end{bmatrix}
$.

Let $\bd{A}$ be the adjacency matrix of the social network of
$n$ users (i.e., nodes) numbered from $1$ to $n$.  Then $a_{ij}$ denotes the entry in the $i$th row and the $j$the entry of $\bd{A}$: $a_{ij}=1$ if there is a link between users $i$ and $j$, and  $a_{ij}=0$ otherwise (we set $a_{ii}=0$ so there are no self-links). We consider undirected networks so $\bd{A}$ is symmetric. For $i=1,2,\ldots,n$, let $p_{i,1}$ denote the probability of node $i$ being in the $I_1$
state (i.e., using product 1). Similarly, we  define $p_{i,2}$ as the probability of node $i$ being in the $I_2$
state (i.e., using product 2), and define $s_{i}$ as the probability of node $i$ being in the $S$
state (i.e., healthy). Clearly, $s_i = 1-p_{i,1}-p_{i,2}$.

For product $1$, its spreading has two different kinds of sources: 1) users adopting product $1$ honestly spread product $1$ (for each user, this happens with probability $r_{11}$); and 2) users adopting product $2$ pretend using product $1$ and disseminate product $1$ (for each user, this happens with probability $r_{21}$). Hence, for each user $i$, another user $j$ contributes to $i$'s adoption of product $1$ with a rate of $ \beta_1 \big[a_{ij} (r_{11}p_{j,1} + r_{21} p_{j,2})\big]$, where $a_{ij}$ is multiplied so that the contribution is non-zero only if $j$ is $i$'s  neighbor in the network, and $\beta_1$ is multiplied since it is the infection rate of product $1$. Then we take a summation for $j$ from $1$ to $n$ and have $ \beta_1 \sum_{j=1}^{n}\big[a_{ij} (r_{11}p_{j,1} + r_{21} p_{j,2})\big]$ (note that $a_{ij}=1$ for neighboring nodes $i$ and $j$,  $a_{ij}=0$ for non-neighboring nodes $i$ and $j$, and also $a_{ii}=0$ so this summation essentially considers only $i$'s neighbors). We further multiply this summation with $s_i$ (the probability that node $i$ is susceptible) to obtain the rate that contributes to the increase of $p_{i,1}$ (i.e., the probability that node $i$ adopts product $1$) over time. On the other hand, since the healing rate of product $1$ is $\delta_1$, we multiply $\delta_1$
with $p_{i,1}$ to obtain the rate that contributes to the decrease of $p_{i,1}$ over time. Given these, we finally obtain \vspace{1pt} that $\frac{\de p_{i,1}}{\de t}$ is the result of  the rate $\beta_1 s_i  \bigg\{  \sum_{j=1}^{n}\big[a_{ij} (r_{11}p_{j,1} + r_{21} p_{j,2})\big] \bigg\}$ minus \vspace{1pt} the rate $\delta_1 p_{i,1}$; i.e., we have (\ref{dp1idt}) below. Similarly, by analyzing the change of $p_{i,2}$ (i.e., the probability that node $i$ adopts product $2$) over time,  we obtain (\ref{dp2idt}) below. Hence, the dynamical system characterizing   our \mbox{privacy-aware} $SI_1I_2S$ competitive diffusion is given by
\begin{subnumcases}{\hspace{-14pt}}
\frac{\de p_{i,1}}{\de t} \\ =  -\delta_1 p_{i,1}  +  \beta_1 s_i  \bigg\{  \sum_{j=1}^{n} \big[a_{ij} (r_{11}p_{j,1} + r_{21} p_{j,2})\big] \bigg\}
,\label{dp1idt} \\ \frac{\de p_{i,2}}{\de t}  \\ =  -\delta_2 p_{i,2}  +  \beta_2 s_i   \bigg\{  \sum_{j=1}^{n} \big[a_{ij} (r_{12}p_{j,1} + r_{22} p_{j,2})\big] \bigg\}
. \label{dp2idt}
\end{subnumcases}




\textbf{Equilibria.} With the dynamical system given above, it is straightforward to characterize possible equilibria.

At an equilibrium, $\frac{\de p_{i,1}}{\de t} =0$ and $\frac{\de p_{i,2}}{\de t} =0$ hold. Applying these to (\ref{dp1idt}) and (\ref{dp2idt}), and recalling $\sigma_1 \da \beta_1 / \delta_1$ and $\sigma_2 \da \beta_2 / \delta_2$, we obtain
\begin{align}
{p}_{i,1}  &=   \sigma_1 {s}_i  \bigg\{  \sum_{j=1}^{n}\big[a_{ij} (r_{11}{p}_{j,1} + r_{21} {p}_{j,2})\big] \bigg\}
,\label{dp1idt-equilibrium}
\end{align}
and
\begin{align}
 {p}_{i,2} & =   \sigma_2  {s}_i   \bigg\{  \sum_{j=1}^{n}\big[a_{ij} (r_{12}{p}_{j,1} + r_{22} {p}_{j,2})\big] \bigg\}
. \label{dp2idt-equilibrium}
\end{align}

To write (\ref{dp1idt-equilibrium}) and (\ref{dp2idt-equilibrium}) conveniently in the vector/matrix form, we let $\bd{P}_1$ (resp., $\bd{P}_2$) be the column vector $[{p}_{1,1},{p}_{2,1},\ldots,{p}_{n,1}]^T$ (resp., $[{p}_{1,2},{p}_{2,2},\ldots,{p}_{n,2}]^T$), where ``$^T$'' means ``transpose''; i.e., the vector $\bd{P}_1$ (resp., $\bd{P}_2$) contains the probability of each user adopting product 1 (resp., product~2). We also define $\bd{S}$ as the column vector $[s_{1},s_{2},\ldots,s_{n}]^T$; i.e., the vector $\bd{S}$ contains the probability of each user not adopting any product.

Writing (\ref{dp1idt-equilibrium}) and (\ref{dp2idt-equilibrium}) in the vector/matrix form, we have
\begin{align}
 \bd{P}_1   &=    \sigma_1 \bd{{S}} \bd{A} (r_{11} \bd{{P}}_1 + r_{21} \bd{{P}}_2)
, \label{dp1idt-equilibrium-matrix}
\end{align}
and
\begin{align}
 \bd{P}_2   &=    \sigma_2 \bd{{S}} \bd{A} (r_{12} \bd{{P}}_1 + r_{22} \bd{{P}}_2)
. \label{dp2idt-equilibrium-matrix}
\end{align}

\subsection{Ideas and Challenges to Establish Theorems \ref{thm:main:generalgraph-aux} and \ref{thm:main:compete-graph}} \label{sec-ideas-thm:main:generalgraph-aux}

 From the statements of Theorem \ref{thm:main:generalgraph-aux} and \ref{thm:main:compete-graph} on Page \pageref{thm:main:generalgraph-aux} and the proof ideas in Section \ref{subsection:basic-ideas}, we will find all equilibria and show that the system has only one stable equilibrium. In this equilibrium, both products co-exist; namely, all elements of $\bd{P}_1$ and $\bd{P}_2$ are positive (i.e., greater than $0$). Compared with finding the equilibrium points, \textit{proving the stability} of the co-existence equilibrium is more challenging here. From stability  theory, we will show that all eigenvalues of the corresponding Jacobian matrix have negative real parts. Due to the introduction of privacy, the dynamics for the two products are coupled together, so the Jacobian matrix is dense in the sense that most (or even all) entries are non-zero. This makes it significantly challenging to analyze the Jacobian matrix's eigenvalues.


\subsection{Ideas and Challenges to Establish Theorem \ref{thm:main:generalgraph-aux-partii}} \label{sec-ideas-thm:main:generalgraph-aux-partii}

 From the statement of Theorem \ref{thm:main:generalgraph-aux-partii}, our goal is to show that the system has only one stable equilibrium, which turns out to be a zero equilibrium; i.e., all elements of $\bd{P}_1$ and $\bd{P}_2$ are zero in this equilibrium. The idea here is to connect our equilibrium with the equilibrium of an $SIS$ diffusion problem, and prove the former  equilibrium being zero by first showing the latter equilibrium being zero. The challenging part is to connect our \mbox{privacy-aware} $SI_1I_2S$ diffusion problem with an $SIS$ diffusion problem, since the privacy scheme considered in Theorem \ref{thm:main:generalgraph-aux-partii} is very general.

\section{Conclusion}
\label{sec:Conclusion}

In competitive diffusion where two products compete for user adoption, prior studies show that the weaker product will soon become
extinct. However, in practice, competing products often co-exist in the market. We find that considering user privacy can address this discrepancy. More specifically, we incorporate user privacy into competitive diffusion to propose the novel problem of {\mbox{privacy-aware} competitive diffusion}, and formally show that privacy can enable the co-existence of competing products.

\normalsize
\setcounter{section}{0}

\appendices

\section{Useful Lemmas} \label{section:Building-Blocks}

To establish our main results, we find it useful to present Lemmas \ref{lem-si}--\ref{lem-matrixSA} below. These lemmas apply for a \mbox{privacy-aware} $SI_1I_2S$ problem with a privacy scheme $
\begin{bmatrix}
    r_{11} & r_{12} & 1 - r_{11} -  r_{12} \\     r_{21} & r_{22}&  1 - r_{21} -  r_{22}
     \end{bmatrix}
$ satisfying $\begin{Bmatrix}
 0 < r_{11} < 1,~~ &  ~~ 0 < r_{12} < 1, \\
 0 < r_{21} < 1,~~ &    ~~ 0 < r_{22} < 1, \\    r_{11} + r_{12} \leq 1,~~&~~    r_{21} + r_{22} \leq 1.
\end{Bmatrix} $. The proofs of Lemmas \ref{lem-si}--\ref{lem-matrixSA} are presented in Appendix A of the online full version \cite{fullversion}, due to space limitation.

\setlist{leftmargin=15pt}

\begin{lem} \label{lem-si}
At an equilibrium, we have $s_i \neq 0$ for any $i=1,2,\ldots,n$.
\end{lem}

\begin{lem} \label{lem-pi1i2-zero-P1P2}
At an equilibrium, we have
\begin{itemize}[leftmargin=45pt]
\item[\textbf{either} (i)]   $p_{i,1}=0$ and $p_{i,2}=0$ for $i=1,2,\ldots,n$ such that $\bd{P}_1 = \bd{0}$ and $\bd{P}_2 = \bd{0}$,
\item[\textbf{or} (ii)]   $p_{i,1}>0$ and $p_{i,2}>0$ for $i=1,2,\ldots,n$ such that $\bd{P}_1$ and $\bd{P}_2$ are both positive vectors (a positive vector means that each dimension is positive).
\end{itemize}
\end{lem}

\begin{lem} \label{lem-matrixSA}
At an equilibrium, we have the following properties for the matrix $\bd{{S}} \bd{A}$.
\begin{itemize}
\item[(i)] $\bd{{S}} \bd{A}$ is non-negative and irreducible.
\item[(ii)] $\bd{{S}} \bd{A}$ has a unique positive real number (say $\lambda
(\bd{{S}} \bd{A})$) as its largest eigenvalue (in magnitude). Furthermore,
the algebraic multiplicity of $\lambda
(\bd{{S}} \bd{A})$ is 1, and it has a positive   eigenvector.
\item[(iii)] Except the largest eigenvalue $\lambda
(\bd{{S}} \bd{A})$, no other eigenvalue of $\bd{{S}} \bd{A}$ has a positive   eigenvector.
\end{itemize}

\end{lem}

\section{{Proving Theorem \ref{thm:main:generalgraph-aux} and its Remark \ref{rem-thm1-cond}}}
 \label{sec:Establishing:thm:main:generalgraphA}

This section is organized as follows. We prove Remark \ref{rem-thm1-cond} of  Theorem \ref{thm:main:generalgraph-aux} in Appendix \ref{sec:Establishing:thm:main:generalgraphC}, and establish Theorem \ref{thm:main:generalgraph-aux}
 in Appendix \ref{sec:Establishing:thm:main:generalgraphC-aux}.

\subsection{{Proving Remark \ref{rem-thm1-cond} of  Theorem \ref{thm:main:generalgraph-aux}}} \label{sec:Establishing:thm:main:generalgraphC}


The proof of Remark \ref{rem-thm1-cond} is straightforward and is still presented here for clarity.
We discuss the three cases of Remark \ref{rem-thm1-cond}.
\begin{itemize}
\item[(a)] Under $\sigma_1>1/\lambda$ and $\sigma_2>1/\lambda$, we will show that
\begin{align}
\hspace{-14pt}\begin{Bmatrix}
\text{(i): }\sigma_1 \gamma_1+\sigma_2 \gamma_2>1/\lambda, \\ \text{(ii): } 0 < \gamma_{1} < 1,~~\text{(iii): }  0 < \gamma_{2} < 1 ,~~\text{(iv): }   \gamma_{1} + \gamma_{2} \leq 1.
\end{Bmatrix} \label{rconditions-perfect-newcond-remrep}
\end{align}
is equivalent to
\begin{align}
\begin{Bmatrix}\text{(i): }0<\gamma_{1}<1,\\ \text{(ii): } \max\big\{0,\frac{\lambda^{-1}-\sigma_1\gamma_{1}}{\sigma_2}\big\}<\gamma_{2}\leq 1 - \gamma_{1}.
\end{Bmatrix}  \label{rconditions-perfect-newcond1}
\end{align}
To see (\ref{rconditions-perfect-newcond1}) $\Longrightarrow$ (\ref{rconditions-perfect-newcond-remrep}), we obtain (\ref{rconditions-perfect-newcond-remrep})-(i) from $\gamma_{2}>\frac{\lambda^{-1}-\sigma_1\gamma_{1}}{\sigma_2}$ of (\ref{rconditions-perfect-newcond1})-(ii), while the inequalities (ii) (iii) (iv) in the second row of (\ref{rconditions-perfect-newcond-remrep}) follow clearly from (\ref{rconditions-perfect-newcond1}). To see (\ref{rconditions-perfect-newcond-remrep}) $\Longrightarrow$ (\ref{rconditions-perfect-newcond1}), we have (\ref{rconditions-perfect-newcond1})-(i) from (\ref{rconditions-perfect-newcond-remrep})-(ii) and  obtain (\ref{rconditions-perfect-newcond1})-(ii) from (\ref{rconditions-perfect-newcond-remrep})-(i) (iii) (iv).
\item[(b)] Under $\sigma_1 >1/\lambda \geq \sigma_2$, we will show that (\ref{rconditions-perfect-newcond-remrep}) is equivalent to
\begin{align}
\begin{Bmatrix} \text{(i): }\frac{\lambda^{-1}-\sigma_2}{\sigma_1-\sigma_2}<\gamma_{1}<1,\\ \text{(ii): }\max\big\{0,\frac{\lambda^{-1}-\sigma_1\gamma_{1}}{\sigma_2}\big\}<\gamma_{2}\leq 1 - \gamma_{1}.
\end{Bmatrix}. \label{rconditions-perfect-newcond2}
\end{align}
To see (\ref{rconditions-perfect-newcond2}) $\Longrightarrow$ (\ref{rconditions-perfect-newcond-remrep}), we obtain (\ref{rconditions-perfect-newcond-remrep})-(i) from $\gamma_{2}>\frac{\lambda^{-1}-\sigma_1\gamma_{1}}{\sigma_2}$ of (\ref{rconditions-perfect-newcond2})-(ii), obtain \vspace{1pt} (\ref{rconditions-perfect-newcond-remrep})-(ii) from (\ref{rconditions-perfect-newcond2})-(i) and the condition $1/\lambda \geq \sigma_2$ here, obtain (\ref{rconditions-perfect-newcond-remrep})-(iii) from (\ref{rconditions-perfect-newcond2})-(ii) and the just-proved result $ \gamma_{1} > 0$ of (\ref{rconditions-perfect-newcond-remrep})-(ii), and   obtain (\ref{rconditions-perfect-newcond-remrep})-(iv) from (\ref{rconditions-perfect-newcond2})-(ii). To see (\ref{rconditions-perfect-newcond-remrep}) $\Longrightarrow$ (\ref{rconditions-perfect-newcond2}), we obtain (\ref{rconditions-perfect-newcond2})-(i) from (\ref{rconditions-perfect-newcond-remrep})-(i) (ii) (iv), obtain (\ref{rconditions-perfect-newcond2})-(ii) from (\ref{rconditions-perfect-newcond-remrep})-(i) (iii) (iv).
\item[(c)] Under $\sigma_2 >1/\lambda \geq \sigma_1$, we will show that (\ref{rconditions-perfect-newcond}) is equivalent to
\begin{align}
 \begin{Bmatrix}\text{(i): } 0<\gamma_{1}<\frac{\sigma_2-\lambda^{-1}}{\sigma_2-\sigma_1},\\ \text{(ii): } \frac{\lambda^{-1}-\sigma_1\gamma_{1}}{\sigma_2}<\gamma_{2}\leq 1 - \gamma_{1}.
\end{Bmatrix}. \label{rconditions-perfect-newcond3}
\end{align}
To see (\ref{rconditions-perfect-newcond3}) $\Longrightarrow$ (\ref{rconditions-perfect-newcond-remrep}), we obtain  (\ref{rconditions-perfect-newcond-remrep})-(i) from $\gamma_{2}>\frac{\lambda^{-1}-\sigma_1\gamma_{1}}{\sigma_2}$ of (\ref{rconditions-perfect-newcond3})-(ii), obtain (\ref{rconditions-perfect-newcond-remrep})-(ii) from (\ref{rconditions-perfect-newcond3})-(i) \vspace{1pt} and the condition $1/\lambda \geq \sigma_1$ here, obtain (\ref{rconditions-perfect-newcond-remrep})-(iii) from (\ref{rconditions-perfect-newcond3})-(ii), the condition $1/\lambda \geq \sigma_1$ here and the just-proved result $0<\gamma_{1}<1$ of (\ref{rconditions-perfect-newcond-remrep})-(ii), and   obtain (\ref{rconditions-perfect-newcond-remrep})-(iv) from (\ref{rconditions-perfect-newcond3})-(ii). To see (\ref{rconditions-perfect-newcond-remrep}) $\Longrightarrow$ (\ref{rconditions-perfect-newcond3}), we obtain (\ref{rconditions-perfect-newcond3})-(i) from (\ref{rconditions-perfect-newcond-remrep})-(i) (ii) (iv), obtain (\ref{rconditions-perfect-newcond3})-(ii) from (\ref{rconditions-perfect-newcond-remrep})-(i) (iii) (iv).
\end{itemize}


\subsection{{Proof of  Theorem \ref{thm:main:generalgraph-aux}}} \label{sec:Establishing:thm:main:generalgraphC-aux}

We now prove Theorem \ref{thm:main:generalgraph-aux} in detail. Specifically, based on (\ref{dp1idt}) and (\ref{dp2idt}), we will discuss equilibrium points in Appendix \ref{thm:main:generalgraph-aux-Equilibrium} and their stability in Appendix \ref{thm:main:generalgraph-aux-Stability}. Although Theorem \ref{thm:main:generalgraph-aux} considers   privacy schemes in the form of $
\begin{bmatrix}
    \gamma_{1} & \gamma_{2} & 1 - \gamma_{1} -  \gamma_{2} \\     \gamma_{1} & \gamma_{2}&  1 - \gamma_{1} -  \gamma_{2}
     \end{bmatrix}
$, we will often use a general privacy scheme $
\begin{bmatrix}
    r_{11} & r_{12} & 1 - r_{11} -  r_{12} \\     r_{21} & r_{22}&  1 - r_{21} -  r_{22}
     \end{bmatrix}
$ to \vspace{2pt} present a general analysis, which will be useful later for proving other theorems.


\subsubsection{\textbf{Equilibrium points}} \label{thm:main:generalgraph-aux-Equilibrium}~

From Lemma \ref{lem-pi1i2-zero-P1P2}, an equilibrium
\begin{itemize}
\item[(i)] \textbf{either} satisfies $p_{i,1}=0$ and $p_{i,2}=0$ for $i=1,2,\ldots,n$ such that $\bd{P}_1 = \bd{0}$ and $\bd{P}_2 = \bd{0}$,
\item[(ii)] \textbf{or} satisfies $p_{i,1}>0$ and $p_{i,2}>0$ for $i=1,2,\ldots,n$ such that $\bd{P}_1$ and $\bd{P}_2$ are both positive vectors (a positive vector means that each dimension is positive).
\end{itemize}
We further analyze the second case below.

Computing $\textrm{(\ref{dp1idt-equilibrium-matrix})}\times \sigma_2 r_{22}  - \textrm{(\ref{dp2idt-equilibrium-matrix})}\times \sigma_1 r_{21} $, we have
\begin{align}
\sigma_2 r_{22}  \bd{P}_1  -\sigma_1r_{21}  \bd{P}_2 =\sigma_1 \sigma_2 (r_{11}r_{22}-r_{12}r_{21})\bd{S} \bd{A} \bd{P}_1 . \label{SAP1eq}
\end{align}
Computing $\textrm{(\ref{dp1idt-equilibrium-matrix})}\times \sigma_2 r_{12}  - \textrm{(\ref{dp2idt-equilibrium-matrix})}\times \sigma_1 r_{11} $, we have
\begin{align}
 \sigma_2 r_{12}  \bd{P}_1  -\sigma_1r_{11}  \bd{P}_2= \sigma_1 \sigma_2 (r_{12}r_{21}-r_{11}r_{22})\bd{S} \bd{A} \bd{P}_2. \label{SAP2eq}
\end{align}

 Under  $r_{11}r_{22}=r_{12}r_{21}$ (which clearly holds given $r_{11}=r_{21} = \gamma_{1}$ and $r_{12}=r_{22} = \gamma_{2}$ \vspace{1pt} for the privacy scheme $
\begin{bmatrix}
    \gamma_{1} & \gamma_{2} & 1 - \gamma_{1} -  \gamma_{2} \\     \gamma_{1} & \gamma_{2}&  1 - \gamma_{1} -  \gamma_{2}
     \end{bmatrix}
$ \vspace{1pt} considered in Theorem \ref{thm:main:generalgraph-aux} here), we obtain from
(\ref{SAP1eq}) (resp., (\ref{SAP2eq})) that
\begin{align}
\textstyle{\bd{P}_1 = \frac{\sigma_1r_{21}}{\sigma_2r_{22}}\bd{P}_2\text{ (resp., } \bd{P}_1 = \frac{\sigma_1r_{11}}{\sigma_2r_{12}}\bd{P}_2\text{)}}.\label{eqr11r22r12r21-P1P2-ratio}
\end{align}
 The above two results $\bd{P}_1 = \frac{\sigma_1r_{21}}{\sigma_2r_{22}}\bd{P}_2$ and $\bd{P}_1 = \frac{\sigma_1r_{11}}{\sigma_2r_{12}}\bd{P}_2$ \vspace{2pt} are the same given $r_{11}r_{22}=r_{12}r_{21}$.
 Substituting $\bd{P}_1 = \frac{\sigma_1r_{21}}{\sigma_2r_{22}}\bd{P}_2$ into (\ref{dp1idt-equilibrium-matrix}), we obtain
\begin{align}
 \bd{P}_{\ell}   &=     (\sigma_1 r_{11}+\sigma_2 r_{22})  \bd{{S}} \bd{A} \bd{P}_{\ell}, \text{ for }\ell=1,2
. \label{dpellidt-equilibrium-matrix}
\end{align}
With $\bd{Q}$ denoting $\bd{P}_1+\bd{P}_2$, (\ref{dpellidt-equilibrium-matrix}) implies
\begin{align}
 \bd{Q}   &=     (\sigma_1 r_{11}+\sigma_2 r_{22})  \bd{{S}} \bd{A} \bd{Q}
. \label{dpellidt-equilibrium-matrixQ}
\end{align}
From (\ref{dpellidt-equilibrium-matrixQ}), as will be clear soon, it is useful to look at a traditional \mbox{privacy-oblivious} $SIS$ diffusion problem (not our \mbox{privacy-aware} $SI_1I_2S$ competitive diffusion problem, yet still on the network with adjacency matrix $\bd{A}$) with infection rate $\beta_*$ and healing rate $\delta_*$ satisfying
\begin{align}
\textstyle{\beta_*/\delta_*=\sigma_1 r_{11}+\sigma_2 r_{22}.} \label{equi-betastar-deltastarQ}
\end{align}
For $i=1,2,\ldots,n$, with $q_i$ denoting the probability of node $i$ being infected in the above $SIS$ diffusion problem, it is well-known  that the dynamical system characterizing $SIS$ diffusion is given by
\begin{align}
\frac{\de q_{i}}{\de t} & = -\delta_* q_{i} + \beta_* (1-q_i)   \sum_{j=1}^{n}(a_{ij} q_j) . \label{qitequilibrium}
\end{align}
Below we explain (\ref{qitequilibrium}) for the above $SIS$ diffusion problem.
For each user $i$, another user $j$ contributes to $i$'s infection with a rate of $ \beta_* a_{ij} q_j$, where $a_{ij}$ is multiplied so that the contribution is non-zero only if $j$ is $i$'s  neighbor in the network, and $\beta_*$ is multiplied since it is the infection rate. Then we take a summation for $j$ from $1$ to $n$ and have $\beta_*  \sum_{j=1}^{n}(a_{ij} q_j)$ (note that $a_{ij}=1$ for neighboring nodes $i$ and $j$,  $a_{ij}=0$ for non-neighboring nodes $i$ and $j$, and also $a_{ii}=0$ so this summation essentially considers only $i$'s neighbors). We further multiply this summation with $1-q_i$ (the probability that node $i$ is susceptible) to obtain the rate that contributes to the increase of $q_i$ over time. On the other hand, since the healing rate is $\delta_*$, we multiply $\delta_*$
with $q_i$ to obtain the rate that contributes to the decrease of $q_i$ over time. Given these, we   obtain that $\frac{\de q_i}{\de t}$ is the result of  the rate $\beta_* (1-q_i)   \sum_{j=1}^{n}(a_{ij} q_j)$ minus the rate $\delta_* q_{i}$; i.e., we have (\ref{qitequilibrium}).

In the above $SIS$ problem, at an equilibrium, (\ref{equi-betastar-deltastarQ}) (\ref{qitequilibrium}) and $\frac{\de q_{i}}{\de t}=0$ together imply
\begin{align}
q_i = (\sigma_1 r_{11}+\sigma_2 r_{22})  (1-q_i)   \sum_{j=1}^{n}(a_{ij} q_j). \label{qitequilibrium2}
\end{align}
We recall (\ref{dpellidt-equilibrium-matrixQ}), where $\bd{Q}$ is a column vector with elements $p_{i,1}+p_{i,2}$ for $i=1,2,\ldots,n$, and $\bd{S}$ is a diagonal matrix with elements $1-(p_{i,1}+p_{i,2})$ for $i=1,2,\ldots,n$. This and (\ref{qitequilibrium2}) together mean that $p_{i,1}+p_{i,2}$ can be understood as $q_i$ in (\ref{qitequilibrium2}) for $i=1,2,\ldots,n$ so that {$\bd{Q}$ is an equilibrium of the above $SIS$ problem}. As shown in prior work \cite{wang2003epidemic,van2009virus,van2012epidemic}, with $\lambda$ denoting the largest eigenvalue  of the adjacency matrix $\bd{A}$, for the above $SIS$ problem, on the one hand, if $\beta_*/\delta_*<1/\lambda$, the only equilibrium is $\bd{Q}=\bd{0}$, which implies $\bd{P}_1 = \bd{0}$ and $\bd{P}_2 = \bd{0}$ from $\bd{Q} = \bd{P}_1 + \bd{P}_2$ and $\bd{P}_1 = \frac{\sigma_1r_{21}}{\sigma_2r_{22}}\bd{P}_2$; on the other hand, if
\begin{align}
\beta_*/\delta_*>1/\lambda, \label{betastardeltastarlambda}
\end{align}
  an equilibrium of positive $\bd{Q}$ exists and can be  determined by $\bd{A}$ and $\beta_*/\delta_*$. We denote this   equilibrium by $\bd{V}_{\text{SIS}}(\bd{A},\beta_*/\delta_*)$.
   Then if
  \begin{align}
\sigma_1 r_{11}+\sigma_2 r_{22}>1/\lambda\label{betastardeltastarlambda2}
\end{align}
which holds from (\ref{equi-betastar-deltastarQ}) and (\ref{betastardeltastarlambda}), we use
  $\bd{P}_1 = \frac{\sigma_1r_{21}}{\sigma_2r_{22}}\bd{P}_2$, $\bd{Q}=\bd{P}_1+\bd{P}_2$, and  $\bd{Q}=\bd{V}_{\text{SIS}}(\bd{A},\beta_*/\delta_*)$ to obtain an equilibrium of positive vectors $\bd{P}_1$ and $\bd{P}_2$:
\begin{align}
\bd{P}_1 &= \frac{\sigma_1r_{21}}{\sigma_1r_{21}+\sigma_2r_{22}}\cdot \bd{V}_{\text{SIS}}(\bd{A},\beta_*/\delta_*), \label{positiveP1} \\ \bd{P}_2 &= \frac{\sigma_2r_{22}}{\sigma_1r_{21}+\sigma_2r_{22}}\cdot \bd{V}_{\text{SIS}}(\bd{A},\beta_*/\delta_*). \label{positiveP2}
\end{align}

To summarize, under $r_{11}=r_{21} = \gamma_{1}$, $r_{12}=r_{22} = \gamma_{2}$ and (\ref{betastardeltastarlambda2}), we have the following.
\begin{itemize}[leftmargin=5pt]
\item An equilibrium of positive $\bd{P}_1$ and $\bd{P}_2$ is given by (\ref{positiveP1}) and (\ref{positiveP2}). In Appendix \ref{thm:main:generalgraph-aux-Stability} below, we will show that this equilibrium is stable under (\ref{betastardeltastarlambda2}).
\item Another equilibrium is the zero equilibrium: $\bd{P}_1 = \bd{0}$ and $\bd{P}_2 = \bd{0}$. In Appendix \ref{thm:main:generalgraph-aux-Instability} below, we will show that this equilibrium is unstable under (\ref{betastardeltastarlambda2}).
\end{itemize}



\subsubsection{\textbf{Stability of the equilibrium given by (\ref{positiveP1}) and (\ref{positiveP2})}}\label{thm:main:generalgraph-aux-Stability}~


To prove Theorem \ref{thm:main:generalgraph-aux}, we will show that the equilibrium of positive $\bd{P}_1$ and $\bd{P}_2$ given by (\ref{positiveP1}) and (\ref{positiveP2}) is stable given   $r_{11}=r_{21}$, $r_{12}=r_{22}$, and (\ref{betastardeltastarlambda2}). To this end, we will derive the Jacobian matrix based on (\ref{dp1idt-equilibrium}) and (\ref{dp2idt-equilibrium}), and prove that all of its eigenvalues have negative real parts.

The Jacobian matrix has four parts $\begin{bmatrix} \bd{J}_{11} & \bd{J}_{12} \\
       \bd{J}_{21} & \bd{J}_{22} \end{bmatrix}
$, where \vspace{1pt} each $\bd{J}_{ab}$ for row index $a=1,2$ and column index $b=1,2$ is an $n\times n$ matrix comprising $\frac{1}{\partial p_{j,b}}\partial\frac{\de p_{i,a}}{\de t}$ for row index $i=1,2,\ldots,n$ and column index $j=1,2,\ldots,n$. In other words,
\begin{itemize}
\item $\frac{1}{\partial p_{j,1}}\partial\frac{\de p_{i,1}}{\de t}$ is in the $i$th row and $j$th column of $\bd{J}_{11}$;
\item  $\frac{1}{\partial p_{j,2}}\partial\frac{\de p_{i,1}}{\de t}$ is in the $i$th row and $j$th column of $\bd{J}_{12}$;
\item $\frac{1}{\partial p_{j,1}}\partial\frac{\de p_{i,2}}{\de t}$ is in the $i$th row and $j$th column of $\bd{J}_{21}$;
\item $\frac{1}{\partial p_{j,2}}\partial\frac{\de p_{i,2}}{\de t}$ is in the $i$th row and $j$th column of $\bd{J}_{22}$.
\end{itemize}

To compute the matrix $\bd{J}_{11}$ which contains \vspace{1pt} $\frac{1}{\partial p_{j,1}}\partial\frac{\de p_{i,1}}{\de t}$ for $i=1,2,\ldots,n$ and $j=1,2,\ldots,n$, from (\ref{dp1idt}) and $s_i = 1-p_{i,1}-p_{i,2}$, we obtain
\begin{align}
 &\frac{1}{\partial p_{i,1}}\partial\frac{\de p_{i,1}}{\de t} \nonumber \\  &  = -\delta_1  + \beta_1  r_{11}\cdot s_ia_{ii}\nonumber \\  & ~~~~~~~~ - \beta_1   \bigg[ r_{11} \sum_{j=1}^{n}(a_{ij}p_{j,1}) + r_{21} \sum_{j=1}^{n}(a_{ij}p_{j,2}) \bigg], \label{dp1idt-par-p1i}
\end{align}
and for $j \neq i$,
\begin{align}
\frac{1}{\partial p_{j,1}}\partial\frac{\de p_{i,1}}{\de t} = \beta_1  r_{11}\cdot s_ia_{ij}.  \label{dp1idt-par-p1j}
\end{align}
To express $\bd{J}_{11}$ using (\ref{dp1idt-par-p1i}) and (\ref{dp1idt-par-p1j}), we now introduce some notation. Let $\bd{I}$ be the $n\times n$ unit matrix.  We define an $n\times n$ diagonal matrix $\textrm{diag}(\bd{A} \bd{P}_1)$ as follows: if $\bd{A} \bd{P}_1 = [w_1,w_2, \ldots,w_n]^T$, then $\textrm{diag}(\bd{A} \bd{P}_1)$
is the diagonal matrix with elements
$w_1,w_2, \ldots,w_n$ (from the upper left to the lower right).
       Similarly, we also define an $n\times n$ diagonal matrix $\textrm{diag}(\bd{A} \bd{P}_2)$. With the above notation, we use (\ref{dp1idt-par-p1i}) and (\ref{dp1idt-par-p1j}) to  express the matrix $\bd{J}_{11}$ comprising \vspace{1pt} $\frac{1}{\partial p_{j,1}}\partial\frac{\de p_{i,1}}{\de t}$ for row index $i=1,2,\ldots,n$ and column index $j=1,2,\ldots,n$ as follows:
\begin{align}
\bd{J}_{11}=&-\delta_1\bd{I}  + \beta_1 r_{11} \bd{S} \bd{A}\nonumber \\ &- \beta_1 r_{11} \textrm{diag}(\bd{A} \bd{P}_1)-  \beta_1 r_{21} \textrm{diag}(\bd{A} \bd{P}_2). \label{Jacobian11}
\end{align}

To compute the matrix $\bd{J}_{12}$ which contains \vspace{1pt} $\frac{1}{\partial p_{j,2}}\partial\frac{\de p_{i,1}}{\de t}$ for $i=1,2,\ldots,n$ and $j=1,2,\ldots,n$, from (\ref{dp1idt}) and $s_i = 1-p_{i,1}-p_{i,2}$, we obtain
\begin{align}
 & \frac{1}{\partial p_{i,2}}\partial\frac{\de p_{i,1}}{\de t}  \nonumber\\  & =   \beta_1  r_{21}   \cdot  s_i a_{ii} - \beta_1 \bigg[  r_{11} \sum_{j=1}^{n}(a_{ij}p_{j,1}) + r_{21} \sum_{j=1}^{n}(a_{ij}p_{j,2}) \bigg], \label{dp1idt-par-p2i}
\end{align}
and for $j \neq i$,
\begin{align}
\frac{1}{\partial p_{j,2}}\partial\frac{\de p_{i,1}}{\de t} = \beta_1 r_{21} \cdot s_i a_{ij} .  \label{dp1idt-par-p2j}
\end{align}
Combining (\ref{dp1idt-par-p2i}) and (\ref{dp1idt-par-p2j}), we express the matrix $\bd{J}_{12}$ comprising $\frac{1}{\partial p_{j,2}}\partial\frac{\de p_{i,1}}{\de t}$ for \vspace{1pt} row index $i=1,2,\ldots,n$ and column index $j=1,2,\ldots,n$ as follows:
\begin{align}
\bd{J}_{12}=  \beta_1 r_{21} \bd{S} \bd{A}    - \beta_1 r_{11} \textrm{diag}(\bd{A} \bd{P}_1) -  \beta_1 r_{21} \textrm{diag}(\bd{A} \bd{P}_2) . \label{Jacobian12}
\end{align}

To compute the matrix $\bd{J}_{21}$ which contains \vspace{1pt} $\frac{1}{\partial p_{j,1}}\partial\frac{\de p_{i,2}}{\de t}$ for $i=1,2,\ldots,n$ and $j=1,2,\ldots,n$, from (\ref{dp2idt}) and $s_i = 1-p_{i,1}-p_{i,2}$, we obtain
\begin{align}
&\frac{1}{\partial p_{i,1}}\partial\frac{\de p_{i,2}}{\de t}  \nonumber\\  & = \beta_2 r_{12} \cdot  s_i a_{ii} -  \beta_2  \bigg[ r_{12} \sum_{j=1}^{n}(a_{ij}p_{j,1}) + r_{22} \sum_{j=1}^{n}(a_{ij}p_{j,2}) \bigg], \label{dp2idt-par-p1i}
\end{align}
and for $j \neq i$,
\begin{align}
\frac{1}{\partial p_{j,1}}\partial\frac{\de p_{i,2}}{\de t} =  \beta_2 r_{12} \cdot s_i a_{ij}.  \label{dp2idt-par-p1j}
\end{align}
Combining (\ref{dp2idt-par-p1i}) and (\ref{dp2idt-par-p1j}), we express the matrix $\bd{J}_{21}$ comprising $\frac{1}{\partial p_{j,1}}\partial\frac{\de p_{i,2}}{\de t}$ \vspace{1pt} for row index $i=1,2,\ldots,n$ and column index $j=1,2,\ldots,n$ as follows:
\begin{align}
\bd{J}_{21}=   \beta_2 r_{12} \bd{S} \bd{A} -  \beta_2 r_{12} \textrm{diag}(\bd{A} \bd{P}_1)  - \beta_2 r_{22} \textrm{diag}(\bd{A} \bd{P}_2) . \label{Jacobian21}
\end{align}

To compute the matrix $\bd{J}_{22}$ which contains \vspace{1pt} $\frac{1}{\partial p_{j,2}}\partial\frac{\de p_{i,2}}{\de t}$ for $i=1,2,\ldots,n$ and $j=1,2,\ldots,n$, from (\ref{dp2idt}) and $s_i = 1-p_{i,1}-p_{i,2}$, we obtain
\begin{align}
 &\frac{1}{\partial p_{i,2}}\partial\frac{\de p_{i,2}}{\de t} \nonumber\\  &=  -\delta_2 + \beta_2 r_{22} \cdot  s_i a_{ii} \nonumber\\  & ~~~~~~~~ -  \beta_2 \bigg[r_{12}  \sum_{j=1}^{n}(a_{ij}p_{j,1}) + r_{22} \sum_{j=1}^{n}(a_{ij}p_{j,2})\bigg], \label{dp2idt-par-p2i}
\end{align}
and for $j \neq i$,
\begin{align}
\frac{1}{\partial p_{j,2}}\partial\frac{\de p_{i,2}}{\de t} = \beta_2 r_{22}  \cdot s_i a_{ij} .  \label{dp2idt-par-p2j}
\end{align}
Combining (\ref{dp2idt-par-p2i}) and (\ref{dp2idt-par-p2j}), we express the matrix $\bd{J}_{22}$ comprising $\frac{1}{\partial p_{j,2}}\partial\frac{\de p_{i,2}}{\de t}$ \vspace{1pt} for row index $i=1,2,\ldots,n$ and column index $j=1,2,\ldots,n$ as follows:
\begin{align}
\bd{J}_{22}= &  - \delta_2\bd{I}  + \beta_2 r_{22} \bd{S} \bd{A} \nonumber\\  & -  \beta_2 r_{12} \textrm{diag}(\bd{A} \bd{P}_1) - \beta_2 r_{22} \textrm{diag}(\bd{A} \bd{P}_2) . \label{Jacobian22}
\end{align}

Our goal is to show that all eigenvalues of the Jacobian matrix  $\begin{bmatrix} \bd{J}_{11} & \bd{J}_{12} \\
       \bd{J}_{21} & \bd{J}_{22} \end{bmatrix}
$ have \vspace{1pt} negative real parts. By definition, with $\bd{I}_{2n}$ denoting the $2n\times 2n$ unit matrix, $x$ is an
eigenvalue of   $\begin{bmatrix} \bd{J}_{11} & \bd{J}_{12} \\
       \bd{J}_{21} & \bd{J}_{22} \end{bmatrix}
$ if $\left|\begin{bmatrix} \bd{J}_{11} & \bd{J}_{12} \\
       \bd{J}_{21} & \bd{J}_{22} \end{bmatrix} - x\bd{I}_{2n}\right|=0$, where \vspace{1pt}  $|\cdot|$ denotes the determinant. From (\ref{Jacobian11}) (\ref{Jacobian12}) (\ref{Jacobian21})   (\ref{Jacobian22}), $r_{11}=r_{21}$ and $r_{12}=r_{22}$, it follows that
\begin{align}
& \begin{bmatrix} \bd{J}_{11} & \bd{J}_{12} \\
       \bd{J}_{21} & \bd{J}_{22} \end{bmatrix} - x\bd{I}_{2n} \nonumber \\ & = \begin{bmatrix}\hspace{-2pt} \begin{array}{l}  -(\delta_1+x)\bd{I}  + \beta_1 r_{11} \bd{S} \bd{A} \\- \beta_1 r_{11} \textrm{diag}(\bd{A} \bd{P}_1) \\-  \beta_1 r_{11} \textrm{diag}(\bd{A} \bd{P}_2)\end{array} & \hspace{-12pt} \begin{array}{l}  - \beta_1 r_{11} \textrm{diag}(\bd{A} \bd{P}_1) \\+ \beta_1 r_{11} \bd{S} \bd{A} \\-  \beta_1 r_{11} \textrm{diag}(\bd{A} \bd{P}_2)\end{array} \hspace{-4pt}\\[3
       em] \hspace{-2pt}\begin{array}{l} - \beta_2 r_{22} \textrm{diag}(\bd{A} \bd{P}_2)\\ + \beta_2 r_{22} \bd{S} \bd{A}\\ -  \beta_2 r_{22} \textrm{diag}(\bd{A} \bd{P}_1)\end{array} & \hspace{-12pt} \begin{array}{l}  -(\delta_2+x)\bd{I}  + \beta_2 r_{22} \bd{S} \bd{A} \\ - \beta_2 r_{22} \textrm{diag}(\bd{A} \bd{P}_2)\\ -  \beta_2 r_{22}\textrm{diag}(\bd{A} \bd{P}_1)\end{array} \hspace{-4pt} \end{bmatrix}\hspace{-3pt} . \label{JminusxI2n}\end{align}


Without loss of generality, we assume $\delta_1 \leq \delta_2$ below.
In the matrix $\begin{bmatrix} \bd{J}_{11} & \bd{J}_{12} \\
       \bd{J}_{21} & \bd{J}_{22} \end{bmatrix} - x\bd{I}_{2n}$ given by (\ref{JminusxI2n}), we add the $(\ell+n)_{\text{th}}$ column to $\ell_{\text{th}}$ column for $\ell=1,2,\ldots,n$. This does not change the determinant of $\begin{bmatrix} \bd{J}_{11} & \bd{J}_{12} \\
       \bd{J}_{21} & \bd{J}_{22} \end{bmatrix} - x\bd{I}_{2n}$; i.e.,     it holds that
\begin{align}
 & \left|\begin{bmatrix} \bd{J}_{11} & \bd{J}_{12} \\
       \bd{J}_{21} & \bd{J}_{22} \end{bmatrix} - x\bd{I}_{2n}\right|\nonumber \\ &=  \begin{vmatrix} \begin{array}{l}  -(\delta_1+x)\bd{I}  \end{array} & \begin{array}{l}  - \beta_1 r_{11} \textrm{diag}(\bd{A} \bd{P}_1) \\+ \beta_1  r_{11} \bd{S} \bd{A} \\- \beta_1  r_{11} \textrm{diag}(\bd{A} \bd{P}_2)\end{array} \\[3
       em] \begin{array}{l}(\delta_2+x)\bd{I}\end{array} & \begin{array}{l}  -(\delta_2+x)\bd{I}  + \beta_2 r_{22} \bd{S} \bd{A} \\ - \beta_2 r_{22} \textrm{diag}(\bd{A} \bd{P}_2)\\ -  \beta_2 r_{22}\textrm{diag}(\bd{A} \bd{P}_1)\end{array} \end{vmatrix}. \label{JminusxI2nx}\end{align}
To prove the result that any $x$ satisfying $\left|\begin{bmatrix} \bd{J}_{11} & \bd{J}_{12} \\
       \bd{J}_{21} & \bd{J}_{22} \end{bmatrix} - x\bd{I}_{2n}\right|=0$  has a negative real part \textbf{by contradiction}, we assume that there exists  $x$ with a \mbox{non-negative}  real part (note $x$ could be real or imaginary). To analyze (\ref{JminusxI2nx}), we use the following result: for $n\times n$ matrices $A$, $B$, $C$ and $D$, if $A$ is invertible, then $\left|\begin{bmatrix} A & B \\
       C & D \end{bmatrix} \right|=|A|\times |D-C A^{-1} B|$. Since $x$ has a non-negative real part, the matrix $-(\delta_1+x)\bd{I}$ is  invertible. Hence,  we obtain from (\ref{JminusxI2nx}) that
\begin{align}
 & \left|\begin{bmatrix} \bd{J}_{11} & \bd{J}_{12} \\
       \bd{J}_{21} & \bd{J}_{22} \end{bmatrix} - x\bd{I}_{2n}\right|\nonumber \\ & = \begin{vmatrix}-(\delta_1+x)\bd{I} \end{vmatrix}\times \begin{vmatrix}\begin{array}{l}\left(\begin{array}{l}  -(\delta_2+x)\bd{I}  + \beta_2  r_{22} \bd{S} \bd{A} \\ - \beta_2  r_{22} \textrm{diag}(\bd{A} \bd{P}_2)\\ -  \beta_2  r_{22}\textrm{diag}(\bd{A} \bd{P}_1) \end{array}\right) \\ - (\delta_2+x)\bd{I} \times [-(\delta_1+x)^{-1}\bd{I} ]\\ \quad \times  \left(\begin{array}{l} - \beta_1  r_{11} \textrm{diag}(\bd{A} \bd{P}_1) \\+ \beta_1  r_{11} \bd{S} \bd{A} \\- \beta_1  r_{11} \textrm{diag}(\bd{A} \bd{P}_2)\end{array}\right) \end{array}\end{vmatrix}\nonumber \\ & = (-1)^n (\delta_1+x)^n \times \nonumber \\ &  \quad\times \begin{vmatrix}\begin{array}{l}\left(\begin{array}{l}  -\delta_2\bd{I}  + y  \bd{S} \bd{A} \\ - y \textrm{diag}(\bd{A} \bd{P}_2)  -  y \textrm{diag}(\bd{A} \bd{P}_1) \end{array}\right)   - x\bd{I} \end{array}\end{vmatrix} \label{JminusxI2nxy}
\end{align}
for
\begin{align}
y \da \beta_2  r_{22} +\frac{\delta_2+x}{\delta_1+x}\beta_1  r_{11}. \label{xdefy}
\end{align}

 Since $x$ has a non-negative real part, we write $x=a+bi$, where $i$ is the imaginary unit, $a$ and $b$ are real numbers, and $a$ is non-negative (if $b=0$, then $x$ is a real number).
Substituting $x=a+bi$ into  (\ref{xdefy}), we can express $y$ as $c + di$, where
\begin{align}
c & \da \beta_2  r_{22} +\frac{(\delta_1+a)(\delta_2+a)+b^2}{(\delta_1+a)^2+b^2}\cdot\beta_1r_{11},\label{cdefbeta2}
\end{align}
and $
d   \da \frac{b(\delta_1-\delta_2)}{(\delta_1+a)^2+b^2}\cdot\beta_1r_{11}.$ From (\ref{JminusxI2nxy}), since $x$ is assumed to have a non-negative real part, we have \mbox{$(\delta_1+x)^n \neq 0$}  so that $\left|\begin{bmatrix} \bd{J}_{11} & \bd{J}_{12} \\
       \bd{J}_{21} & \bd{J}_{22} \end{bmatrix} - x\bd{I}_{2n}\right|=0$ implies $\begin{vmatrix}\begin{array}{l}\left(\begin{array}{l}  -\delta_2\bd{I}  + y  \bd{S} \bd{A} \\ - y \textrm{diag}(\bd{A} \bd{P}_2)  -  y \textrm{diag}(\bd{A} \bd{P}_1) \end{array}\right)   - x\bd{I} \end{array}\end{vmatrix}=0$; i.e., $x$ is an eigenvalue of $\bd{D}$ for
       \begin{align}
\bd{D} \da   -\delta_2\bd{I}  + y  \bd{S} \bd{A}   - y \textrm{diag}(\bd{A} \bd{P}_2)  -  y \textrm{diag}(\bd{A} \bd{P}_1) . \label{Dmatrixdef}
\end{align}
For any matrix $\bd{M}$, we define $\lambda_{\max}^{\mathbb{R}}$ as the maximum among the real parts of the eigenvalues of $\bd{M}$. From (\ref{Dmatrixdef}) and $y=c + di$, it holds  that
\begin{align}
 & \lambda_{\max}^{\mathbb{R}}(\bd{D}+\bd{D}^{T})\nonumber \\ & = \lambda_{\max}^{\mathbb{R}}\left(\begin{array}{l}-2\delta_2\bd{I}+(c+di)\cdot(\bd{S} \bd{A} + \bd{A}\bd{S})   \\   -2(c+di)\cdot\textrm{diag}(\bd{A} \bd{P}_1)  \\  -2(c+di)\cdot\textrm{diag}(\bd{A} \bd{P}_2) \end{array}\right),\label{lambdaDDT} \end{align}
  where we use $(\bd{S} \bd{A})^T=\bd{A}^T\bd{S}^T=\bd{A}\bd{S}$ since  matrices $\bd{A}$ and $\bd{S}$ are both symmetric.

 From standard linear algebra \cite{horn2012matrix}, if matrices $\bd{X}$ and $\bd{Y}$ are symmetric, then $\lambda_{\max}^{\mathbb{R}}(\bd{X}+\bd{Y}) \leq \lambda_{\max}^{\mathbb{R}}(\bd{X}) + \lambda_{\max}^{\mathbb{R}}(\bd{Y}) $. Hence, we obtain from (\ref{lambdaDDT}) that
\begin{align}
 & \lambda_{\max}^{\mathbb{R}}(\bd{D}+\bd{D}^{T})\nonumber \\ &\leq -2\delta_2 + \lambda_{\max}^{\mathbb{R}}\big((c+di)\cdot(\bd{S} \bd{A} + \bd{A}\bd{S})\big)\nonumber \\ &\quad + \lambda_{\max}^{\mathbb{R}}\big(-2(c+di)\cdot\textrm{diag}(\bd{A} \bd{P}_1)\big)\nonumber \\ &\quad+ \lambda_{\max}^{\mathbb{R}}\big(-2(c+di)\cdot\textrm{diag}(\bd{A} \bd{P}_2)\big). \label{boundlambdamaxDDT}
 \end{align}
 From Lemma \ref{lem-pi1i2-zero-P1P2}, it holds that $p_{j,1}>0$ for $j=1,2,\ldots,n$, implying that $\sum_{j=1}^{n}(a_{ij}p_{j,1})>0$ for $i=1,2,\ldots,n$. Then $\textrm{diag}(\bd{A} \bd{P}_1)$ is a diagonal matrix with all positive entries and hence has all positive
eigenvalues. This means that $\lambda_{\min}\big(\textrm{diag}(\bd{A} \bd{P}_1)\big)$ denoting the minimal eigenvalue of $\textrm{diag}(\bd{A} \bd{P}_1)$ is negative. With $c>0$ from (\ref{cdefbeta2}), $\lambda_{\max}^{\mathbb{R}}\big(-2(c+di)\cdot\textrm{diag}(\bd{A} \bd{P}_1)\big)$ equals $- 2c\cdot \lambda_{\min}\big(\textrm{diag}(\bd{A} \bd{P}_1)\big)$ and thus is negative; i.e.
\begin{align}
\lambda_{\max}^{\mathbb{R}}\big(-2(c+di)\cdot\textrm{diag}(\bd{A} \bd{P}_1)\big) & <0 .\label{lambdacdiAp1neg}
\end{align}
 Similar to the above analysis, we use Lemma \ref{lem-pi1i2-zero-P1P2} and eventually obtain
 \begin{align}
\lambda_{\max}^{\mathbb{R}}\big(-2(c+di)\cdot\textrm{diag}(\bd{A} \bd{P}_2)\big) & <0 .\label{lambdacdiAp2neg}
\end{align}
 Since $\bd{S} \bd{A} + \bd{A}\bd{S}$ is a symmetric and real matrix, its eigenvalues are all real. Then
 \begin{align}
\lambda_{\max}^{\mathbb{R}}\big((c+di)\cdot(\bd{S} \bd{A} + \bd{A}\bd{S})\big) & = c \lambda_{\max}(\bd{S} \bd{A} + \bd{A}\bd{S}) . \label{lambdamaxSAAS}
\end{align}
 Then it follows from (\ref{dpellidt-equilibrium-matrixQ}) that $ \bd{Q}  =     (\sigma_1 r_{11}+\sigma_2 r_{22}) \bd{S}\bd{A} \bd{Q}$. In addition, $\bd{S}\bd{A} $ is a positive vector from Lemma \ref{lem-si}. Combining the above with Lemma \ref{lem-matrixSA}, $\lambda_{\max}(\bd{S}\bd{A})$ denoting the largest eigenvalue  (in magnitude) of $\bd{S}\bd{A}$ is given by
  \begin{align}
\lambda_{\max}(\bd{S} \bd{A})=1/(\sigma_1r_{11}+\sigma_2 r_{22}) . \label{lambdamaxSAASsblam}
\end{align}
From matrix theory \cite{horn2012matrix}, for any real non-negative matrix $\bd{M}$, if $\lambda_{\max}(\bd{M})$ denoting the largest eigenvalue  (in magnitude) of $\bd{M}$ is positive, then $\lambda_{\max}(\bd{M}+\bd{M}^T) \leq 2  \lambda_{\max}(\bd{M})$. This result along with  (\ref{lambdamaxSAASsblam}) above induces
  \begin{align}
\lambda_{\max}(\bd{S} \bd{A} + \bd{A}\bd{S})  & \leq 2 \lambda_{\max}(\bd{S} \bd{A})=2/(\sigma_1r_{11}+\sigma_2 r_{22}). \label{lambdamaxSAAS2}
\end{align}
Combining (\ref{lambdamaxSAAS}) and (\ref{lambdamaxSAAS2}), we have
 \begin{align}  \lambda_{\max}^{\mathbb{R}}\big((c+di)\cdot(\bd{S} \bd{A} + \bd{A}\bd{S})\big)  &   \leq  2c/(\sigma_1r_{11}+\sigma_2 r_{22}). \label{lambdamaxSAAS3}
\end{align}
To use (\ref{lambdamaxSAAS3}), it is useful to bound $c$ defined by (\ref{cdefbeta2}). Recalling $a \geq 0$, we obtain from (\ref{cdefbeta2}) that
 \begin{align}
c & =\beta_2  r_{22} +\frac{(\delta_1+a)(\delta_2+a)+b^2}{(\delta_1+a)^2+b^2}\cdot\beta_1r_{11} \nonumber \\ & = \beta_2  r_{22} + \beta_1r_{11}  \cdot  \bigg(1+\frac{\delta_2-\delta_1}{\delta_1+a}+\frac{b^2}{(\delta_1+a)^2+b^2}\bigg)\nonumber \\ & \leq \beta_2  r_{22} + \beta_1r_{11}  \cdot  \bigg(1+\frac{\delta_2-\delta_1}{\delta_1}\bigg) \nonumber \\ & = \beta_2  r_{22} + \frac{\beta_1r_{11} \delta_2}{\delta_1} \nonumber \\ & = \delta_2\cdot (\sigma_1r_{11}+\sigma_2 r_{22}), \label{cdefbeta2bound}
\end{align}
where the last step uses $\beta_1 = \delta_1 \sigma_1$ and $\beta_2 = \delta_2 \sigma_2$.

Applying (\ref{lambdacdiAp1neg}) (\ref{lambdacdiAp2neg}) (\ref{lambdamaxSAAS3}) and (\ref{cdefbeta2bound}) to (\ref{boundlambdamaxDDT}), we finally derive $\lambda_{\max}^{\mathbb{R}}(\bd{D}+\bd{D}^{T})<0$. From the Lyapunov
theorem~\cite{hirsch2012differential}, it further follows that all eigenvalues of $\bd{D}$ have negative real parts. Hence,  recalling $x$ is an eigenvalue of $\bd{D}$ (see the sentence containing (\ref{Dmatrixdef})), the real part of $x$ is negative, which contradicts with the assumption that  the real part of $x$ is \mbox{non-negative}. Then the assumption does not hold and hence we have proved that any eigenvalue $x$ of the  Jacobian matrix  $\begin{bmatrix} \bd{J}_{11} & \bd{J}_{12} \\
       \bd{J}_{21} & \bd{J}_{22} \end{bmatrix}
$ has negative real parts so that the  equilibrium given by (\ref{positiveP1}) and (\ref{positiveP2}) is stable.
 \hfill$\blacksquare$

\subsubsection{\textbf{Instability of the equilibrium given by $\bd{P}_1 = \bd{0}$ and $\bd{P}_2 = \bd{0}$ for Theorem \ref{thm:main:generalgraph-aux}}}\label{thm:main:generalgraph-aux-Instability}~

Under $\bd{P}_1 = \bd{0}$ and $\bd{P}_2 = \bd{0}$, which implies $\bd{{S}}=\bd{I}_{n}$, we obtain from   (\ref{JminusxI2nxy}) that
\begin{align}
 & \left|\begin{bmatrix} \bd{J}_{11} & \bd{J}_{12} \\
       \bd{J}_{21} & \bd{J}_{22} \end{bmatrix} - x\bd{I}_{2n}\right| \nonumber \\ & = (-1)^n (\delta_1+x)^n \times \begin{vmatrix}\begin{array}{l}\left(\begin{array}{l}  -\delta_2\bd{I}  + y  \bd{A} \\   \end{array}\right)   - x\bd{I} \end{array}\end{vmatrix} \label{JminusxI2nxyP1P20}
\end{align}
for
\begin{align}
y \da \beta_2  r_{22} +\frac{\delta_2+x}{\delta_1+x}\beta_1  r_{11}. \label{xdefyP1P20}
\end{align}
From (\ref{JminusxI2nxyP1P20}), $x$ is an eigenvalue of $\begin{bmatrix} \bd{J}_{11} & \bd{J}_{12} \\
       \bd{J}_{21} & \bd{J}_{22} \end{bmatrix}$ if $-\delta_2\bd{I}  + y   \bd{A}$ has an eigenvalue $x$. Recall that $\lambda$ denotes the largest eigenvalue  of the adjacency matrix $\bd{A}$. Then $-\delta_2\bd{I}  + y   \bd{A}$ has an eigenvalue $-\delta_2\bd{I}  + y  \cdot \lambda$. Setting $-\delta_2\bd{I}  + y  \lambda= x$, we substitute (\ref{xdefyP1P20}) into this equation and have
\begin{align}
\begin{array}{l}x^2 - [(\delta_1+\delta_2)-(\beta_1 r_{11} + \beta_2 r_{22})  \lambda] x \\ ~~~+ \delta_1\delta_2 - (\beta_1 \delta_2 r_{11} + \beta_2
\delta_1 r_{22}) \lambda=0.\end{array} \label{xequationsolve}
\end{align}
Under the condition $\sigma_1 r_{11}+\sigma_2 r_{22}>1/\lambda$, the constant term in the quadratic equation (\ref{xequationsolve}) of $x$ is negative given
\begin{align}
 & \delta_1\delta_2 - (\beta_1 \delta_2 r_{11} + \beta_2
\delta_1 r_{22})  \lambda\nonumber \\ & = \delta_1\delta_2 - \delta_1\delta_2 \cdot (\sigma_1 r_{11}+\sigma_2 r_{22})  \lambda<0,\nonumber
\end{align}
where $\beta_1=\sigma_1\delta_1$ and $\beta_2=\sigma_2\delta_2$ have been used. Hence, there exists a positive real solution to
(\ref{xequationsolve}), which further means that $\begin{bmatrix} \bd{J}_{11} & \bd{J}_{12} \\
       \bd{J}_{21} & \bd{J}_{22} \end{bmatrix}$ has a positive real eigenvalue under $\bd{P}_1 = \bd{0}$ and $\bd{P}_2 = \bd{0}$. Hence, the equilibrium given by $\bd{P}_1 = \bd{0}$ and $\bd{P}_2 = \bd{0}$ is unstable.    \hfill$\blacksquare$

\section{{Proving Theorem  \ref{thm:main:generalgraph-aux-partii} and its Remark \ref{rem-thm3-cond}}} \label{sec:Establishing:thm:main:generalgraphB}

%



\subsection{{Proving Remark \ref{rem-thm3-cond} of Theorem  \ref{thm:main:generalgraph-aux-partii}}} \label{sec:Establishing:thm:main:generalgraphB-partI}


We will prove that for $\sigma_1<1/\lambda$ and $\sigma_2<1/\lambda$, any privacy scheme $
\begin{bmatrix}
    r_{11} & r_{12} & 1 - r_{11} -  r_{12} \\     r_{21} & r_{22}&  1 - r_{21} -  r_{22}
     \end{bmatrix}
$ will satisfy (\ref{equi-condition-lambda-repeat2}). By symmetry, without loss of generality, we can assume $\sigma_1\geq\sigma_2$. We will show that the left hand side of (\ref{equi-condition-lambda-repeat2}) is no greater than $\sigma_1$. Given $r_{12} \leq 1- r_{11}$ and $r_{21} \leq 1- r_{22}$, we explain that the desired result follows once we show
\begin{align}
&\sqrt{(\sigma_1 r_{11}-\sigma_2 r_{22})^2 + 4 \sigma_1 \sigma_2 (1 - r_{11}) (1 - r_{22})} \nonumber \\ & \leq \sigma_1 (2-r_{11}) - \sigma_2 r_{22}. \label{equi-condition-lambda-repeat2prf}
\end{align}
To see (\ref{equi-condition-lambda-repeat2prf}) $\Longrightarrow$ (\ref{equi-condition-lambda-repeat2}), we note that given (\ref{equi-condition-lambda-repeat2prf}), the left hand side of (\ref{equi-condition-lambda-repeat2}) is no greater than $\frac{1}{2}\big[\sigma_1 r_{11} + \sigma_2 r_{22}+\sigma_1 (2-r_{11}) - \sigma_2 r_{22}\big] $\\$\leq \sigma_1<1/\lambda$.

Note that the right hand side of (\ref{equi-condition-lambda-repeat2prf}) is non-negative from $\sigma_1\geq\sigma_2$, $ r_{11} \leq 1$  and $ r_{22} \leq 1$. To prove (\ref{equi-condition-lambda-repeat2prf}), we have
\begin{align}
 & \big[\text{left hand side of (\ref{equi-condition-lambda-repeat2prf})}\big]^2- \big[\text{right hand side of (\ref{equi-condition-lambda-repeat2prf})}\big]^2\nonumber \\ & = (\sigma_1 r_{11}-\sigma_2 r_{22})^2 + 4 \sigma_1 \sigma_2 (1 - r_{11}) (1 - r_{22})\nonumber \\ &\quad - \big[\sigma_1 (2-r_{11}) - \sigma_2 r_{22}\big]^2\nonumber \\ & =  4 \sigma_1 \sigma_1(r_{11}-1) (\sigma_1 -\sigma_2) \leq 0, \nonumber
\end{align}
where the last step uses $ r_{11} \leq 1$  and the assumption $\sigma_1\geq\sigma_2$. Hence, (\ref{equi-condition-lambda-repeat2prf}) is proved so that the left hand side of (\ref{equi-condition-lambda-repeat2}) is no greater than $2\sigma_1$ and further strictly less than $2/\lambda$
under $\sigma_1<1/\lambda$. This means that if $\sigma_1<1/\lambda$ and $\sigma_2<1/\lambda$, then (\ref{equi-condition-lambda-repeat2}) holds for any privacy scheme $
\begin{bmatrix}
    r_{11} & r_{12} & 1 - r_{11} -  r_{12} \\     r_{21} & r_{22}&  1 - r_{21} -  r_{22}
     \end{bmatrix}
$. Hence, given Theorem \ref{thm:main:generalgraph-aux-partii}, we have proved Remark \ref{rem-thm3-cond}.


\subsection{{Establishing Theorem  \ref{thm:main:generalgraph-aux-partii}}} \label{sec:Establishing:thm:main:generalgraphB-partII}

We discuss the following three cases, respectively: i)   $r_{11}r_{22}=r_{12}r_{21}$, ii)  $r_{11}r_{22}>r_{12}r_{21}$, and iii)  $r_{11}r_{22}<r_{12}r_{21}$. The analysis below is for an equilibrium.

In Case i) of $r_{11}r_{22}=r_{12}r_{21}$, we have shown in (\ref{eqr11r22r12r21-P1P2-ratio}) that $\bd{{P}}_1 = \frac{\sigma_1r_{21}}{\sigma_2r_{22}}\bd{{P}}_2$.
In Case ii) or Case iii) below, given $r_{11}r_{22}\neq r_{12}r_{21}$, (\ref{SAP1eq}) and (\ref{SAP2eq}) yield
\begin{align}
\bd{{S}} \bd{A} \bd{{P}}_1  = a_1 \bd{{P}}_1 + b_1 \bd{{P}}_2\text{ for }\begin{array}{l}a_1 \da \frac{\sigma_2 r_{22}}{\sigma_1 \sigma_2 (r_{11}r_{22}-r_{12}r_{21})}\text{ and }\\b_1 \da \frac{-\sigma_1r_{21}}{\sigma_1 \sigma_2 (r_{11}r_{22}-r_{12}r_{21})},\end{array} \label{SAP1eqSAP}
\end{align}
and
\begin{align}
\bd{{S}} \bd{A} \bd{{P}}_2  = a_2 \bd{{P}}_1 + b_2 \bd{{P}}_2\text{ for }\begin{array}{l}a_2 \da \frac{-\sigma_2 r_{12}}{\sigma_1 \sigma_2 (r_{11}r_{22}-r_{12}r_{21})} \text{ and }\\b_2 \da \frac{\sigma_1r_{11}  }{\sigma_1 \sigma_2 (r_{11}r_{22}-r_{12}r_{21})}. \end{array} \label{SAP2eqSAP}
\end{align}

\textbf{Case ii):} We consider $r_{11}r_{22}>r_{12}r_{21}$ here. If there exist scalars $t$ and $c$ satisfying
\begin{align}
 a_1 + t a_2 = c   \label{t1a1t2a2}
\end{align}
and
\begin{align}
 b_1 + t b_2 = c t  \label{t1b1t2b2},
\end{align}
then (\ref{SAP1eqSAP}) (\ref{SAP2eqSAP}) (\ref{t1a1t2a2}) and (\ref{t1b1t2b2}) together induce
\begin{align}
\bd{{S}} \bd{A} (\bd{{P}}_1+t\bd{{P}}_2) & = c(\bd{{P}}_1+t\bd{{P}}_2). \label{SAP1tP2cinduce}
\end{align}
From (\ref{t1a1t2a2}) and (\ref{t1b1t2b2}), $c$ and $t$ are determined from
\begin{align}
t = \frac{c-a_1}{a_2} \label{tequationcaseii}
\end{align}
and
\begin{align}
c^2 - ( a_1+b_2 ) c + a_1 b_2 - a_2 b_1 = 0. \label{cequationcaseii}
\end{align}
Using the expressions of $a_1, a_2, b_1, b_2$ from (\ref{SAP1eqSAP}) and (\ref{SAP2eqSAP}), we write (\ref{cequationcaseii}) as $c^2-\mu_1 c +\nu_1=0$, where $\mu_1 \da \frac{\sigma_1 r_{11}+\sigma_2 r_{22}}{\sigma_1\sigma_2(r_{11}r_{22}-r_{12}r_{21})}$ and $\nu_1 \da \frac{1}{\sigma_1\sigma_2(r_{11}r_{22}-r_{12}r_{21})}$. Under $r_{11}r_{22}>r_{12}r_{21}$ and the condition that $r_{11}$, $r_{12}$, $r_{21}$, $r_{22}$ are positive, we have $\mu_1>0$, $\nu_1>0$, and ${\mu_1}^2-4\nu_1=\frac{(\sigma_1 r_{11}-\sigma_2 r_{22})^2 + 4 \sigma_1 \sigma_2 r_{12}r_{21}}{[\sigma_1\sigma_2(r_{11}r_{22}-r_{12}r_{21})]^2}>0$. \vspace{1.5pt} Then there are two positive   solutions $c_1<c_2$ to (\ref{cequationcaseii}). From (\ref{tequationcaseii}), we define
   $t_1 \da \frac{b_1}{c_1-b_2}$ and $t_2 \da \frac{b_1}{c_2-b_2}$ so it follows from (\ref{SAP1tP2cinduce}) that $\bd{{S}} \bd{A} (\bd{{P}}_1+t_1\bd{{P}}_2)  = c_1(\bd{{P}}_1+t_1\bd{{P}}_2)$ and $\bd{{S}} \bd{A} (\bd{{P}}_1+t_2\bd{{P}}_2)  = c_2(\bd{{P}}_1+t_2\bd{{P}}_2)$. From Lemma \ref{lem-matrixSA}, except the largest eigenvalue of $\bd{{S}} \bd{A}$, no other eigenvalue of $\bd{{S}} \bd{A}$ has a positive   eigenvector. Then at least one of the following cases occur:
\begin{itemize}
\item $\bd{{P}}_1+t_1\bd{{P}}_2$ equals $\bd{0}$,
\item $\bd{{P}}_1+t_2\bd{{P}}_2$ equals $\bd{0}$,
\item $\bd{{P}}_1+t_1\bd{{P}}_2$ equals some constant times of $\bd{{P}}_1+t_2\bd{{P}}_2$.
\end{itemize}
In any case, $\bd{{P}}_1$ equals some constant times of $\bd{{P}}_2$.

\textbf{Case iii):} We consider $r_{11}r_{22}<r_{12}r_{21}$ here. If there exist scalars $t$ and $c$ satisfying
\begin{align}
 a_1 + t a_2 = c t   \label{t1a1t2a2caseiii}
\end{align}
and
\begin{align}
 b_1 + t b_2 = c  \label{t1b1t2b2caseiii},
\end{align}
then (\ref{SAP1eqSAP}) (\ref{SAP2eqSAP}) (\ref{t1a1t2a2caseiii}) and (\ref{t1b1t2b2caseiii}) together induce
\begin{align}
\bd{{S}} \bd{A} (\bd{{P}}_1+t\bd{{P}}_2) & = c(\bd{{P}}_1+t\bd{{P}}_2). \label{SAP1tP2cinduce2}
\end{align}
From (\ref{t1a1t2a2caseiii}) and (\ref{t1b1t2b2caseiii}), $c$ and $t$ are determined from
\begin{align}
t = \frac{c-b_1}{b_2} \label{tequationcaseiii}
\end{align}
and
\begin{align}
c^2 - (a_2 + b_1) c + a_2 b_1 - a_1 b_2= 0. \label{cequationcaseiii}
\end{align}
Using the expressions of $a_1, a_2, b_1, b_2$ from (\ref{SAP1eqSAP}) and (\ref{SAP2eqSAP}), we write (\ref{cequationcaseiii}) as $c^2-\mu_2 c +\nu_2=0$, where $\mu_2 \da \frac{\sigma_1 r_{21}+\sigma_2 r_{12}}{\sigma_1\sigma_2(r_{12}r_{21}-r_{11}r_{22})}$ and $\nu_2 \da \frac{1}{\sigma_1\sigma_2(r_{12}r_{21}-r_{11}r_{22})}$. Under $r_{11}r_{22}<r_{12}r_{21}$ and the condition that $r_{11}$, $r_{12}$, $r_{21}$, $r_{22}$ are positive, we have $\mu_2>0$, $\nu_2>0$, and ${\mu_2}^2-4\nu_2=\frac{(\sigma_1 r_{21}-\sigma_2 r_{12})^2 + 4 \sigma_1 \sigma_2 r_{11}r_{22}}{[\sigma_1\sigma_2(r_{12}r_{21}-r_{11}r_{22})]^2}>0$. \vspace{1.5pt} Then  there are two positive   solutions $c_1<c_2$ to (\ref{cequationcaseiii}). From (\ref{tequationcaseiii}), we define
   $t_1 \da \frac{c_1-b_1}{b_2}$ and $t_2 \da \frac{c_2-b_1}{b_2}$  so it follows from (\ref{SAP1tP2cinduce2}) that $\bd{{S}} \bd{A} (\bd{{P}}_1+t_1\bd{{P}}_2)  = c_1(\bd{{P}}_1+t_1\bd{{P}}_2)$ and $\bd{{S}} \bd{A} (\bd{{P}}_1+t_2\bd{{P}}_2)  = c_2(\bd{{P}}_1+t_2\bd{{P}}_2)$. From Lemma \ref{lem-matrixSA}, except the largest eigenvalue of $\bd{{S}} \bd{A}$, no other eigenvalue of $\bd{{S}} \bd{A}$ has a positive   eigenvector. Then at least one of the following cases occur:
\begin{itemize}
\item $\bd{{P}}_1+t_1\bd{{P}}_2$ equals $\bd{0}$,
\item $\bd{{P}}_1+t_2\bd{{P}}_2$ equals $\bd{0}$,
\item $\bd{{P}}_1+t_1\bd{{P}}_2$ equals some constant times of $\bd{{P}}_1+t_2\bd{{P}}_2$.
\end{itemize}
In any case, $\bd{{P}}_1$ equals some constant times of $\bd{{P}}_2$.






Summarizing cases i) and ii), we can define some $h$ such that $\bd{P}_1 = h \bd{P}_2$. Then (\ref{dp1idt-equilibrium-matrix}) and (\ref{dp2idt-equilibrium-matrix}) imply
\begin{align}
 \bd{P}_1   &=    \sigma_1 \bd{{S}} \bd{A} (r_{11} h + r_{21} )\bd{{P}}_2
 \label{dp1idt-equilibrium-matrix-new}
\end{align}
and
\begin{align}
 \bd{P}_2   &=    \sigma_2 \bd{{S}} \bd{A} (r_{12} h + r_{22} )\bd{{P}}_2
. \label{dp2idt-equilibrium-matrix-new}
\end{align}
From Lemma \ref{lem-matrixSA}, except the largest eigenvalue of $\bd{{S}} \bd{A}$, no other eigenvalue of $\bd{{S}} \bd{A}$ has a positive   eigenvector. This along with  (\ref{dp1idt-equilibrium-matrix-new}) and (\ref{dp2idt-equilibrium-matrix-new}) implies
  \begin{align}
 h & = \frac{\sigma_1}{\sigma_2}\cdot \frac{r_{11}h + r_{21}}{r_{12}h + r_{22}} ,\nonumber
\end{align}
which reduces to
\begin{align}
{\sigma_2}r_{12} h^2
 + ({\sigma_2}r_{22} - {\sigma_1}r_{11})h
   -{\sigma_1}r_{21} & = 0 . \label{equationhsigmaeq}
\end{align}
The quadratic equation (\ref{equationhsigmaeq}) has a positive solution and a negative solution, so  $h$ equals the positive solution; i.e.,
\begin{align}
h & = \frac{ {\sigma_1}r_{11}-{\sigma_2}r_{22} + \sqrt{({\sigma_1}r_{11} - {\sigma_2}r_{22})^2 + 4{\sigma_1}{\sigma_2}r_{12} r_{21}}}{2{\sigma_2}r_{12}} .\label{labelkabfollowsnewsolution}
\end{align}

With $\bd{Q}$ denoting $\bd{P}_1+\bd{P}_2$, (\ref{dp2idt-equilibrium-matrix-new}) and $\bd{P}_1 = h \bd{P}_2$ imply
\begin{align}
 \bd{Q}   &=    \sigma_2 (r_{12} h+ r_{22})\bd{{S}} \bd{A} \bd{Q}
. \label{dpellidt-equilibrium-matrixQ-newneg}
\end{align}
Similar to the analysis after Equation (\ref{dpellidt-equilibrium-matrixQ}), from (\ref{dpellidt-equilibrium-matrixQ-newneg}), it is useful to look at a traditional \mbox{privacy-oblivious} $SIS$ diffusion problem (not our \mbox{privacy-aware} $SI_1I_2S$ competitive diffusion problem, yet still on the network with adjacency matrix $\bd{A}$)
 with infection rate $\beta_*$ and healing rate $\delta_*$ satisfying
 \begin{align}
\textstyle{\beta_*/\delta_*= \sigma_2 (r_{12} h+ r_{22}).} \label{equi-betastar-deltastarQ-newneg}
\end{align}
For $i=1,2,\ldots,n$, with $q_i$ denoting the probability of node $i$ being infected in the above $SIS$ diffusion problem, it is well-known  that the dynamical system characterizing $SIS$ diffusion is given by
\begin{align}
\frac{\de q_{i}}{\de t} & = -\delta_* q_{i} + \beta_* (1-q_i)   \sum_{j=1}^{n}(a_{ij} q_j) . \label{qitequilibriumst}
\end{align}
Below we explain (\ref{qitequilibriumst}) for the above $SIS$ diffusion problem.
For each user $i$, another user $j$ contributes to $i$'s infection with a rate of $ \beta_* a_{ij} q_j$, where $a_{ij}$ is multiplied so that the contribution is non-zero only if $j$ is $i$'s  neighbor in the network, and $\beta_*$ is multiplied since it is the infection rate. Then we take a summation for $j$ from $1$ to $n$ and have $\beta_*  \sum_{j=1}^{n}(a_{ij} q_j)$ (note that $a_{ij}=1$ for neighboring nodes $i$ and $j$,  $a_{ij}=0$ for non-neighboring nodes $i$ and $j$, and also $a_{ii}=0$ so this summation essentially considers only $i$'s neighbors). We further multiply this summation with $1-q_i$ (the probability that node $i$ is susceptible) to obtain the rate that contributes to the increase of $q_i$ over time. On the other hand, since the healing rate is $\delta_*$, we multiply $\delta_*$
with $q_i$ to obtain the rate that contributes to the decrease of $q_i$ over time. Given these, we   obtain that $\frac{\de q_i}{\de t}$ is the result of  the rate $\beta_* (1-q_i)   \sum_{j=1}^{n}(a_{ij} q_j)$ minus the rate $\delta_* q_{i}$; i.e., we have (\ref{qitequilibriumst}) above.

In the above $SIS$ problem, at an equilibrium, (\ref{equi-betastar-deltastarQ-newneg}) (\ref{qitequilibriumst}) and $\frac{\de q_{i}}{\de t}=0$ together imply
\begin{align}
q_i = (\sigma_1 r_{11}+\sigma_2 r_{22})  (1-q_i)   \sum_{j=1}^{n}(a_{ij} q_j). \label{qitequilibriumvb2v2}
\end{align}
We recall (\ref{dpellidt-equilibrium-matrixQ-newneg}), where $\bd{Q}$ is a column vector with elements $p_{i,1}+p_{i,2}$ for $i=1,2,\ldots,n$, and $\bd{S}$ is a diagonal matrix with elements $1-(p_{i,1}+p_{i,2})$ for $i=1,2,\ldots,n$. This and (\ref{qitequilibriumvb2v2}) together mean that $p_{i,1}+p_{i,2}$ can be understood as $q_i$ in (\ref{qitequilibriumvb2v2}) for $i=1,2,\ldots,n$ so that {$\bd{Q}$ is an equilibrium of the above $SIS$ problem}.

As shown in prior work \cite{wang2003epidemic,van2009virus,van2012epidemic}, with $\lambda$ denoting the largest eigenvalue  of the adjacency matrix $\bd{A}$,  if
\begin{align}
\beta_*/\delta_* < 1/\lambda,  \label{equi-betastar-deltastarQ-newneg2}
\end{align}
the only equilibrium for the above $SIS$ problem is $\bd{Q}=\bd{0}$, which   implies $\bd{P}_1 = \bd{0}$ and $\bd{P}_2 = \bd{0}$ from $\bd{Q}=\bd{P}_1+\bd{P}_2$ and $\bd{P}_1 = h \bd{P}_2$. Since (\ref{labelkabfollowsnewsolution}), (\ref{equi-betastar-deltastarQ-newneg}), and (\ref{equi-betastar-deltastarQ-newneg2}) together induce
\begin{align} \frac{1}{2}\left[\begin{array}{l}
	\sigma_1 r_{11} + \sigma_2 r_{22} \\[2pt] + \sqrt{(\sigma_1 r_{11}-\sigma_2 r_{22})^2 + 4 \sigma_1 \sigma_2 r_{12} r_{21}}
\end{array}\right]<  \lambda^{-1}, \label{equi-condition-lambda-repeat-thmeq-newneg}
\end{align}
we obtain that under (\ref{equi-condition-lambda-repeat-thmeq-newneg}),
the only equilibrium for our  $SI_1I_2S$ problem is $\bd{P}_1 = \bd{0}$ and $\bd{P}_2 = \bd{0}$. \hfill$\blacksquare$

\end{document}

Then the real parts of the eigenvalues $\lambda_{\pm}$ are negative if and only if
  \begin{subnumcases}{\hspace{-27pt}}
\hspace{-9pt}\begin{array}{l}
a_{11} + a_{22} < 0\textrm{ and}\\ a_{11} a_{22}>a_{12}a_{21}
  \end{array} \textrm{ for }(a_{11} - a_{22})^2 + 4a_{12}a_{21} > 0,   \label{eigenvaluesconditions1} \vspace{3pt} \\
 a_{11} + a_{22} < 0  ~~~~~\hspace{1pt}\textrm{ for }(a_{11} - a_{22})^2 + 4a_{12}a_{21} \leq 0. \label{eigenvaluesconditions2}
  \end{subnumcases}
We will prove the following stronger result:?
\begin{subnumcases}{}
a_{11} + a_{22} < 0\textrm{ and}\\ a_{11} a_{22}>a_{12}a_{21}.
  \end{subnumcases}

   Hence, the eigenvalues of the matrix in (\ref{matrixeigenvalue}) will be negative if we show
\begin{align}
-\delta_1  + \beta_1 r_{11} S - \beta_1 r_{11} I_1 -  \beta_1 r_{21} I_2 & < 0 , \label{a11positive1}
\\
-\delta_2  + \beta_2 r_{22} S - \beta_2 r_{22} I_2 -  \beta_2 r_{12} I_1 & < 0 ,\label{a22positive1}
\end{align}
and
\begin{align}
& \quad(-\delta_1  + \beta_1 r_{11} S - \beta_1 r_{11} I_1 -  \beta_1 r_{21} I_2) \nonumber \\ & \quad \times (-\delta_2  + \beta_2 r_{22} S - \beta_2 r_{22} I_2 -  \beta_2 r_{12} I_1)\nonumber \\ & > (- \beta_1 r_{11} I_1  + \beta_1 r_{21} S -  \beta_1 r_{21} I_2)  \nonumber \\ & \quad\times (- \beta_2 r_{22} I_2  + \beta_2 r_{12} S -  \beta_2 r_{12} I_1 ). \label{a12-21}
\end{align}
Given $k_1 \da \frac{I_1}{n}$, $k_2 \da \frac{I_2}{n}$ and $s \da \frac{S}{n}$, we find
\begin{align}
& -\delta_1  + \beta_1 r_{11} S - \beta_1 r_{11} I_1 -  \beta_1 r_{21} I_2 \nonumber \\ & = -\frac{\beta_1n}{\varsigma_1}  + \beta_1 r_{11} s n  - \beta_1 r_{11} k_1 n -  \beta_1 r_{21} k_2 n \nonumber \\ & = n\beta_1\bigg(-\frac{1}{\varsigma_1}+ r_{11} s - r_{11} k_1 -  r_{21} k_2 \bigg) \nonumber \\ & = n\beta_1\bigg[-\frac{1}{\varsigma_1}+ r_{11} (1-2k_1-k_2) - r_{21} k_2 \bigg]. \label{a11}
\\&  -\delta_2    + \beta_2 r_{22} S - \beta_2 r_{22} I_2  -  \beta_2 r_{12} I_1 \nonumber \\ & = -\frac{\beta_2n}{\varsigma_2}  + \beta_2 r_{22} sn - \beta_2 r_{22} k_2 n -  \beta_2 r_{12} k_1 n  \nonumber \\ & = n\beta_2\bigg( -\frac{1}{\varsigma_2} +  r_{22} s - r_{22} k_2  -  r_{12} k_1 \bigg) \nonumber \\ & = n\beta_2\bigg[-\frac{1}{\varsigma_2} +  r_{22} (1-k_1-2k_2)    -  r_{12} k_1 \bigg]. \label{a22}
\\& - \beta_1 r_{11} I_1  + \beta_1 r_{21} S -  \beta_1 r_{21} I_2 \nonumber \\ & = - \beta_1 r_{11} k_1 n  + \beta_1 r_{21} s n -  \beta_1 r_{21} k_2 n \nonumber \\ & = n \beta_1 (-r_{11} k_1 +  r_{21} s -r_{21} k_2) \nonumber  \\ & =  n \beta_1 \big[-r_{11} k_1 +  r_{21} (1-k_1-2k_2) \big]
, \label{a12}
\end{align}
and
\begin{align}
&  - \beta_2 r_{22} I_2  + \beta_2 r_{12} S -  \beta_2 r_{12} I_1  \nonumber \\ & =  - \beta_2 r_{22} k_2 n  + \beta_2 r_{12} s n -  \beta_2 r_{12} k_1 n  \nonumber \\ & = n \beta_2 (-r_{22} k_2 +  r_{12} s - r_{12} k_1 ) \nonumber  \\ & =  n \beta_2 \big[ -r_{22} k_2 +  r_{12} (1-2k_1-k_2)  \big]
. \label{a21}
\end{align}

The Jacobian matrix in (\ref{matrixeigenvalue}) equals
\begin{align}
\begin{bmatrix}
\hspace{-3pt}\begin{array}{l} n\beta_1\big[-{\varsigma_1}^{-1} - r_{21} k_2\\ ~~~~~~~\hspace{-1pt}+r_{11} (1-2k_1-k_2) \big] \end{array} & \hspace{-7pt}\begin{array}{l} n \beta_1 \big[-r_{11} k_1 \\~~~~~~~\hspace{-1pt}+  r_{21} (1-k_1-2k_2)  \end{array} \\[2
       em]
\hspace{-3pt}\begin{array}{l} n \beta_2 \big[ -r_{22} k_2\\ ~~~~~~~\hspace{-1pt}+  r_{12} (1-2k_1-k_2)  \big] \end{array} & \hspace{-7pt}\begin{array}{l}n\beta_2\big[- {\varsigma_2}^{-1} -  r_{12} k_1\\~~~~~~~\hspace{-1pt}+  r_{22} (1-k_1-2k_2) \big] \end{array}
\end{bmatrix}. \label{matrixeigenvaluesimple}
\end{align}
As proved above, given the conditions $r_{11}>0,r_{12}>0,r_{21}>0,r_{22}>0$ enforced in a privacy scheme, there are only two possible equilibrium points: $\langle k_1 = 0, k_2 = 0\rangle$, and $\langle k_1 > 0, k_2 > 0\rangle$. To complete proving Theorem \ref{thm:main:compete-graph}, we show that i) under the equilibrium $\langle k_1 > 0, k_2 > 0\rangle$, for the Jacobian matrix in (\ref{matrixeigenvaluesimple}), the real parts of its eigenvalues are all negative given (\ref{labelk9}); ii) under the equilibrium $\langle k_1 = 0, k_2 = 0\rangle$, for the Jacobian matrix in (\ref{matrixeigenvaluesimple}), there exists an eigenvalue whose real part is non-negative given (\ref{labelk9}). Due to space limitation, we provide the remaining details in the full version \cite{fullversion}.

\end{document}

\begin{itemize}
  \item[i)]
If $k_1=0 $ and $ k_2=0$, then the Jacobian matrix in (\ref{matrixeigenvaluesimple}) becomes
\begin{align}
\begin{bmatrix}
N\beta_1\big(-{\varsigma_1}^{-1}+r_{11}\big) & N\beta_1r_{21} \\[0.7
       em] N\beta_2r_{12} &  N\beta_2\big(-{\varsigma_2}^{-1}+r_{22}\big)
\end{bmatrix}. \label{matrixeigenvaluesimple1}
\end{align}
Given $\big[N\beta_1\big(-{\varsigma_1}^{-1}+r_{11}\big)
-N\beta_2\big(-{\varsigma_2}^{-1}+r_{22}\big)\big]^2+
4N\beta_1r_{21} \cdot N\beta_2r_{12} \geq 0,$ we use (\ref{eigenvaluesconditions1}) to obtain that the real parts of the eigenvalues in  (\ref{matrixeigenvaluesimple1}) are negative if and only if
\begin{align}
\begin{cases}
N\beta_1\big(-{\varsigma_1}^{-1}+r_{11}\big) + N\beta_2\big(-{\varsigma_2}^{-1}+r_{22}\big) <0 ,
\\
N\beta_1\big(-{\varsigma_1}^{-1}+r_{11}\big) \cdot N\beta_2\big(-{\varsigma_2}^{-1}+r_{22}\big)   \\       \quad > N\beta_1r_{21}\cdot N\beta_2r_{12},
\end{cases}\nonumber
\end{align}
which clearly holds if
\begin{align}
\begin{cases}
\varsigma_1 < {r_{11}}^{-1} ,
\\
\varsigma_2 < {r_{22}}^{-1} ,\\
\big( {\varsigma_1}^{-1}-r_{11}\big)
\big( {\varsigma_2}^{-1}-r_{22}\big) > r_{12} r_{21}.
\end{cases}\nonumber
\end{align}
  \item[ii)] If $k_1=0 $, $ k_2=1- \frac{1}{ \varsigma_2 r_{22}}$ and $r_{12} = 0$, then the Jacobian matrix in (\ref{matrixeigenvaluesimple}) becomes
\begin{align}
\begin{bmatrix}
\begin{array}{l}N\beta_1\big[- {\varsigma_1}^{-1} +  \frac{r_{11}}{\varsigma_2 r_{22}}\vspace{3pt} \\~~~~~~~\hspace{-1pt}-  r_{21} \big(1- \frac{1}{ \varsigma_2 r_{22}})  \big] \end{array}& N\beta_1r_{21}\big( \frac{2}{ \varsigma_2 r_{22}}-1\big) \\[0.7
       em] N\beta_2\big( {\varsigma_2}^{-1}-r_{22}\big) &  N\beta_2\big( {\varsigma_2}^{-1}-r_{22}\big)
\end{bmatrix}. \nonumber
\end{align}
  \item[iii)] If $k_1=1- \frac{1}{ \varsigma_1 r_{11}} $, $ k_2=0$ and $r_{21} = 0$, then the Jacobian matrix in (\ref{matrixeigenvaluesimple}) becomes
\begin{align}
\begin{bmatrix}
N\beta_1\big( {\varsigma_1}^{-1}-r_{11}\big) & N\beta_1\big( {\varsigma_1}^{-1}-r_{11}\big) \\[0.7
       em] N\beta_2r_{12}\big( \frac{2}{ \varsigma_1 r_{11}}-1\big) &  \begin{array}{l}N\beta_2\big[- {\varsigma_2}^{-1} +  \frac{r_{22}}{\varsigma_1 r_{11}} \vspace{3pt}\\~~~~~~~\hspace{-1pt}-  r_{12} \big(1- \frac{1}{ \varsigma_1 r_{11}})  \big] \end{array}
\end{bmatrix}.\nonumber
\end{align}
\item If $k_1 > 0$ and $k_2 > 0$, we obtain from (\ref{delta1k1}) and (\ref{delta2k2}) that
\begin{align}
  {1}/{\varsigma_1}   & =  {1}/{k_1}\cdot (1-k_1-k_2) ( r_{11}k_1 + r_{21}k_2) , \label{delta1k1abbrvsigma1} \\
 {1}/{\varsigma_2}   & =  {1}/{k_2} \cdot (1-k_1-k_2) ( r_{12} k_1 + r_{22} k_2). \label{delta2k2abbrvsigma2}
\end{align}
Then the Jacobian matrix in (\ref{matrixeigenvaluesimple}) becomes
\begin{align}
\begin{bmatrix}
N \beta_1 \cdot \frac{-r_{11}{k_1}^2 - k_2 + r_{21}{k_2}^2}{k_1} & N \beta_1 \cdot \begin{array}{l}[-r_{11}{k_1} \\+ r_{21}(1-k_1-2k_2)]\end{array} \\[0.7
       em] N \beta_2 \cdot \begin{array}{l}[-r_{22}{k_2} \\+ r_{12}(1-2k_1-k_2)]\end{array} &  N \beta_2 \cdot \frac{-r_{22}{k_2}^2 - k_1 + r_{12}{k_1}^2}{k_2}
\end{bmatrix}.\nonumber
\end{align}
\end{itemize}
In each case, we compute the discriminant of the quadratic  characteristic equation of the  Jacobian matrix.

\end{document}